\documentclass[%
aps,
jmp,%
amsmath,amssymb,
reprint,%
]{revtex4-2}
\usepackage{graphicx}% Include figure files
\usepackage{subfigure}
\usepackage{dcolumn}% Align table columns on decimal point
\usepackage{bm}% bold math
\usepackage[mathlines]{lineno}% Enable numbering of text and display math
\usepackage{graphicx}
\usepackage{graphicx}% Include figure files
\usepackage{subfigure}
\usepackage{dcolumn}% Align table columns on decimal point
\usepackage{bm}% bold math
\usepackage[mathlines]{lineno}% Enable numbering of text and display math
\graphicspath{{figure/}}
\usepackage{array}
\usepackage{braket}
\usepackage{diagbox}
\usepackage{amsmath}
\usepackage{appendix}
\usepackage{multirow}
\usepackage{bm,color,bbm}

\usepackage{float}
\newcolumntype{.}{D{.}{.}{-1}}
\usepackage{amsopn}
\usepackage[colorlinks,linkcolor=blue, anchorcolor=blue, urlcolor=blue, citecolor=blue]{hyperref}
\usepackage{epstopdf}
\allowdisplaybreaks[1]
\usepackage{xcolor}
\usepackage{amsmath}

\begin{document}

	\title {Practical asynchronous measurement-device-independent quantum key distribution with advantage distillation}% Force line breaks with \\
	\author{Di Luo$^{1}$, Xin Liu$^{1}$, Kaibiao Qin$^1$, Zhenrong Zhang$^{2}$, and Kejin Wei$^{1,*}$}
	\address{
		$^1$Guangxi Key Laboratory for Relativistic Astrophysics, School of Physical Science and Technology,
		Guangxi University, Nanning 530004, China\\		$^2$Guangxi Key Laboratory of Multimedia Communications and Network Technology, School of Computer Electronics and Information, Guangxi University, Nanning 530004, China\\
		$^*$Corresponding author: kjwei@gxu.edu.cn
	}

		\begin{abstract}

The advantage distillation (AD) method has proven effective in improving the performance of quantum key distribution (QKD). In this paper, we introduce the AD method into a recently proposed asynchronous measurement-device-independent (AMDI) QKD protocol, taking finite-key effects into account. Simulation results show that the AD method significantly enhances AMDI-QKD, e.g., extending the transmission distance by 16 km with a total pulse count of $N=7.24\times10^{13}$, and enables AMDI-QKD, previously unable to generate keys, to generate keys with a misalignment error rate of 10$\%$. As the AD method can be directly integrated into the current system through refined post-processing, our results facilitate the practical implementation of AMDI-QKD in various applications, particularly in scenarios with high channel losses and misalignment errors.
			
    	\end{abstract}
	    \maketitle
	    \section{INTRODUCTION}

	    Quantum key distribution (QKD) enables two legitimate users to share an unconditionally secure key, even in the presence of eavesdroppers with unlimited computational power and storage capacity. Since the proposal of the first QKD by Bennett and Brassard in 1984~\cite{1984bennett}, and through the extensive efforts of researchers, QKD has achieved significant milestones both theoretically and experimentally~\cite{2005lo_decoy,2005wang_beating,2014Lim,2018Boaron,2019Dynes,2020Gr,2021Madi-BB84,2022Scalcon,2023LI_Wei,2023wei}. It is poised to become a crucial technology ensuring secure communication in the future.

		However, due to the gap between the ideal model and practical devices, especially at the detector side~\cite{2014Lo,2020Xufeihu-Review}, eavesdroppers can exploit these gaps to steal information~\cite{2010Lydersen,2019Wei-Attack,2020Ye,2023Acheva,2014Tamaki,2022Anastasiya,2022chen_ye,2023Acheva}. Fortunately, measurement-device-independent quantum key distribution (MDI-QKD) was proposed~\cite{2012Lo,2012Braunstein}, addressing detection loopholes by utilizing Bell state measurements. Currently, most MDI-QKD systems ~\cite{2016Yin,2018Liu,2020Wei-MDI,2020Cao,2021Fan,2021Woodward,2022Lu_MDI,2022Gu,2023Liu_MDI} require strict coincidence detection for key generation, limiting the key rate and transmission distance of MDI-QKD and preventing it from surpassing the repeaterless rate-transmittance bound~\cite{2014Takeoka,2017Pirandola}.
	    
		The twin-field quantum key distribution (TF-QKD)~\cite{2018Lucamarini}, based on the single-photon interference concept, eliminates the need for coincidence detection, surpassing the Pirandola-Laurenza-Ottaviani-Banchi (PLOB) bound~\cite{2017Pirandola}. Various TF-QKD protocol variants have been proposed~\cite{2018Ma,2018Wang,2019Curty,2019Wangshuang-TF}, with experimental demonstrations reported~\cite{2020fang_PM,2021Pittaluga,2022wang,2022Chen,2022Chen,2023Zhou_TF,2023Li_TF,2023Liu}. However, TF-QKD necessitates stringent phase and frequency locking techniques to stabilize the fluctuation between two coherent states, significantly increasing the system's cost and complexity.

 		To address these challenges, two recent works, namely asynchronous MDI-QKD (AMDI-QKD)~\cite{2022Xie} or mode-pairing QKD (MP-QKD)~\cite{2022Zeng}, were proposed almost simultaneously. Using an asynchronous coincidence method, both protocols can surpass the PLOB bound without requiring phase locking and phase tracking techniques. The practicality of these MDI-type QKD protocols has been demonstrated~\cite{2023Zhu,2023Zhou}. In particular, Zhou et al.~\cite{2023Zhou} achieved a breakthrough in the transmission distance of MDI-QKD, extending it from 404 km~\cite{2016Yin} to 508 km over fiber. Furthermore, theoretical developments have been reported~\cite{2023wang,2023Bai,2023Xie,2023Liu_x}. Much effort has been put into further extending the distance~\cite{2016Yin,2022wang,2023Liu,2021Chen_Yu}, which is a Holy Grail in communication.
 		
		 Advantage distillation (AD), proposed by Maurer~\cite{1993Maurer}, is a classical two-way communication protocol used to enhance the error tolerance~\cite{2003Gottesman,2006Ma,2017Khatri}. The core step of the AD method involves dividing raw keys into a few blocks to extract highly correlated keys from weakly correlated keys, thereby increasing the correlation between the raw keys. The AD method has been applied to various QKD protocols~\cite{2008Renner,2020tan,2023Hu,2023ZhuRFI,2023Jiang,2022Li,2022LITF,2022Wang_R,2023Liu_x,2023zhang,2024ZhouYao} to extend the transmission distance and increase the maximum tolerance of background noise. Importantly, the AD method has been shown to improve performance when considering  statistical fluctuations~\cite{2023Hu,2023ZhuRFI,2023Jiang}.  Recently, Liu et al.~\cite{2023Liu_x} demonstrated that the AD method can significantly enhance the performance of MP-QKD.  However, whether AD can improve the performance of AMDI-QKD, especially when accounting for finite-key effects, remains unknown.
  
	\begin{figure*}[hbt]
	\centering
	\includegraphics[width=1.6\columnwidth]{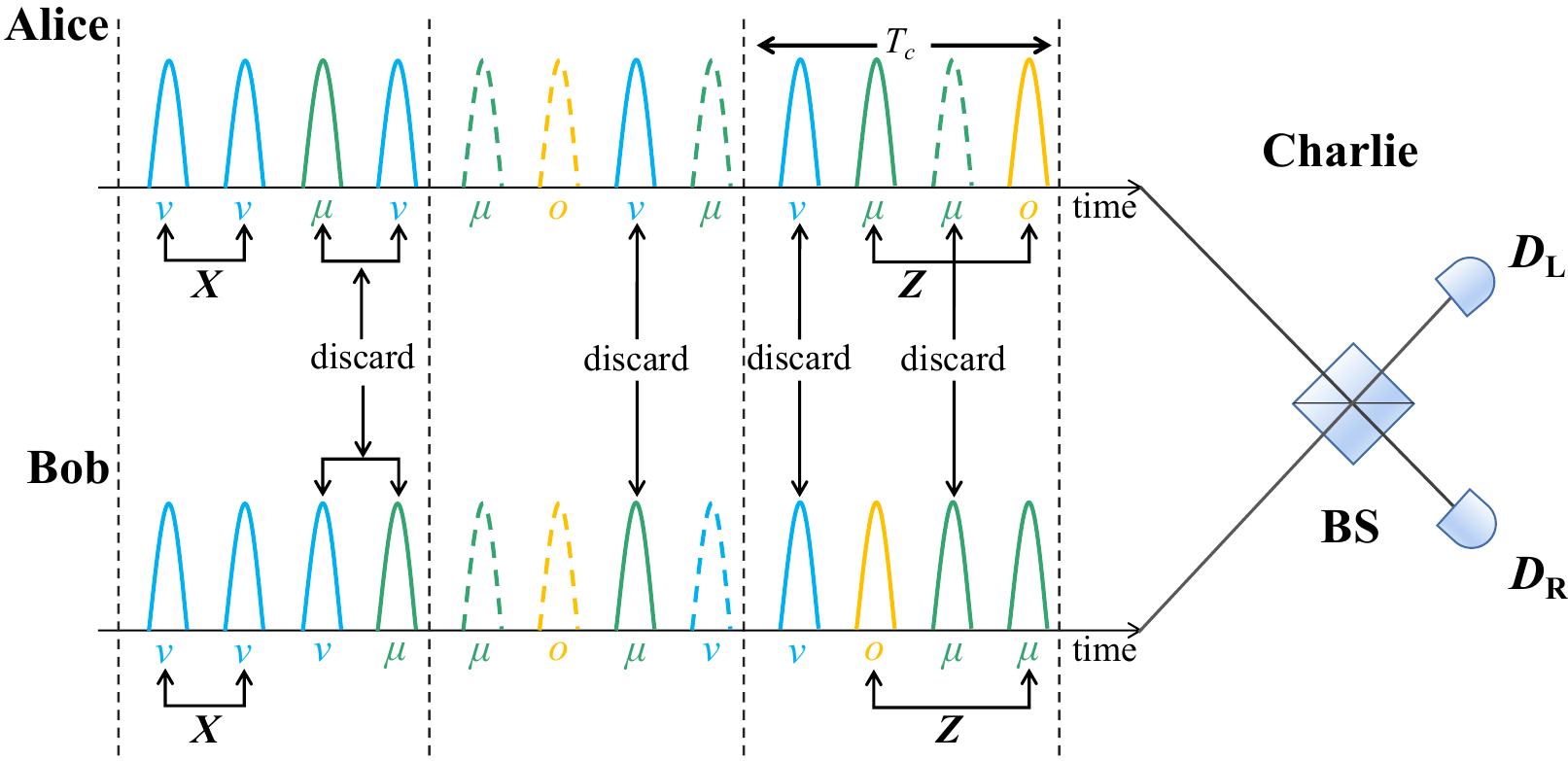}
	\caption{Schematic diagram of the AMDI-QKD protocol by using click filtering operation. Alice and Bob send the prepared state to Charlie for interference measurement. In the maximum pairing interval $T_{c}$, for any successful coincidence, if the total pairing intensity $\kappa^{tot}_{a(b)}$ = $\mu$ as the $Z$ basis, if the total pairing intensity $\kappa^{tot}_{a(b)}$= $2\nu$ as the $X$ basis, and if the total pairing intensity  $\kappa^{tot}_{a(b)}\ge\mu+\nu$, then coincidence is discarded. If the click events are $(\nu_{a},\mu_{b})$ or $(\mu_{a},\nu_{b})$, these click events are discarded. If one or both Alice and Bob do not click in the same time bin, the corresponding data is discarded. BS is beam splitter. $D_{L}\text{ }(D_{R})$ represents left (right) detector.
	}
	\label{liu}
\end{figure*}

In this paper, building upon the work in Ref.~\cite{2023Liu_x}, we incorporate the AD method into AMDI-QKD~\cite{2023Zhou}. Additionally, we analyze its performance while considering finite-key effects. Using typical experimental parameters of AMDI-QKD, the simulation results demonstrate a significant improvement in the key rate, transmission distance, and enhance tolerance to misalignment errors when accounting for finite-key effects.

The remainder of the article is structured as follows: In Sec.~\ref{AMDI protocol}, we briefly introduce the original AMDI-QKD protocol. Moving on to Sec.~\ref{AMDI-QKD with Advantage distillation}, we analyze the post-processing step using the AD method. Subsequently, in Sec.~\ref{simulation}, we present simulation results for a more practical AMDI-QKD model, comparing the performance of AMDI-QKD with and without AD. Finally, we discuss and summarize our findings in Sec.~\ref{conclusion}.
    	
    	\section{AMDI-QKD protocol}
    	\label{AMDI protocol}

    Here, we provide a brief overview of the AMDI-QKD protocol~\cite{2022Xie,2023Zhou}, using a three-intensity decoy-state scheme as an example. The schematic diagram of the scheme is shown in Fig.~\ref{liu}, and the detailed process is as follows:

   {$Step$} $\emph{1}$. Preparation and Measurement: This step is repeated for $N$ rounds to accumulate sufficient data. For each round or time bin $i\in\left\{1,2,\dots, N\right\}$, Alice randomly prepares a weak coherent state pulse $|{e}^{\textbf{i}\theta_{a}^i}\sqrt{\kappa_{a}^i}\rangle$ with intensities  $\kappa_{a}^i\in \left\{\mu_{a},\nu_{a},o_{a} \right\}$ ($\mu_{a}>\nu_{a}>o_{a}$=0), and the corresponding probabilities are $p_{\mu_{a}}$, $p_{\nu_{a}}$ and 1-$p_{\mu_{a}}$-$p_{\nu_{a}}$. In this context, the random phase $\theta_{a}^i=2\pi M^{i}_{a}/M$ with $M^{i}_{a}\in \left\{0,1,\dots,M-1\right\}$. Similarly, Bob prepares weak coherent pulses $|{e}^{\textbf{i}\theta^{i}_{b}}\sqrt{\kappa^{i}_{b}}\rangle$, where $\kappa^{i}_{b}\in \left\{\mu_{b},\nu_{b},o_{b} \right\}$ using the same operation as Alice. Next, Alice and Bob send their prepared optical pulses to Charlie through the quantum channel. Charlie performs the interference measurement and then announces which detector ($D_{L}$ or $D_{R}$) clicked and the corresponding time-bin $i$.

   {$Step$} $\emph{2}$. Click Filtering: For each click, Alice and Bob announce whether they had sent a decoy-state pulse of intensity $\nu_{a}$ ($\nu_{b}$). If the click event is ({$\mu$}$_{a}$$|${$\nu$}$_{b}$) and ({$\nu$}$_{a}$$|${$\mu$}$_{b}$), a click filtering operation is applied to discard this click event. All other click events are kept. Here, we use  $(\kappa_{a}$$|$ $\kappa_{b})$  to denote a successful click given that Alice and Bob sent pulse intensities of $\kappa_{a}$ and $\kappa_{b}$.

   	{$Step$} $\emph{3}$. For all kept clicks, Alice and Bob pair two adjacent clicks within the time interval $T_c$ to form a successful coincidence. If the partner cannot be found within the maximum pairing interval $T_{c}$, Alice and Bob discard the corresponding lone click. For all successful pairing coincidences, Alice and Bob calculate the total intensity [{$\kappa$}${^e_a}$$+${$\kappa$}${^l_a}$=${\kappa}{^{tot}_a}$, {$\kappa$}${^e_b}$$+${$\kappa$}$^{l}_{b}$={$\kappa$}${^{tot}_b}$] and the phase difference $\varphi_{a(b)}=\theta^{l}_{a(b)}-\theta^{e}_{a(b)}$ between two time bins, where superscript  `$e$' (`$l $') denotes the early (late) time bin, and {$\kappa$}${^{tot}_a}$ ({$\kappa$}${^{tot}_b}$) denotes the total intensity of the pairing coincidence,  the notation [$\kappa^{tot}_{a}$,$\kappa^{tot}_{b}$]  denotes paring coincidence where the total intensity in the two time bins Alice (Bob) is  $\kappa{^{tot}_a}$ ($\kappa{^{tot}_b}$).

   {$Step$} $\emph{4}$. Sifting: Alice and Bob publish the computational results of all successful coincidences and discard the events if {$\kappa$}${^{tot}_{a(b)}}\geq{\mu}_{a(b)}+{\nu}_{a(b)}$. For [$\mu_{a}$,$\mu_{b}$] coincidence, if Alice (Bob) sends intensity $\mu_{a}$ ($o_{b}$) in the early time-bin and sends intensity $o_{a}$ ($\mu_{b}$) in the late time bin, they obtain $Z$ basis bit 0. Otherwise, bit 1 is obtained. For [$2\nu_{a}$,$2\nu_{b}$] coincidence, Alice and Bob calculate the relative phase difference $\varphi_{ab}=(\varphi_{a}-\varphi_{b})$ mod $2\pi$. If $\varphi_{ab}=0$ or $\varphi_{ab}=\pi$, they extract $X$ basis bit 0. As an extra step on the $X$ basis, if $\varphi_{ab}=0$ and both detectors are clicked or $\varphi_{ab}=\pi$ and the same detector clicked twice, Bob flips his bit value. The coincidence with other phase differences $(\varphi_{ab}\ne 0 \text{ or }  \pi)$ is discarded.

    	{$Step$} $\emph{5}$. Postprocessing: Alice and Bob assign their data to different datasets $S_{[{\kappa}{^{tot}_a},{\kappa}{^{tot}_b}]}$ and count the corresponding amount $n_{[{\kappa}{^{tot}_a},{\kappa}{^{tot}_b}]}$.
    	Then, they respectively generate raw keys $\mathcal{Z}$ by using data $n_{[\mu_{a},\mu_{b}]}$ from $S_{[{\mu_a},{\mu_b}]}$.
    	The secret keys are obtained through error correction and privacy amplification with a security bound $\varepsilon _{sec}$.
    	
     After the above steps, the final key rate of AMDI-QKD with considering finite-key effects can be expressed as:
    		\begin{equation}
    			\begin{aligned} 
    		R&=\frac{1}{N}\left\{\underline{s}^{z}_{0}+\underline{s}^{z}_{11}\left [1-H(\overline{\phi}^{z}_{11})\right ]-\lambda_{EC}  \right. \\
    		&\left.-\text{log}_{2}\frac{2}{\varepsilon_{\text{cor}}}-2\text{log}_{2}\frac{2}{\varepsilon^{\prime}\hat{\varepsilon}}-2\text{log}_{2}\frac{1}{2\varepsilon_{\text{PA}}}   \right\},
    \end{aligned}
	\end{equation} 
     where $N$ is the total number of pulses, $\underline{s}^{z}_{0}$ is the lower bound of the estimated number of vacuum events, ${s}^{z}_{11}$ is event number of single-photon pair,  ${\phi}^{z}_{11}$ is phase error rate of single-photon pair in the $Z$ basis. The underline and  overline of the parameters denote the lower  and upper bounds, respectively. $\lambda_{EC}=n_{[\mu_{a},\mu_{b}]}fH(E_{z})$ represents the maximum amount of information stolen by Eve during error correction step, $f$ is  correction efficiency, $E_{z}$ is quantum bit error rate (QBER), $n_{[\mu_{a},\mu_{b}]}$ is total successful paring number in $Z$ basis. $\varepsilon_{\text{cor}}$ implies the failure probability in the error correction, $\varepsilon_{\text{PA}}$  denotes the failure probability in the privacy amplification. $\varepsilon^{\prime}$ and  $\hat{\varepsilon}$ are coefficients after using smooth entropies. The specific calculation of the parameters is in Appendix ~\ref{Appendix A}.   	
          \begin{figure*}[hbt]  	
     	\vspace{0.5cm} 
     	\subfigbottomskip=2pt
     	\subfigcapskip=1pt 
     	\subfigure[]{
     		\label{AM1}
     		\includegraphics[width=0.475\linewidth]{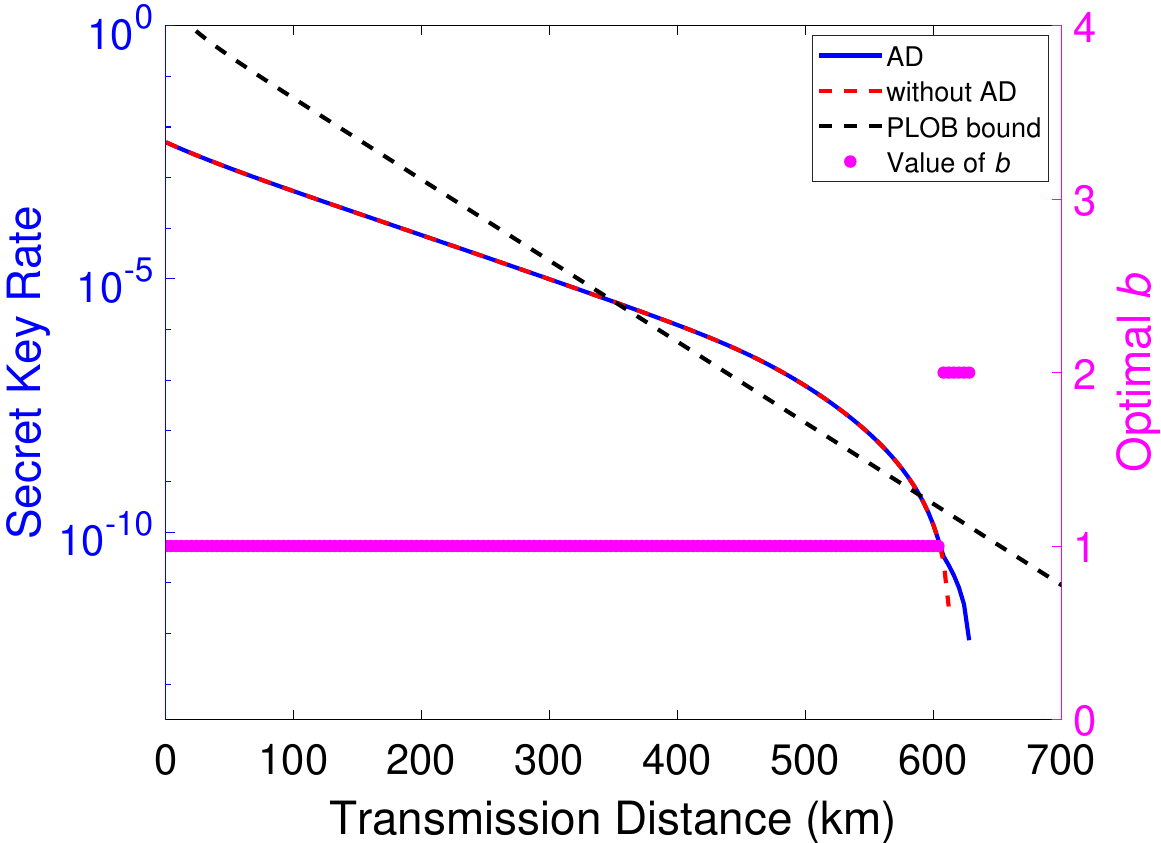}}
     	\quad 
     	\subfigure[]{
     		\label{AM2}
     		\includegraphics[width=0.485\linewidth]{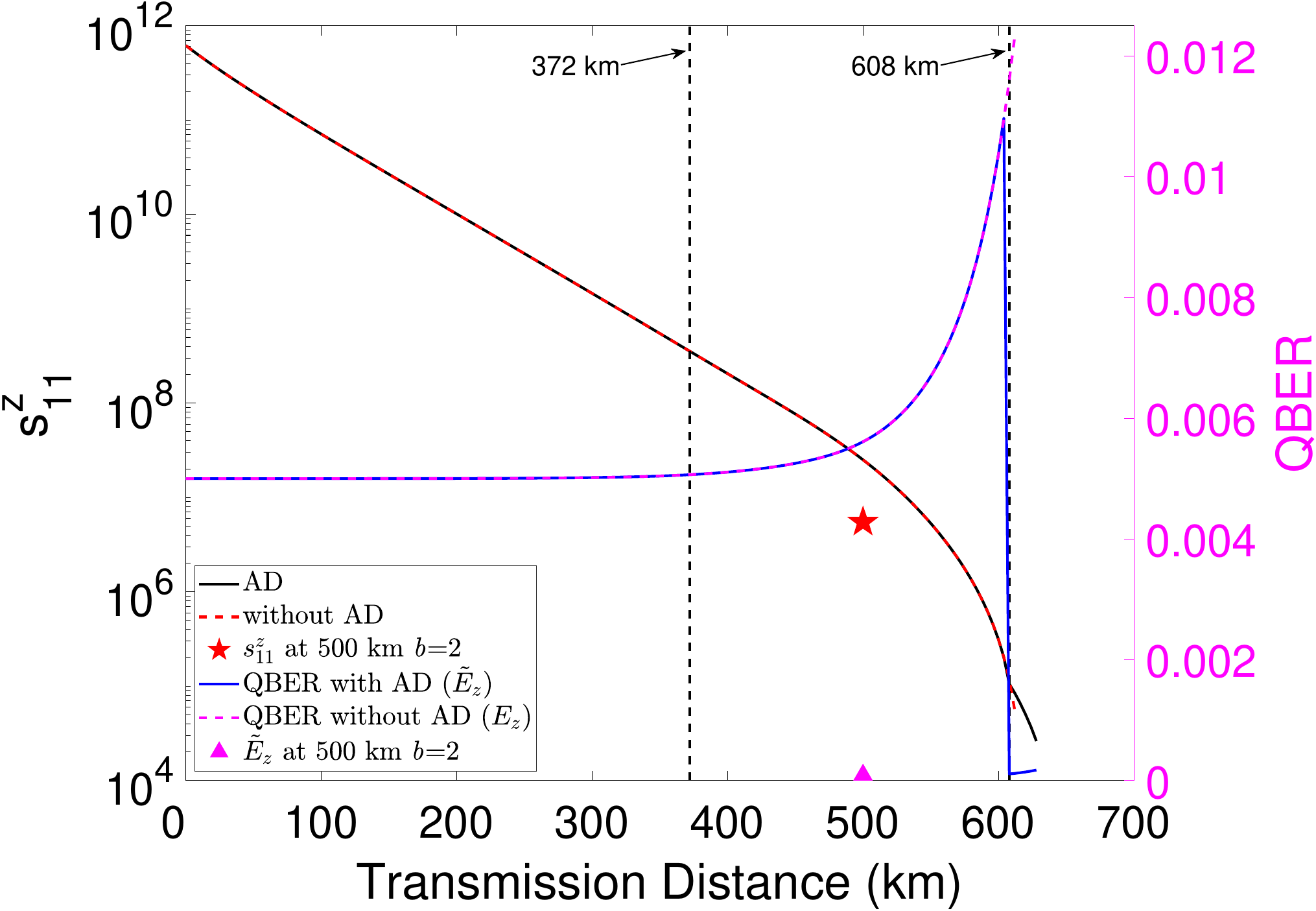}}   	     	
     	\caption{Comparison of AMDI-QKD performance with and without AD.  (a) The relationship between secret key rate and transmission distance, and the corresponding optimal value of $b$. The blue solid line and red dotted line represent the secret key rate of AMDI-QKD with AD and without AD, respectively. The black dashed line is the PLOB bound, and the pink scatter represents the optimal $b$ value. (b) The relationship between $s_{11}^z$ and transmission distance, and the corresponding QBER. The black solid and red dotted lines represent the  $s_{11}^z$ of AMDI-QKD with AD and without AD, respectively. The blue solid  and  pink dotted  lines represent the QBER of AMDI-QKD with AD and without AD, respectively. The red five-pointed star and pink  triangles are $s_{11}^z$ and QBER at 500 km and $b$=2, respectively. Here, we set  $ N=7.24\times10^{13} $, $ e^{z}_{d}=0.5\% $, and $ E_{\text{Hom}}=4\% $.}
     	\label{AM}     	
     \end{figure*}

    	\section{AMDI-QKD with Advantage distillation}
    	\label{AMDI-QKD with Advantage distillation}
     As a post-processing method, AD only changes the step of the data processing to improve the performance, Thus AMDI-QKD with AD is different from the original AMDI-QKD only in $step$ $\emph{5}$. The new $step$  $\emph{5}$ is as follows:
  
  New $step$ $\emph{5}$: After obtaining the raw key, Alice and Bob divide their raw key into $b$ bit blocks $\left\{x_{1}, x_{2}, \cdots,  x_{b}\right\}$ and $\left\{y_{1}, y_{2}, \cdots,  y_{b}\right\}$. Alice chooses a local random bit $c \in \left\{0, 1\right\}$ and sends the messages $m=\left\{m_{1}, m_{2}, \cdots, m_{b}\right\}=\left\{x_{1}\oplus c, x_{2}\oplus c, \cdots, x_{b}\oplus c\right\}$ to Bob. Alice and Bob only keep the block if Bob calculates the results of $\left\{m_{1}\oplus y_{1}, m_{2}\oplus y_{2}, \cdots, m_{b}\oplus y_{b}\right\}=\left\{0, 0, \cdots, 0\right\}$ or $\left\{1, 1, \cdots, 1\right\}$, then retain the first bit, $x_{1}$ and $y_{1}$, as the raw key. It is noteworthy that if the block size 
 $ b $ is 1, it means that the AD step has not been executed.  Finally, the final keys are obtained through error correction and privacy amplification.
      \begin{figure}[hbt]
  	\centering
  	\includegraphics[width=\linewidth]{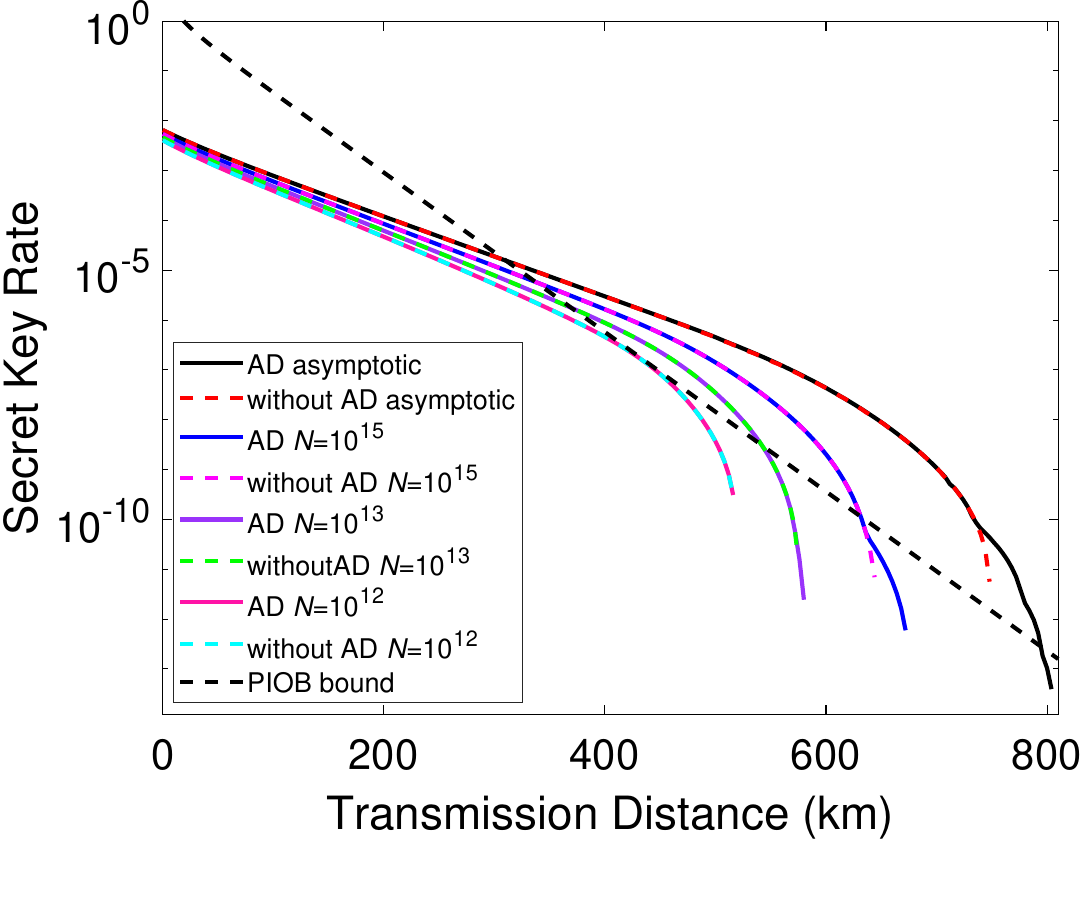}
  	\caption{Performance comparison between AMDI-QKD with AD and without AD under different pulses. We simply fix numbers of pulse $10^{12}$, $10^{13}$, $10^{15}$ and asymptotic case, the misalignment error rates are $E_\text{{Hom}}$=4$\%$, $e^{z}_{d}=0.05\%$. The solid and dotted colorful curves represent the results of AMDI-QKD with and without AD, respectively. The optimal value of $b$ of the overlapping part by the solid line and the dotted line is 1, and the remaining solid line part is $b > 1$.
  	}
  	\label{N.jpg}
  \end{figure}

  To gain deeper insights into the enhanced key rate achieved through the AD method, we employ the information security theoretical analysis method to reassess the key rate of AMDI-QKD. The detailed analysis is provided in Appendix~\ref{Appendix B}. The key rate of AMDI-QKD is reformulated as follows:
  
  \begin{equation}
  	\begin{aligned}      	    		
  		R&\geq\min _{\lambda_{1}, \lambda_{2}, \lambda_{3}, \lambda_{4}}
  		\frac{1}{N}n_{z}\left\{\frac{\underline{s}^{z}_{0}}{n_{z}}+\frac{\underline{s}^{z}_{11}}{n_{z}}\bigg[1-\right. \bigg.  \\
  		&\bigg. \left(\lambda_{1}+\lambda_{2}\right) H\left(\frac{\lambda_{1}}{\lambda_{1}+\lambda_{2}}\right)-\left(\lambda_{3}+\lambda_{4}\right) H\left(\frac{\lambda_{3}}{\lambda_{3}+\lambda_{4}}\right)\bigg ]\\ 
  		&\left. -fH\left(E_{z}\right)-\frac{1}{n_{z}}\left (\text{log}_{2}\frac{2}{\varepsilon_{\text{cor}}}+2\text{log}_{2}\frac{2}{\varepsilon^{\prime}\hat{\varepsilon}}+2\text{log}_{2}\frac{1}{2\varepsilon_{\text{PA}}}\right )\right\},
  	\end{aligned}	
  \end{equation}
  where $\sum_{i=1}^{4} \lambda_{i}=1$ and $\lambda_{i} (i=1, 2, 3, 4)$ satisfy following relationships with the error rates:  $\underline{\phi}^{z}_{11}\le\lambda_{2}+\lambda_{4}\le	\overline{\phi}^{z}_{11}$,  $\underline{e}^{z}_{11}\le\lambda_{3}+\lambda_{4}\le\overline{e}^{z}_{11}$.
  
       After postprocessing using the AD method (new step 5),  and the key rate of AMDI-QKD can be rewritten  as follows (The detailed can be found in Appendix~\ref{Appendix C}):
      	\begin{equation}
    	\begin{aligned}
    		\tilde{R} \geq & \max _{b} \min _{\lambda_{1}, \lambda_{2}, \lambda_{3}, \lambda_{4}} \frac{1}{N}\frac{n_{z}}{b}q_{\text {succ}}\\ &\left\{{\left(\frac{\underline{s}^{z}_{0}}{n_{z}}\right)}^{b}+{\left(\frac{\underline{s}^{z}_{11}}{n_{z}}\right)}^{b}\left[1-\left(\tilde{\lambda}_{1}+
    		\tilde{\lambda}_{2}\right) 
    		H\left(\frac{\tilde{\lambda}_{1}}{\tilde{\lambda}_{1}+\tilde{\lambda}_{2}}\right)\right.\right. \\
    	&  \left. -\left(\tilde{\lambda}_{3}+\tilde{\lambda}_{4}\right) H\left(\frac{\tilde{\lambda}_{3}}{\tilde{\lambda}_{4}+\tilde{\lambda}_{4}}\right)\right]-fH\left(\tilde{E}_{z}\right)\\
    	&\left.-\frac{b}{n_{z}q_{\text{succ}}}\left (\text{log}_{2}\frac{2}{\varepsilon_{\text{cor}}}+2\text{log}_{2}\frac{2}{\varepsilon^{\prime}\hat{\varepsilon}}+2\text{log}_{2}\frac{1}{2\varepsilon_{\text{PA}}}\right )\right\},
    	\end{aligned}
    	\label{SKR}
    \end{equation}
    subject to  
     				\begin{equation}
     				\begin{split}	  
     {\underline{\phi}^{z}_{11}}& {\le\lambda_{2}+\lambda_{4}\le 	\overline{\phi}^{z}_{11}},  \\  
     			{\underline{e}^{z}_{11}}& {\le\lambda_{3}+\lambda_{4}\le\overline{e}^{z}_{11}},\\ 
     			q_{\text {succ }}&={\left(E_{z}\right) }^{b}+{\left(1-E_{z}\right)}^{b},\\
     				\tilde{E}_{z}&=\frac{\left(E_{z}\right )^{b}}{{\left(E_{z}\right) }^{b}+{\left(1-E_{z}\right)}^{b}},
     			  	\end{split}
     		\end{equation}
     		and
	\begin{equation}
	\begin{split} 		
		\tilde{\lambda}_{1}&=\frac{\left(\lambda_{1}+\lambda_{2}\right)^{b}+\left(\lambda_{1}-\lambda_{2}\right)^{b}}{2p_{\text{succ}}}, \\
		\tilde{\lambda}_{2}&=\frac{\left(\lambda_{1}+\lambda_{2}\right)^{b}-\left(\lambda_{1}-\lambda_{2}\right)^{b}}{2p_{\text{succ}}},\\  	
		\tilde{\lambda}_{3}&=\frac{\left(\lambda_{3}+\lambda_{4}\right)^{b}+\left(\lambda_{3}-\lambda_{4}\right)^{b}}{2 p_{\text{succ}}}, \\
		\tilde{\lambda}_{4}&=\frac{\left(\lambda_{3}+\lambda_{4}\right)^{b}-\left(\lambda_{3}-\lambda_{4}\right)^{b}}{2p_{\text{succ}}},
	\end{split}
\end{equation}
	where $\underline{\phi}^{z}_{11}$ ($\underline{e}^{z}_{11}$) and $\overline{\phi}^{z}_{11}$ ($\overline{e}^{z}_{11}$) denote the lower and upper bounds of the error rates of $\phi^{z}_{11}$ ($e^{z}_{11}$)~\cite{2023Zhou,2023Xie}, which can be estimated by the decoy state, $ p_{\text{succ}}=\left(\lambda_{1}+\lambda_{2}\right)^{b}+\left(\lambda_{3}+\lambda_{4}\right)^{b}$, $q_{\text{succ}}$ and $\tilde{E}_{z}$ respectively represent successful probability to perform the AD method and total error rate after AD postprocessing in $Z$ basis. 

		\begin{table}[h] \tiny
		\centering
		\caption{ Numerical simulation parameters. $\eta^{D_{L}}_{d}$ $(\eta^{D_{R}}_{d}$) and $p^{D_{L}}_{d}$ $(p^{D_{R}}_{d})$ respectively denotes the detection efficiency and dark count rate, $D_{L}\text{ }(D_{R})$ is left (right) detector. $\alpha(dB/km)$ and $f$ denotes loss coefficient of the fiber and error-correction efficiency, respectively. $\epsilon$ is the failure probability. The insert loss is 1.50 dB on Charlie’s side.}
		\resizebox{\linewidth}{!}{
			\begin{tabular}{cccccccccc}
				\hline\hline 
				$\eta^{D_{L}}_{d}$ &$\eta^{D_{R}}_{d}$ & $p^{D_{L}}_{d}$ & $p^{D_{R}}_{d}$ & $\alpha$ & $\epsilon $ & $f$     \\  \hline 
				78.1\% & 77\%  & 3.03$\times 10^{-9}$ & 3.81$\times 10^{-9}$    & 0.16 & $10^{-10}$  & 1.1    \\  
				\hline\hline 
			\end{tabular}
			\label{table1}
		}
	\end{table}
	
 Equation~(\ref{SKR}) can be understood in two perspectives. Firstly, the quantum channel is manipulated by Eve, who can select the optimal parameter $\lambda _{i}\left ( i=1,2,3,4 \right ) $ to diminish the key rate. Secondly, the AD method is governed by Alice and Bob, affording them the capability to choose the optimal value of $b$ to increase the key rate. Furthermore, Alice and Bob can optimize value of $b$ in order to enhance  the successful probability $ q_{\text{succ}}$. Consequently, the error rate in the $Z$ basis can be changed from $E_z$ to $ \tilde{E}_{z}=\frac{(E_z)^b}{q_{\text{succ}}}$, the number of raw keys and single-photon bits retained by Alice and Bob are $n_{z}q_{\text{succ}}/b$ and $\left (\underline{s}^{z}_{11}/n_{z}\right )^{b} n_{z}q_{\text{succ}}/b$, respectively. 

  \begin{figure*}[hbt]
  	\vspace{0.5cm} 
  	\centering
  	\subfigbottomskip=2pt
  	\subfigcapskip=1pt 
  	\subfigure[]{
  		\label{ed}
  		\centering
  		\includegraphics[width=0.48\linewidth]{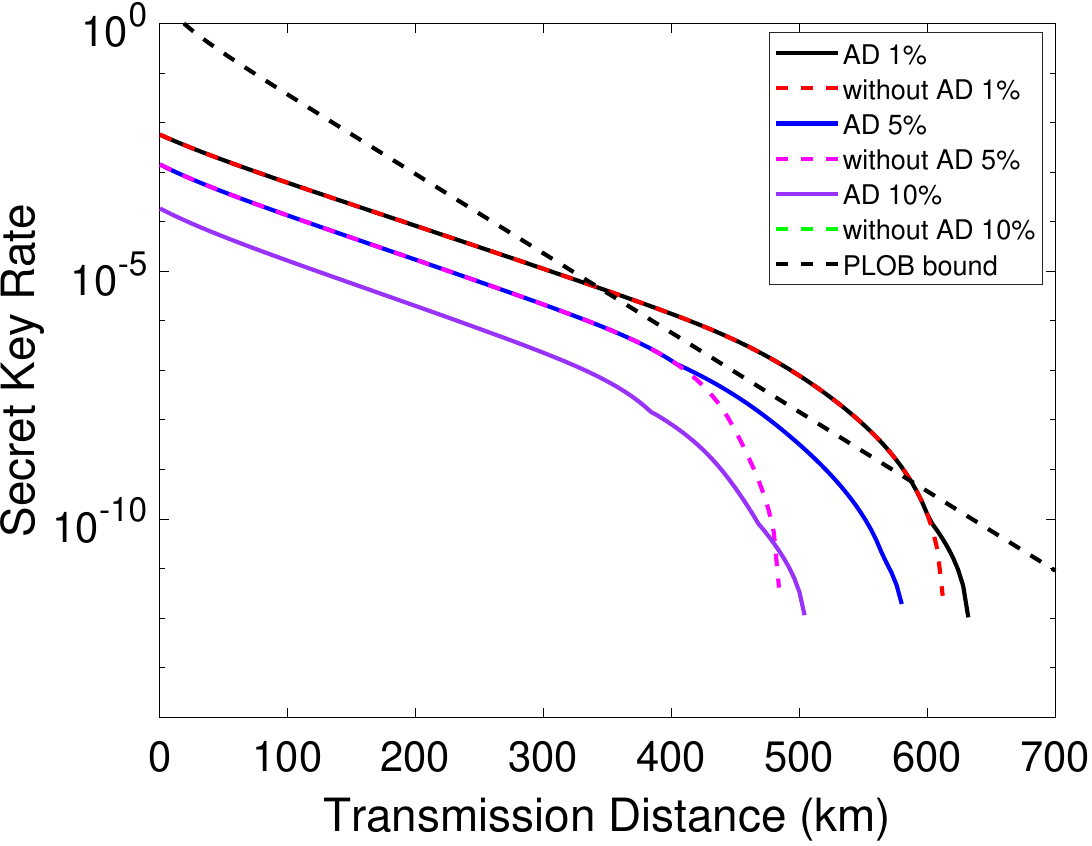}}
  	\quad 
  	\subfigure[]{
  		\label{b}
  		\centering
  		\includegraphics[width=0.45\linewidth]{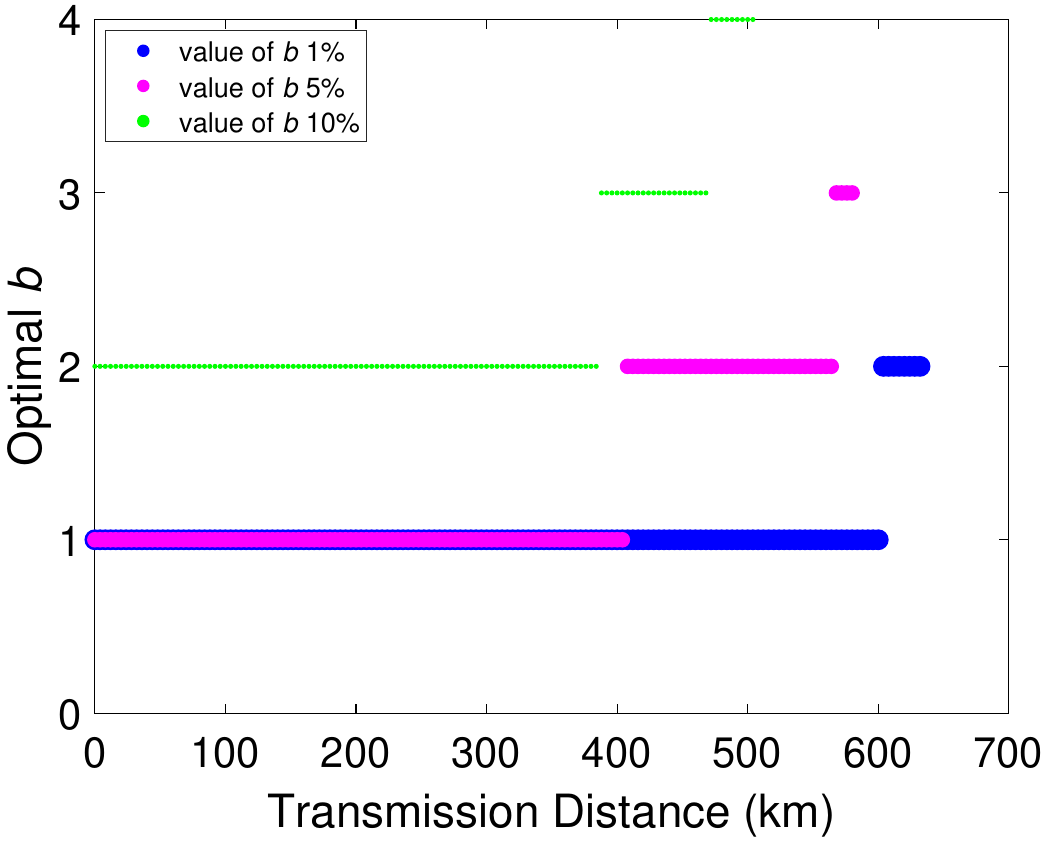}}\caption{Performance  of AMDI-QKD with AD  under different  misalignment error rate. Here, we assume $e^{z}_{d}=E_\text{{Hom}}$=$ 1\%$ ($5\%$, $10 \%$) and $N=7.24\times10^{13}$. (a)  The relationship between key rate and distance under different  misalignment error rate $e^{z}_{d}$ and $E_\text{{Hom}}$. The solid line indicates AMDI-QKD with AD, and the dashed line indicates AMDI-QKD without AD. (b)  The relationship between the optimal $b$ value and transmission distance.}
  	\quad 
  	\label{ded.jpg}     	   	
  \end{figure*}

\begin{figure*}[hbt]  	
	\vspace{0.5cm} 
	\subfigbottomskip=2pt
	\subfigcapskip=1pt 
	\subfigure[]{
		\label{Different1}
		\includegraphics[width=0.482\linewidth]{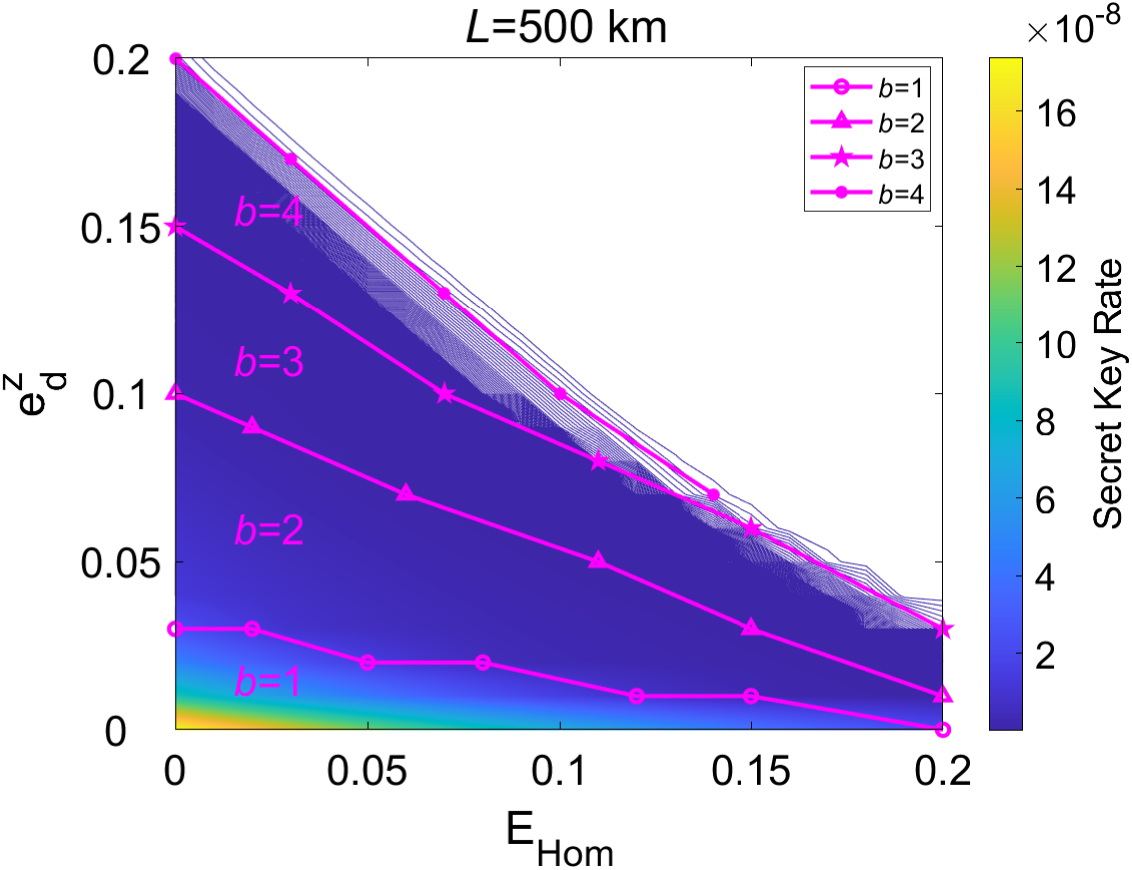}}
	\quad 
	\subfigure[]{
		\label{Different2}
		\includegraphics[width=0.479\linewidth]{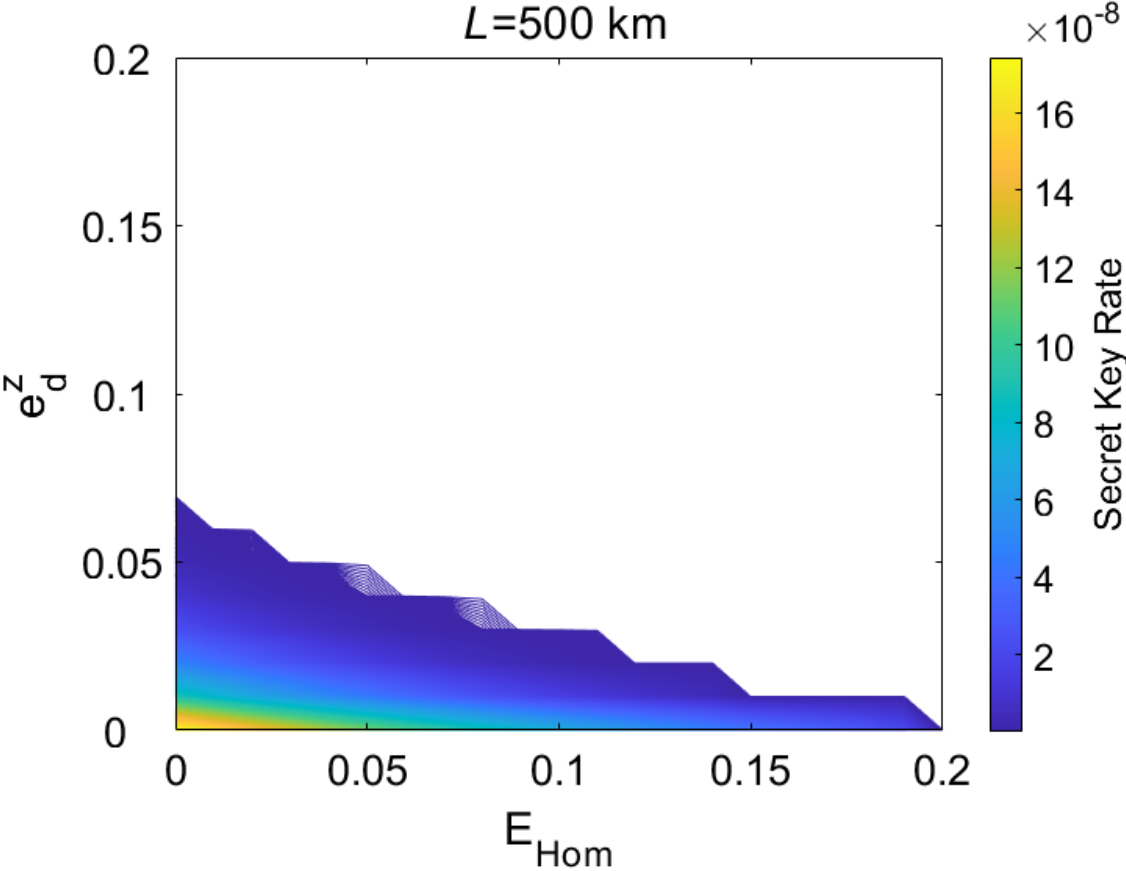}}   	     	
	\caption{The relationship between arbitrary $e^{z}_{d}$ and $E_\text{{Hom}}$ combinations and key rates. (a) The results of AMDI-QKD with AD, where the hollow circle for $b$=1, the triangle for $b$=2,  the five-pointed star for $b$=3 and the solid circle for $b$=4. (b) The results of AMDI-QKD without AD. We assume $L$=500 km, $N$=$7.24\times10^{13}$, the misalignment error rate ranges from [0, 20$\%$]. The color depth represents the distribution of key rates. }	
	\label{AD}     	
\end{figure*}  
    	\section{simulation}
    	\label{simulation} 
   	 
    	 In this section, we present simulation results that detail the performance of the proposed scheme, taking into account finite-key effects. The simulations use a standard symmetric quantum channel model and include practical experimental parameters. The specific numerical values for the simulations are outlined in Table~\ref{table1}, taken from the experimental data in Ref.~\cite{2023Zhou}. For each distance, we optimize implementation parameters using a numerical simulation tool. This includes the intensities of the signal and decoy states, the probabilities of sending them. The optimization routine resembled that in Ref.~\cite{2022Lizijian}. Furthermore, the variable $ b$ is restricted to the interval [1, 4].

   	     First, we evaluate the performance of AMDI-QKD with and without the AD method under the given parameters: $e^{z}_{d}=0.5\%$, $E_\text{{Hom}}=4\%$, and a total pulse count of $N=7.24\times 10^{13}$. As shown in Fig.~\ref{AM1}, at a transmission distance exceeding 608 km, the key rate of AMDI-QKD with AD surpasses that of AMDI-QKD without AD, resulting in a maximum distance increase of 16 km. 
   	    
   	    To further explain the essential reasons for the AD method increasing the key rate over long distances, we present Fig. ~\ref{AM2}. At 0-372 km, the influence of noise on the correlation between Alice and Bob is minimal. The AD method cannot significantly increase $s^{z}_{11}$ and reduce QBER. At 372-608 km, the QBER begins to increase significantly. If the AD method is applied and the raw key is divided into $b$=2 blocks at 500 km, the QBER is reduced, but the finite-key effect leads to a significant reduction in $s^{z}_{11}$. The impact of the reduction in $s^{z}_{11}$ is greater, and the key rate is lower, Therefore, AD is not effective at the distances. At 608-628 km, noise significantly disrupts the correlation of the raw keys between Alice and Bob, and the QBER is already close to the maximum error tolerated by the original protocol. The QBER using AD method will be changed from $E_z$ to $\left(E_{z}\right)^{2} /\left[\left(E_{z}\right)^{2}+\left(1-E_{z}\right)^{2}\right]$, and the reduction in QBER will contribute more to the key rate than the finite-key effect, so the use of the AD method can obtain a higher key rate. 
   	   
    Then, we delve into the finite-key effects and scrutinize the distinct impact of the AD method on AMDI-QKD under varying pulse counts: $N$=$10^{12}$, $10^{13}$, $10^{15}$, and in the asymptotic case. In Fig.~\ref{N.jpg}, it is evident that when the pulse count is $10^{15}$, the AD method extends the maximum transmission distance of AMDI-QKD by 28 km. Nevertheless, with a reduced pulse count of $10^{13}$, the maximum transmission distance improvement for AMDI-QKD with AD is limited to 8 km.  Further reducing the  pulse count to $10^{12}$,  the optimal value of $b$ is all 1, which means that the AD step is not executed.  These results suggest that AD becomes sensitive to pulse reduction when subjected to a  finite-key analysis.
      	  	      	   
   Furthermore, we investigate the impact of the misalignment error rate on the performance of AMDI-QKD with AD, and the results are illustrated in Fig.~\ref{ded.jpg}.  For a relatively small misalignment error rate of $e^{z}_{d}$=$E_\text{{Hom}}$=1$\%$ (5$\%$), the optimal value of $b$ exceeds 1 at 604 km (408 km), leading to a substantial increase in the maximum transmission distance of AMDI-QKD with AD by 20 km (96 km).  However, with a further increase in the misalignment error rate to 10$\%$, AMDI-QKD without AD is unable to generate a key, whereas AMDI-QKD with AD exhibits a substantial key rate and transmission distance of up to  504 km. 
   	        	     
   Finally, we assess the performance for arbitrary combinations of misalignment error rates $e^{z}_{d}$ and $E_\text{{Hom}}$. We fix the misalignment error rate within the range [0, 20$\%$], set $L$=500 km, and $N$=$7.24\times10^{13}$. The simulation results are shown in Fig.~\ref{AD}. The comparison reveals that AMDI-QKD with AD exhibits a higher tolerance for misalignment error rates in the $Z$ basis used for key generation compared to AMDI-QKD without AD. For instance, at $e^{z}_{d}=6\%$ and $E_\text{{Hom}}$=6$\%$, AMDI-QKD with AD can generate a key rate of 1.35$\times 10^{-9}$ bit/pulse, while AMDI-QKD without AD is unable to generate a key. Objectively, AMDI-QKD with AD outperforms AMDI-QKD without AD, particularly at high misalignment error rates.

         \section{ CONCLUSION}
        \label{conclusion}

 In conclusion, we explore the performance of AMDI-QKD with AD, considering a  finite-key effect. Simulation results demonstrate the feasibility and significant impact of the AD method in enhancing the secure key rate and transmission distance of AMDI-QKD. Meanwhile, with a misalignment error rate of 10$\%$, AMDI-QKD without AD fails to generate the key at 0 km, whereas AMDI-QKD with AD can still generate secure bits.
 
 In future research, it will be interesting to further analyze the performance of AMDI-QKD with AD under a limited number of modulated phases~\cite{2024zhang}.

        \section*{Acknowledgment}
       This study was supported by the National Natural Science Foundation of China (Nos. 62171144 and 11905065), Guangxi Science Foundation (Nos.2021GXNSFAA220011 and 2021AC19384), Open Fund of IPOC (BUPT) (No. IPOC2021A02), and Innovation Project of Guangxi Graduate Education (No.YCSW2022040).

   	  \appendix 
   	 \section{SECRET KEY RATE CALCULATION}
   	 \label{Appendix A}
   	 \subsection{SECRECY ANALYSIS}
   	 
   	 	The secrecy analysis follows the idea of Ref.~\cite{2014curty,2014Lim}. If the protocol successfully passes error correction step, then it is $\varepsilon_{\text{cor}}$-correct. If the extracted key length of the protocol does not surpass a certain length, then it is $\varepsilon_{sec}$-secret. Specifically, if the protocol fulfills both conditions $\varepsilon_{\text{cor}}$-correct and $\varepsilon_{sec}$-secret, and $\varepsilon_{tol }$=$\varepsilon_{cor }$+$\varepsilon_{sec }$, it is $\varepsilon_{tol }$-secure.
   	 	  	 
   	   Using a random universal$_2$ hash function~\cite{2011_tomamichel}, the communication parties can extract a  $\varepsilon_{sec }$-secret key of length $\ell$ from the raw key $\mathcal{Z}$ \cite{2012tomamichel},
   	  		\begin{equation}
   	 	\varepsilon_{sec}=2\varepsilon+\frac{1}{2}\sqrt{2^{\ell-H^{\varepsilon}_{\text{min}}(\mathcal{Z}|E^{\prime })}},
   	 	\label{A1}
   	 \end{equation}
   	  where $E^{\prime }$ denotes all information of Eve learned from raw key after error correction, $H^{\varepsilon}_{\text{min}}(\mathcal{Z}|E^{\prime })$ denotes the smooth min-entropy, which quantifies the average probability that Eve guesses $\mathcal{Z}$ correctly using the optimal strategy with access to $E^{\prime }$.
   	
   	 According to a chain-rule inequality for smooth entropies~\cite{2008Renner}, we obtain 
   	  		\begin{equation}
   		H^{\varepsilon}_{\text{min}}(\mathcal{Z}|E^{\prime })\ge 	H^{\varepsilon^{}}_{\text{min}}(\mathcal{Z}|E)-\lambda_{EC}-\text{log}_{2}\left (2/\varepsilon_{\text{cor}}\right ),
   		\label{A2}
   	\end{equation}
   	 where $E$ denotes the information of Eve before error correction, and $ \lambda_{\text{EC}}$ + log$_{2}(2/\varepsilon_{\text{cor}} )$ is the amount of bit information about  that are leaked to during the error correction step. The bits of $\mathcal{Z}$ can be distributed among three different strings:  $\mathcal{Z}_{0}$, $\mathcal{Z}_{11}$ and  $\mathcal{Z}_{\text{rest}}$, where the $\mathcal{Z}_{0}$ is the bits where Alice sent a vacuum state, $\mathcal{Z}_{11}$ is the bits where both Alice and Bob sent a single photon, and $\mathcal{Z}_{\text{rest}}$ is the rest of bits. Using a generalized chain-rule for smooth entropies~\cite{2013Vitanov}, we have
   			\begin{equation}
   		\begin{split} 
   			H^{\varepsilon}_{\text{min}}(\mathcal{Z}|E)&\ge 	H^{\varepsilon^{\prime}+2\varepsilon_{e}+\left(\hat{\varepsilon}+2\hat{\varepsilon}^{\prime}+\hat{\varepsilon}^{\prime\prime}\right) }_{\text{min}}(\mathcal{Z}_{0}\mathcal{Z}_{11}\mathcal{Z}_{\text{rest}}|E)\\
   			&\ge H^{\hat{\varepsilon}^{\prime\prime}}_{\text{min}}(\mathcal{Z}_{\text{0}}|E)+ H^{\varepsilon_{e} }_{\text{min}}(\mathcal{Z}_{11}|\mathcal{Z}_{0}\mathcal{Z}_{\text{rest}}E)\\
   			&+H^{\hat{\varepsilon}^{\prime}}_{\text{min}}(\mathcal{Z}_{\text{rest}}|\mathcal{Z}_{0}E)-2\text{log}_{2}\frac{2}{\varepsilon^{\prime}\hat{\varepsilon}},\\
   			&\ge s^{z}_{0}+H^{\varepsilon_{e} }_{\text{min}}(\mathcal{Z}_{11}|\mathcal{Z}_{0}\mathcal{Z}_{\text{rest}}E)-2\text{log}_{2}\frac{2}{\varepsilon^{\prime}\hat{\varepsilon}},
   		\end{split}
   		\label{A3}
   	\end{equation}
   	 where $\varepsilon=\varepsilon^{\prime}+2\varepsilon_{e}+\left(\hat{\varepsilon}+2\hat{\varepsilon}^{\prime}+\hat{\varepsilon}^{\prime\prime}\right)$, and we have used the fact that  $H^{\hat{\varepsilon}^{\prime}}_{\text{min}}(\mathcal{Z}_{\text{rest}}|\mathcal{Z}_{0}E)\ge0$, 
   	 since all multiphoton events are considered insecure due to the risk of photon-number-splitting attacks, $H^{\hat{\varepsilon}^{\prime\prime}}_{\text{min}}(\mathcal{Z}_{\text{0}}|E)\ge H_{\text{min}}(\mathcal{Z}_{0})=\text{log}_{2}2^{s^{z}_{0}}=s^{z}_{0}$. Here, we consider that the bit values of the vacuum states are uniformly distributed and contain no information. 
   	  
    In addition, we use $\left | 01  \right \rangle$ and $\left | 10  \right \rangle $ for the $Z$ basis, $\frac{1}{\sqrt{2} } \left ( \left | 10  \right \rangle \pm e^{\textbf{i}\varphi}\left | 10  \right \rangle  \right ) $ for the $X$ basis. We let Alice and Bob to use $\chi_{11}$ and $\chi^\prime_{11}$ of length $ s^z_{11}$ instead of the raw keys $\mathcal{Z}_{11}$ and $\mathcal{Z}^\prime_{11}$ if they had measured on the $X$ basis. According to the uncertainty relation of smooth min-entropy and max-entropy, we can get a lower bound for $H^{\varepsilon_{e} }_{\text{min}}(\mathcal{Z}_{11}|\mathcal{Z}_{0}\mathcal{Z}_{\text{rest}}E)$ as follows:   
        \begin{equation}
    	\begin{split} 
    	\label{A4}
    	H^{\varepsilon_{e} }_{\text{min}}(\mathcal{Z}_{11}|\mathcal{Z}_{0}\mathcal{Z}_{\text{rest}}E)
    	&\ge s^z_{11}-H^{\varepsilon_{e}}_{\text{max}}(\chi _{11}|\chi^{\prime} _{11})\\
    	&\ge s^z_{11}[1-H(\phi^{z}_{11})],
    	\end{split}
    \end{equation}
 where the first inequality is derived from the uncertainty relation in Ref.~\cite{2011Tomamichel}. The second inequality utilizes the Lemma 3 from  Ref.~\cite{2012tomamichel}, $H(\phi^{z}_{11})$ quantifies the number of bits Bob needs to reconstruct $\chi _{11}$ using bit string $\chi ^{\prime }_{11}$ , $\phi^{z}_{11}$ is phase error of single-photon in $Z$ basis.
 
   Combined with Eqs.~(\ref{A1}), we obtain the expression for the key length $\ell $ as follows:
\begin{equation}
	\begin{aligned} 
		\ell&=	H^{\varepsilon}_{\text{min}}(\mathcal{Z}|E^{\prime })-2\text{log}_{2}\frac{1}{2\varepsilon_{PA}},
\end{aligned}
\label{A5}
\end{equation}
and combining Eqs.~(\ref{A2})-~(\ref{A5}), the specific expression for $\ell $ is given as follows:

\begin{equation}
	\begin{aligned} 
	\ell&
	\ge s^{z}_{0}+s^{z}_{11}\left [1-H(\phi^{z}_{11})\right ]-\lambda_{EC}   \\
		&-\text{log}_{2}\frac{2}{\varepsilon_{\text{cor}}}-2\text{log}_{2}\frac{2}{\varepsilon^{\prime}\hat{\varepsilon}}-2\text{log}_{2}\frac{1}{2\varepsilon_{\text{PA}} },
	\end{aligned}
\end{equation} 
 where 
 $\varepsilon_{sec}=2(\varepsilon^{\prime}+2\varepsilon_{e}+\hat{\varepsilon}+2\hat{\varepsilon}^{\prime}+\hat{\varepsilon}^{\prime\prime})+\varepsilon_{{PA}}$, we assume $\hat{\varepsilon}^{\prime}=\hat{\varepsilon}^{\prime\prime}=0$ without compromising security. $\varepsilon _{\text{PA} }=\frac{1}{2} \sqrt{2^{\ell -H^{\varepsilon }_{\text{min} }(\mathcal{Z}|E^{\prime })}} $ is a security parameter that involves privacy amplification. 
 Finally, the parameters $s^{z}_{0}$, $s^{z}_{11}$ and $e^{x}_{11}$ are estimated by using the failure probabilities $\varepsilon_{0}$, $\varepsilon_{1}$ and $\varepsilon_{\beta }$, respectively, we have 
  $\varepsilon_{sec}=2(\varepsilon^{\prime}+2\varepsilon_{e}+\hat{\varepsilon})+\varepsilon_{0}+ \varepsilon_{1}+ \varepsilon_{\beta }+\varepsilon_{\text{PA}}$.
 
   	  \subsection{Parameter estimation}
   	  By using decoy state estimation, the key rate can be written as~\cite{2023Zhou}	  
   	  	\begin{equation}
   	  		\begin{split}  
   	  			R&=\frac{1}{N}\left\{\underline{s}^{z}_{0}+\underline{s}^{z}_{11}\left [1-H(\overline{\phi}^{z}_{11})\right ]-\lambda_{EC}  \right. \\
   	  			&\left.-\text{log}_{2}\frac{2}{\varepsilon_{\text{cor}}}-2\text{log}_{2}\frac{2}{\varepsilon^{\prime}\hat{\varepsilon}}-2\text{log}_{2}\frac{1}{2\varepsilon_{\text{PA}}}   \right\}.
   	  			\end{split}
     		\end{equation}
     	
     	We assume failure parameters $\varepsilon_{\text{cor}}$, $\varepsilon^{\prime}, \hat{\varepsilon}$ and $\varepsilon_{\text{PA}}$ to be the equal $\epsilon $.    	
    	As described in the protocol, Alice and Bob publish information about the decoy states when the click filtering operation is used, so only [$\mu_{a},\mu_{b}$] can be used to generate the key. The amount of data consumed during error correction is
    		\begin{equation}
    	\lambda_{EC}=n_{[\mu_{a}, \mu_{b}]}fH\left (E_{z}\right ) ,
           \end{equation}
    	 where $E_{z}=\frac{m_{[\mu_{a}, \mu_{b}]}}{n_{[\mu_{a}, \mu_{b}]}}$ is QBER in $Z$ basis. The total error paring numbers $m^{\prime}_{[\mu_{a}, \mu_{b}]}$ don't considering $e^{z}_{d}$ in $Z$ basis, which include two click events with $(\mu_{a}|\mu_{b})$ and $(o_{a}|o_{b})$ to pair.
    	 
    When we take $e^{z}_{d}$ into account~\cite{2023wang}, the count of error detection can be given
    	\begin{equation}
    		\begin{aligned}
   m_{[\mu_{a},\mu_{b}]}=e^{z}_{d}(n_{[\mu_{a},\mu_{b}]}-m^{\prime}_{[\mu_{a},\mu_{b}]})+(1-e^{z}_{d})m^{\prime}_{[\mu_{a},\mu_{b}]}.
    		\end{aligned}    	
    		\end{equation}	
    	
    	The lower bound of single-photon pair  events in $Z$ basis estimated by the decoy states method can be written as follows:
    	\begin{equation}
    		\begin{split} 
    			\underline{s}^{z\ast }_{11}\geq\frac{e^{-\mu_{a}-\mu_{b}}p_{[\mu_{a},\mu_{b}]}}{\nu_{a}
    				\nu_{b}(\mu^{'}-\nu^{'})}(S_{1}-S_{2}),
    		\end{split}
    \end{equation}   	
      where
			 \begin{equation}
			 	\begin{aligned}
			 	S_{1}&=\mu_{a}\mu_{b}\mu^{\prime}\left (e^{\nu_{a}+\nu_{b}}\frac{\underline{n}^{*}
			 		_{[\nu_{a},\nu_{b}]}}{p_{[\nu_{a},\nu_{b}]}}-e^{\nu_{b}}\frac{\overline{n}^{*}
			 		_{[o_{a},\nu_{b}]}}{p_{[o_{a},\nu_{b}]}}-e^{\nu_{a}}\frac{\overline{n}^{*}
			 		_{[\nu_{a},o_{b}]}}{p_{[\nu_{a},o_{b}]}} \right .\\
		 		&\left.+\frac{\underline{n}^{*}_{[o_{a},o_{b}]}}{p_{[o_{a}o_{b}]}}\right),\\
		 		S_{2}&=\nu_{a}\nu_{b}\nu^{\prime}\left (e^{\mu_{a}+\mu_{b}}\frac{\overline{n}^{*}
		 			_{[\mu_{a},\mu_{b}]}}{p_{[\mu_{a},\mu_{b}]}}-e^{\mu_{b}}\frac{\underline{n}^{*}
		 			_{[o_{a},\mu_{b}]}}{p_{[o_{a}\mu_{b}]}}-e^{\mu_{a}}\frac{\underline{n}^{*}
		 			_{[\mu_{a},o_{b}]}}{p_{[\mu_{a},o_{b}]}} \right .\\
	 			&\left .+\frac{\underline{n}^{*}_{[o_{a},o_{b}]}}{p_{[o_{a}o_{b}]}}\right ),\\	 		
			 	\end{aligned}    	
			 \end{equation}
		 and
		 	\begin{equation}	
		 	\begin{aligned}
		 	p_{[\kappa^{tot}_{a},\kappa^{tot}_{b}]}=\sum_{\kappa^{e}_{a}+\kappa^{l}_{a}=\kappa^{tot}_{a}}\sum_{\kappa^{e}_{b}+\kappa^{l}_{b}=\kappa^{tot}_{b}}\frac{p_{\kappa ^{e}_{a}}p_{\kappa ^{e}_{b}}}{p_{s}}\frac{p_{\kappa ^{l}_{a}}p_{\kappa ^{l}_{b}}}{p_{s}},
		 		\end{aligned}
	 	\end{equation}
		   where $p_{[\kappa^{tot}_{a},\kappa^{tot}_{b}]}$ is coincidence pairing $[\kappa^{tot}_{a},\kappa^{tot}_{b}]$ successful probability, $p_{s}$=1$-$$p_{\mu_{a}}p_{\nu_{b}}$$-$$p_{\nu_{a}}p_{\mu_{b}}$ by using click filtering, $e$ ($l$) denotes early (late) time bin. We set {$o$}$_{a}$={$o$}$_{b}$=0, {$\mu$}$_{a}$={$\mu$}$_{b}$=$\mu^{\prime}$ and  {$\nu$}$_{a}$={$\nu$}$_{b}$=$\nu^{\prime }$ with symmetric channel. The lower bound of vacuum number is given by
		    \begin{equation}	
		    	\begin{aligned}
      			\underline{s}^{z\ast }_{0}=\frac{e^{-\mu_{a}}p_{[\mu_{a},\mu_{b}]}}{p_{[o_{a},\mu_{b}]}}\underline{n}^
      			{\ast}_{[o_{a},\mu_{b}]}.	
		    	\end{aligned}
		    \end{equation}
	    		     		     
		     The upper bound on the number of errors of single photon pairs in $X$ basis is given:
		     	\begin{equation}	
		     	\begin{aligned}
		     	\overline{t}^{x}_{11}\le m_{[2\nu_{a},2\nu_{b}]}-\underline{m}^{0}_{[2\nu_{a},2\nu_{b}]},
		     	\end{aligned}
		     \end{equation}
	     where  $m_{[2\nu_{a},2\nu_{b}]}$ observed error pairing number in $X$ basis,  $\underline{m}^{0}_{[2\nu_{a},2\nu_{b}]}$ is lower bound of error pairing numbers when there is at least one sending vacuum state between communication parties in $X$ basis can be given by
	      \begin{equation}
	     	\begin{aligned}
	    	\underline{m}^{0*}_{[2\nu_{a},2\nu_{b}]}&=e^{-2\nu_{a}}\frac{p_{[2\nu_{a},2\nu_{b}]}\underline{n}^{*}_{[o_{a},2\nu_{b}]}}{2p_{[o_{a},2\nu_{b}]}}+e^{-2\nu_{b}}\frac{p_{[2\nu_{a},2\nu_{b}]}\underline{n}^{*}_{[2\nu_{a},o_{b}]}}{2p_{[2\nu_{a},o_{b}]}}\\
	    	&-e^{-2\nu_{a}-2\nu_{b}}\frac{p_{[2\nu_{a},2\nu_{b}]}\overline{n}^{*}_{[o_{a},o_{b}]}}{2p_{[o_{a},o_{b}]}},\\	     	 		
	     	\end{aligned}    	
	     \end{equation}
     where $p_{[2\nu_{a},2\nu_{b}]}=\frac{2}{M}\frac{p^2_{\nu_{a}}p^2_{\nu_{b}}}{p^2_{s}}$ denotes probability of coincidence $[2\nu_{a},2\nu_{b}]$ in $X$ basis.
     
     The single photon pairing error numbers in $X$ basis is
    	   \begin{equation}
    	   	\begin{aligned}  	   	
    	   	  \overline{e}^{x}_{11}=\frac{\overline{t}^{x}_{11}}{\underline{s}^{x}_{11}},    	 		
    	   	\end{aligned}    	
    	   \end{equation}
       where $\underline{s}^{x}_{11}$ is the number of single photon pairing in $X$ basis, whose lower bound is estimated by decoy state, expression is given as follow:
    	    \begin{equation}	
    		\begin{aligned}
    		\underline{s}^{x*}_{11}\le\frac{e^{-2\nu_{a}-2\nu_{b}}4p_{[2\nu_{a},2\nu_{b}]}}{\mu_{a}
    	  \mu_{b}(\mu^{'}-\nu^{'})}(S_{1}-S_{2}).
    		\end{aligned}
    	\end{equation}
    
    The upper bound on the phase error rate of single photon pair in $Z$ basis is
    	\begin{equation}	
    	\begin{aligned}
    		\overline{\phi}^{z}_{11}\le  \overline{e}^{x}_{11}+\gamma (\underline{s}^{z}_{11},\underline{s}^{x}_{11}, \overline{e}^{x}_{11}, \varepsilon_e).
    	\end{aligned}
    \end{equation}

	     \subsection{Simulation formula}
	     We assume that Alice and Bob respectively send pulses of intensity $\kappa_{a}$ and $\kappa_{b}$ with phase difference  $\theta$,  the gain of only one detector response can be given
	   \begin{equation}	
	   	\begin{aligned}
	   	q^{\theta,D_{L}}_{(\kappa_{a}\mid\kappa_{b})}&=y^{D_{R}}_{(\kappa_{a}\mid\kappa_{b})}e^{\eta^{D_{R}}_{d}\sqrt{\eta_{a}\kappa_{a}\eta_{b}\kappa_{b}}cos\theta}\\
	   	&\left [1-y^{D_{L}}_{(\kappa_{a}\mid\kappa_{b})}e^{-\eta^{{D_L}}_{d}\sqrt{\eta_{a}\kappa_{a}\eta_{b}\kappa_{b}}cos\theta}\right ],\\
	   \end{aligned}
   \end{equation}
	   and,
	   \begin{equation}	
	   	\begin{aligned}
	      q^{\theta,D_{R}}_{(\kappa_{a}\mid\kappa_{b})}&=y^{{D_L}}_{(\kappa_{a}\mid\kappa_{b})}e^{-\eta^{D_{L}}_{d}\sqrt{\eta_{a}\kappa_{a}\eta_{b}\kappa_{b}}cos\theta}\\
	      &\left [1-y^{D_{R}}_{(\kappa_{a}\mid\kappa_{b})}e^{\eta^{D_{R}}_{d}\sqrt{\eta_{a}\kappa_{a}\eta_{b}\kappa_{b}}cos\theta}\right ],\\	      	
	   	\end{aligned}
	   \end{equation}
	 where  $y^{D_{L}(D_{R})}_{(\kappa_{a}\mid\kappa_{b})}=(1-p^{D_{L}(D_{R})}_{d})e^{-\eta^{D_{L}(D_{R})}_{d}\frac{(\eta_{a}\kappa_{a}+\eta_{b}\kappa_{b})}{2}}$, $D_{L}\text{ }(D_{R})$ is left (right) detector, $\eta_{a(b)}=10^{-\frac{\alpha L_{a(b)}}{10}}$. By calculating the phase $\theta$ from 0 to $2\pi$, the total gain of Alice sends $\kappa_{a}$ and Bob sends $\kappa_{b}$  can be given	 
	 \begin{equation}	
	 	\begin{aligned}
	 		q_{(\kappa_{a}\mid\kappa_{b})}&=\frac{1}{2\pi}\int_{0}^{2\pi} (	q^{\theta,D_{L}}_{(\kappa_{a}\mid\kappa_{b})}+q^{\theta,D_{R}}_{(\kappa_{a}\mid\kappa_{b})})d \theta\\
	 		&=y^{D_{L}}_{(\kappa_{a}\mid\kappa_{b})}I_{0}({\eta^{D_{L}}_{d}\sqrt{\eta_{a}\kappa_{a}\eta_{b}\kappa_{b}}})\\
	 		&+y^{D_{R}}_{(\kappa_{a}\mid\kappa_{b})}I_{0}({\eta^{D_{R}}_{d}\sqrt{\eta_{a}\kappa_{a}\eta_{b}\kappa_{b}}})\\
	 		&-2y^{D_{L}}_{(\kappa_{a}\mid\kappa_{b})}y^{D_{R}}_{(\kappa_{a}\mid\kappa_{b})}I_{0}\left [(\eta^{D_{L}}_{d}-\eta^{D_{R}}_{d})\sqrt{\eta_{a}\kappa_{a}\eta_{b}\kappa_{b}}\right],\\	 		 	 		
	 	\end{aligned}
	 \end{equation}
	       where $I_{0}(x)\approx 1+\frac{x^{2}}{4}$ represents the first kind of zero-order modified Bessel function, $q^{\theta}_{(\kappa_{a}\mid\kappa_{b})}=q^{\theta,D_{L}}_{(\kappa_{a}\mid\kappa_{b})}+q^{\theta,D_{R}}_{(\kappa_{a}\mid\kappa_{b})}$. The total number of valid successful pairing is $n_{tot}=\frac{Nq_{tot}}{1+1/q_{T_{c}}}$, $q_{T_{c}}=1-(1-q_{tot})^{N_{T_{c}}}$ is probability of having at least a click at the time interval $T_{c}$ if detector has a click in a time bin. Thus, an average of $1+1/q_{T_{c}}$ valid event is required to form an valid pairing.
	   The average time it takes to form a valid paring is $T_{\text{mean}}=\frac{1-N_{T_{c}}q_{tot}(1/q_{T_{c}}-1)}{Fq_{tot}}$. $F$=1 GHz is system repetition rate, $N_{T_{c}}=FT_{c}$ is total time bins number in interval $T_{c}$.
	       	      	       
	      When we use the three intensities decoy state protocol, there are nine independent and random events, which are $(\mu_{a}|\mu_{b})$, $(\mu_{a}|\nu_{b})$, $(\mu_{a}|o_{b})$, $(\nu_{a}|\nu_{b})$, $(\nu_{a}|\mu_{b})$, $(\nu_{a}|o_{b})$, $(o_{a}|o_{b})$, $(o_{a}|\mu_{b})$ and $(o_{a}|\nu_{b})$. By using  click filtering, $(\mu_{a}|\nu_{b})$ and $(\nu_{a}|\mu_{b})$ are discarded. The probability of having the click event is $q_{tot}=\sum_{\kappa_{a}, \kappa_{b}}p_{\kappa_{a}}p_{\kappa_{b}}q_{(\kappa_{a}|\kappa_{b})}-p_{\mu_{a}}p_{\nu_{b}}q_{(\mu_{a}|\nu_{b})}-p_{\nu_{a}}p_{\mu_{b}}q_{(\nu_{a}|\mu_{b})}$.	      
	       Besides, the number of successful pairing $S_{[{\kappa}{^{tot}_a},{\kappa}{^{tot}_b}]}$ (except the set $S_{[{2\nu}_{a},{2\nu}{_b}]}$) is counted   
	          
	    \begin{equation}
	      		\begin{split} 
	      		\label{AZ}
	      		n_{[\kappa^{tot}_{a},\kappa^{tot}_{b}]}&=n_{tot}\sum_{\kappa^{e}_{a}+\kappa^{l}_{a}=\kappa^{tot}_{a}}\sum_{\kappa^{e}_{b}+\kappa^{l}_{b}=\kappa^{tot}_{b}}\\
	      	&	\left (\frac{p_{\kappa ^{e}_{a}}p_{\kappa ^{e}_{b}}q_{(\kappa^{e}_{a}\mid\kappa^{e}_{b})}}{q_{tot}}\right.
	      		\left.\frac{p_{\kappa ^{l}_{a}}p_{\kappa ^{l}_{b}}q_{(\kappa^{l}_{a}\mid\kappa^{l}_{b})}}{q_{tot}}\right).  
	  	\end{split}
        \end{equation} 

        The set  $S_{[{2\nu}_{a},{2\nu}{_b}]}$ which need to take into account phase difference is counted:       
        \begin{equation}	
        	\begin{aligned}
        		n_{[2\nu_{a},2\nu_{b}]}=\frac{n_{tot}}{M\pi}\int_{0}^{2\pi} \left (\frac{p_{\nu _{a}}p_{\nu _{b}}q^{\theta}_{(\nu_{a}\mid\nu_{b})}}{q_{tot}}
        		 \frac{p_{\nu_{a}}p_{\nu _{b}}q^{\theta }_{(\nu_{a}\mid\nu_{b})}}{q_{tot}}\right)d \theta.
        	\end{aligned}
        \end{equation}
    
     In the experiment, we encode the quantum states by randomly selecting phase $\{0,\text{ }(2 \pi / M),\text{ }(4 \pi / M), \ldots,\text{ }[2 \pi(M-1)] / M\}\text{ }(M=16)$ to fulfill the phase randomization requirement. The total number of errors in the $X$ basis can be given as
	  \begin{equation}	
	  	\begin{aligned}
	  		&m_{[2\nu_{a},2\nu_{b}]}=\frac{n_{tot}}{M\pi}\int_{0}^{2\pi} \\
	  &\left\{(1-E_{\text{Hom}})\frac{p^{2}_{\nu _{a}}p^{2}_{\nu _{b}}\left [q^{\theta,D_{L}}_{(\nu_{a}\mid\nu_{b})}q^{\theta+\delta,D_{R}}_{(\nu_{a}\mid\nu_{b})}+q^{\theta,D_{R}}_{(\nu_{a}\mid\nu_{b})}q^{\theta+\delta,D_{L}}_{(\nu_{a}\mid\nu_{b})}\right ]}{q^{2}_{tot}}\right.\\
	  &\left. + E_{\text{Hom}}\frac{p^{2}_{\nu _{a}}p^{2}_{\nu _{b}}\left [q^{\theta,D_{L}}_{(\nu_{a}\mid\nu_{b})}q^{\theta+\delta,D_{L}}_{(\nu_{a}\mid\nu_{b})}+q^{\theta,D_{R}}_{(\nu_{a}\mid\nu_{b})}q^{\theta+\delta,D_{R}}_{(\nu_{a}\mid\nu_{b})}\right]}{q^{2}_{tot}}\right\}d\theta, 
	  	\end{aligned}
	  \end{equation}
      where $E_{\text{Hom}}$ is interference misalignment error rate, $\delta=T_{\text{mean}}(2\pi\Delta\nu+\omega_{\text{fib}})$ is light pulse phase drift cause by laser frequency difference $\Delta\nu$=10 Hz and fibre drift rate $ \omega_{\text{fib}}$=5900 rad/s.  
      \subsection{Statistical fluctuation}    
       We use the Chernoff bound~\cite{2020Yin} to calculate statistical fluctuation. Assuming a failure probability $\epsilon$ and expectation  $x^{*}$ value, we can estimate upper and lower bounds of the observed value $x$ by Chernoff bound:
       \begin{equation}	
       	\begin{aligned}
       	\overline{x}&=x^{*}+\frac{\beta}{2}+\sqrt{2\beta x^{*}+\frac{\beta^{2}}{4}},
       \end{aligned}
       \end{equation}
and
       \begin{equation}	
    	\begin{aligned}       	
       	\underline{x}&=x^{*}-\sqrt{2\beta x^{*}},
       	\end{aligned}
       \end{equation}
   where $\beta=\text{ln}\epsilon^{-1}$. similarly, the variant of Chernoff bound can be used to estimate the upper and lower bounds of the expected value $x^{*}$ from observed value $x$:
    \begin{equation}	
   	\begin{aligned}
   		\overline{x}^{*}&=x+\beta+\sqrt{2\beta x+\beta^{2}},   	
   	\end{aligned}
   \end{equation}
   and
   \begin{equation}	
   	\begin{aligned}       	
   		\underline{x}^{*}&=\text{max}\left\{x-\frac{\beta}{2}-\sqrt{2\beta x+\frac{\beta^{2}}{4}},0\right\}.
   	\end{aligned}
   \end{equation}

   The upper bound of phase error rate in $Z$ basis is estimated by random sampling theorem, specific expression is as follow~\cite{2020Yin}:
   \begin{equation}	
   	\begin{aligned}       	
   		\overline{\chi}\le \lambda+\gamma ^{U}(n, k, \lambda, \epsilon),
   	\end{aligned}
   \end{equation}
   where
   \begin{equation}	
   	\begin{aligned}       	
   		\gamma ^{U}(n, k, \lambda, \epsilon)=\frac{\frac{(1-2\lambda)AG}{n+k}+\sqrt{\frac{A^{2}G^{2}}
   				{(n+k)^{2}}+4\lambda(1-\lambda)G}}{2+2\frac{A^{2}G}{(n+k)^{2}}},\\
				A=\max \left\{n, k\right\} \text{and } 
				G=\frac{n+k}{nk}\text{ln}\frac{n+k}{2\pi n k\lambda (1-\lambda ) \epsilon^{2}  }.
   	\end{aligned}
   \end{equation}
	   \section{SECURITY OF AMDI-QKD BASED
	   	ON QUANTUM INFORMATION THEORY}
	  \label{Appendix B} 
	  In this brief description security of AMDI-QKD based on quantum information theory, more detailed analysis in ~\cite{2008Renner,2023Zhou,2023Liu}.    	
    The key rate based on quantum information theory as fellows~\cite{2008Renner}:
    	\begin{equation}
    		\begin{split} 
    		R=\min_{\sigma _{AB}\in \Gamma}S\left ( X\mid E \right ) -H\left ( X\mid Y \right ),
    		\end{split}
    \end{equation}     
    where $\Gamma$ represents the set of all density operators $\sigma _{AB}$ within the Hilbert space $\mathcal{H}_{A}\otimes\mathcal{H}_{B} $ that satisfy the requirements, $S\left ( X\mid E \right )$ represents the uncertainty of the eavesdropper's auxiliary state $E$ for Alice's measurement outcome $X$, quantified by von Neumann entropy, $H\left ( X|Y \right ) $ represents the uncertainty of Bob's measurement outcome $Y$ to Alice's measurement outcome $X$, quantified by classical Shannon entropy.
   
 Security analysis similar to entanglement purification protocol,
  Alice and Bob randomly prepare the quantum states $|1,0\rangle^{i,j}$ and $|0,1\rangle^{i,j}$ as $Z$ basis, $ \left(|1,0\rangle^{i,j}\pm |0,1\rangle^{i,j}\right )\sqrt{2}$ as $X$ basis, and send to Charlie for Bell-state measurement. The $|1,0\rangle^{i,j}$ indicates $|1\rangle^{i}\otimes  |0\rangle^{j}$ and is tensor product time bin $i$ and $j$, quantum states $|1\rangle$ and $|0\rangle$ respectively indicate single photon and vacuum states. Before the communication parties measurement, the whole system consisting of Alice, Bob and Eve can be described by quantum state as fellows  	    	
    	\begin{equation}
    		|\Phi\rangle_{ABE} :=\sum_{i=1}^{4} \sqrt{\lambda_{i}}\left|\varphi _{i}\right\rangle_{A B} \otimes\left|\upsilon _{i}\right\rangle_{E},
    	\end{equation}
    	where
    		\begin{equation}
    			\begin{split} 
		    \left|\varphi_{1}\right\rangle=\frac{1}{\sqrt{2}}\left (|1,0\rangle^{i,j}_{A}|1,0\rangle^{i,j}_{B}+|0,1\rangle^{i,j}_{A}|0,1\rangle^{i,j}_{B}\right ),\\
		   	\left|\varphi_{2}\right\rangle=\frac{1}{\sqrt{2}}\left (|1,0\rangle^{i,j}_{A}|1,0\rangle^{i,j}_{B}-|0,1\rangle^{i,j}_{A}|0,1\rangle^{i,j}_{B}\right ),\\
		    \left|\varphi_{3}\right\rangle=\frac{1}{\sqrt{2}}\left (|1,0\rangle^{i,j}_{A}|0,1\rangle^{i,j}_{B}+|0,1\rangle^{i,j}_{A}|1,0\rangle^{i,j}_{B}\right ),\\
		    \left|\varphi_{4}\right\rangle=\frac{1}{\sqrt{2}}\left (|1,0\rangle^{i,j}_{A}|0,1\rangle^{i,j}_{B}-|0,1\rangle^{i,j}_{A}|1,0\rangle^{i,j}_{B}\right ),
    	    	\end{split}
        \end{equation}
    	 and $\sum_{i=1}^{4} \lambda_{i}=1$,  the single-photon error rate  satisfy the following equation $\lambda_{2}+\lambda_{4}=\phi ^{z}_{11}$ and $\lambda_{3}+\lambda_{4}=e^{z}_{11}$.  The subscripts $A$ and $B$ represent Alice and Bob, respectively. $\left | \upsilon_{i}  \right \rangle_{E} $  denotes the orthonormal basis in a Hilbert space $ \mathcal{H}_{E}$. 
    	 
    	Because Eve controls the channel, if the measurement outcomes of communication parties are  $xy\in\left \{00, 11, 01, 10\right \}$,   
    the quantum states that Eve obtains include
    	\begin{equation}
    		\begin{split} 
    			\left|\phi\right\rangle^{0,0}=\frac{1}{\sqrt{2}}\left(\sqrt{\lambda_{1}}\left|\upsilon_{1}\right\rangle+\sqrt{\lambda_{2}}\left|\upsilon_{2}\right\rangle\right), \\
    			\left|\phi\right\rangle^{1,1}=\frac{1}{\sqrt{2}}\left(\sqrt{\lambda_{1}}\left|\upsilon_{1}\right\rangle-\sqrt{\lambda_{2}}\left|\upsilon_{2}\right\rangle\right), \\
    			\left|\phi\right\rangle^{0,1}=\frac{1}{\sqrt{2}}\left(\sqrt{\lambda_{3}}\left|\upsilon_{3}\right\rangle+\sqrt{\lambda_{4}}\left|\upsilon_{4}\right\rangle\right), \\
    			\left|\phi\right\rangle^{1,0}=\frac{1}{\sqrt{2}}\left(\sqrt{\lambda_{3}}\left|\upsilon_{3}\right\rangle-\sqrt{\lambda_{4}}\left|\upsilon_{4}\right\rangle\right).
    			\end{split}
    	\end{equation} 	
        	
    	 After disturbance from Eve on the quantum channel, Alice and Bob gain the density operators for the whole system, as follows:
    	\begin{equation}
    		\sigma_{X Y E}=\sum_{xy}|x\rangle\langle x|\otimes| y\rangle\langle y| \otimes | \phi^{xy}\rangle\langle \phi^{xy}|.
    	\end{equation}   
    
     	 Taking all the above analysis into account, we can get
    	\begin{equation}
    			\begin{split} 
    		S\left(\sigma_{X E}\right)=&1+H\left(\lambda_{1}+\lambda_{2}\right), \\
    		S\left(\sigma_{E}\right)=&H\left(\lambda_{1}+\lambda_{2}\right)+\left(\lambda_{1}+\lambda_{2}\right) H\left(\frac{\lambda_{1}}{\lambda_{1}+\lambda_{2}}\right)\\
    		&+\left(\lambda_{3}+\lambda_{4}\right) H\left(\frac{\lambda_{3}}{\lambda_{3}+\lambda_{4}}\right), \\
    		H(X \mid Y)=&H\left(\lambda_{1}+\lambda_{2}\right).
    			\end{split}
    	\end{equation}

    Finally, we obtain the final key rate in the asymptotic case as follow:

    	\begin{equation}
    		\label{B7}	
    		\begin{split} 
    			R \ge & \min _{\lambda_{1}, \lambda_{2}, \lambda_{3}, \lambda_{4}} S(X \mid E)-H(X \mid Y) \\
    			= & \min _{\lambda_{1}, \lambda_{2}, \lambda_{3}, \lambda_{4}} H\left(\sigma_{X E}\right)-H\left(\sigma_{E}\right)-H(X \mid Y) \\
    			= & \min _{\lambda_{1}, \lambda_{2}, \lambda_{3}, \lambda_{4}} 1-\left(\lambda_{1}+\lambda_{2}\right) H\left(\frac{\lambda_{1}}{\lambda_{1}+\lambda_{2}}\right) \\
    			&-\left(\lambda_{3}+\lambda_{4}\right) H\left(\frac{\lambda_{3}}{\lambda_{3}+\lambda_{4}}\right)-H\left(\lambda_{1}+\lambda_{2}\right).
    		\end{split}
    \end{equation}

   	Suppose that the single-photon bit error rate and phase error rate are $ e^{z}_{11}$ and $\phi^{z}_{11}$, respectively, we have 
   	    	\begin{equation}
   		\begin{split} 
  		\lambda _{2}+\lambda _{4}&=\phi _{11}^{z},\\
  		\lambda _{3}+\lambda _{4}&=e _{11}^{z},\\
  		\lambda _{1}+\lambda _{2}+\lambda _{3}+\lambda _{4}&=1,
   		\end{split}
   	\end{equation}
  by simplifying the above formula, we can get
     	  	\begin{equation}
     	  		\label{B9}		
     		\begin{split} 
  			\lambda _{1}&=1-e _{11}^{z}-\phi _{11}^{z}-\lambda _{4},\\
  			\lambda _{2}&=\phi _{11}^{z}-\lambda _{4},\\
  			\lambda _{3}&=e _{11}^{z}-\lambda _{4}.
     		\end{split}
     	\end{equation}
     
     By combining Eq.~(\ref{B9}) and Eq.~(\ref{B7}), we obtain a new key rate formula
        	\begin{equation}
    	\label{B10}	
    	\begin{split} 
    		R \ge 
    		& \min _{\lambda_{1}, \lambda_{2}, \lambda_{3}, \lambda_{4}} 1-\left(1-e^{z}_{11}\right) H\left(\frac{1-\phi_{11}^{z}-e^{z}_{11}+\lambda_{4}}{1-e^{z}_{11}}\right) \\
    		&-e^{z}_{11} H\left(\frac{e^{z}_{11}-\lambda_{4}}{e^{z}_{11}}\right)-H\left(1-e^{z}_{11}\right).
    	\end{split}
    \end{equation}

The minimum value of the above key rate formula, by computing the partial derivative with respect to Eq.~(\ref{B10}), we can obtain the satisfying condition $\lambda_{4}=e^{z}_{11}\phi^{z}_{11} $.

In order to find the minimum value of Eq.~(\ref{B10}), $\lambda _{1}$, $\lambda _{2}$, $\lambda _{3}$ and $\lambda _{4}$ should satisfy the following conditions
	  	\begin{equation}
	\label{B11}		
	\begin{split} 
		\lambda _{1}&=1-e _{11}^{z}-\phi _{11}^{z}-\lambda _{4},\\
		\lambda _{2}&=\phi _{11}^{z}-\lambda _{4},\\
		\lambda _{3}&=e _{11}^{z}-\lambda _{4},\\
		\lambda_{4}&=e^{z}_{11}\phi^{z}_{11}.
	\end{split}
\end{equation}

In the practical AMDI-QKD protocol~\cite{2023Zhou}, we utilize discrete phase random modulation to fulfill the phase randomization requirements, use Chernoff bound to calculate statistical fluctuations. After obtaining the raw key generated in the $Z$ basis, Alice and Bob perform error correction on the raw key. So the uncertainty between Alice and Bob $H(X|Y)\le fH(E_{z})$, where $H(E_{z})$ is the maximum amount of information leaked during error correction. The key rate formula of AMDI-QKD is as follows:  
		  \begin{equation}
		  	  	\label{B12}
		   \begin{aligned}      	    		
		   		R&\geq\min _{\lambda_{1}, \lambda_{2}, \lambda_{3}, \lambda_{4}}
		   		\frac{1}{N}n_{z}\left\{\frac{\underline{s}^{z}_{0}}{n_{z}}+\frac{\underline{s}^{z}_{11}}{n_{z}}\bigg [1-\right.\bigg . \\
		   		&\bigg .(\lambda_{1}+\lambda_{2}) H\left(\frac{\lambda_{1}}{\lambda_{1}+\lambda_{2}}\right)-\left(\lambda_{3}+\lambda_{4}\right) H\bigg(\frac{\lambda_{3}}{\lambda_{3}+\lambda_{4}}\bigg)\bigg ]\\ 
		   		&\left.-fH\left(E_{z}\right)-\frac{1}{n_{z}}\left(\text{log}_{2}\frac{2}{\varepsilon_{\text{cor}}}+2\text{log}_{2}\frac{2}{\varepsilon^{\prime}\hat{\varepsilon}}+2\text{log}_{2}\frac{1}{2\varepsilon_{\text{PA}}}\right )\right\},
		\end{aligned}
		   \end{equation}  
	   	 where $N$ is the total pulses number sent by Alice, $n_{z} $ is total bits number in $Z$ basis, $\underline{s}^{z}_{0}$ is lower bounds of vacuum state, $\underline{s}^{z}_{11}$ represents the single photon pair successful coincidence number, $E_{z}$ is QBER in $Z$ basis. 
	   	 
		   \section{ SECURITY OF AMDI-QKD WITH AD}  
		  \label{Appendix C} 
		  Next, we analyze security of AMDI with AD, some similar analysis in~\cite{2023Liu_x,2022Li,2024ZhouYao}. In AMDI-QKD, Alice and Bob respectively divide the raw keys that they get into blocks of size $b$, that is $\left\{x_{1}, x_{2}, \cdots,  x_{b}\right\}$ and $\left\{y_{1}, y_{2}, \cdots,  y_{b}\right\}$. And then, Alice randomly selects a privately generated bit $c\in \left\{0,1\right\} $, sends the messages $m=\left\{m_{1}, m_{2}, \cdots, m_{b}\right\}$=$\left\{x_{1}\oplus c, x_{2}\oplus c, \cdots, x_{b}\oplus c\right\}$ to Bob through a public authenticated classical channel. Alice and Bob acquire blocks when Bob  calculates the results $\left\{m_{1}\oplus y_{1}, m_{2}\oplus y_{2}, \cdots, m_{b}\oplus y_{b}\right\}$=$\left\{0, 0, \cdots, 0\right\}$ or $\left\{1, 1, \cdots, 1\right\}$, they retain $x_{1}$ and $y_{1}$ as raw key. Moreover, if Eve acquaints arbitrary measurement results  $m_i$ $(1\le i\le b)$, she has the ability to know all the $b$ measurement results.  
		  Thus, only when all the $b$ pulses are single-photon states can they be employed for key generation. The successful probability of advantage distillation can be calculated as
      
     \begin{equation}   	    		
   	q_{\text{succ}}=(E_{z})^{b}+\left(1-E_{z}\right)^b.
   \end{equation}
     
     When performing the AD step, the quantum state of the whole system composed of Alice, Bob and Eve can be described as~\cite{2024ZhouYao,2008Renner}
     \begin{equation}
     	|	\tilde{\Phi}\rangle_{ABE} :=\sum_{i=1}^{4} \sqrt{	\tilde{\lambda}_{i}}\left|\varphi _{i}\right\rangle_{A B} \otimes\left|\upsilon _{i}\right\rangle_{E},
     \end{equation}
 where 
\begin{equation}
	\begin{split} 		
		\tilde{\lambda}_{1}&=\frac{\left(\lambda_{1}+\lambda_{2}\right)^{b}+\left(\lambda_{1}-\lambda_{2}\right)^{b}}{2p_{\text{succ}}}, \\
		\tilde{\lambda}_{2}&=\frac{\left(\lambda_{1}+\lambda_{2}\right)^{b}-\left(\lambda_{1}-\lambda_{2}\right)^{b}}{2p_{\text{succ}}},\\  	
		\tilde{\lambda}_{3}&=\frac{\left(\lambda_{3}+\lambda_{4}\right)^{b}+\left(\lambda_{3}-\lambda_{4}\right)^{b}}{2 p_{\text{succ}}}, \\
		\tilde{\lambda}_{4}&=\frac{\left(\lambda_{3}+\lambda_{4}\right)^{b}-\left(\lambda_{3}-\lambda_{4}\right)^{b}}{2p_{\text{succ}}},
	\end{split}
\end{equation}
and $ p_{\text{succ}}=\left(\lambda_{1}+\lambda_{2}\right)^{b}+\left(\lambda_{3}+\lambda_{4}\right)^{b}$.  Based on quantum state $|	\tilde{\Phi}\rangle_{ABE}$ and the optimal value of $b$, the Eq.~(\ref{B7}) is amended as follows  	
  	\begin{equation} 
	\label{C4}	
	\begin{split} 
		\tilde{R} \ge 
		 & \max_{b}\min _{\lambda_{1}, \lambda_{2}, \lambda_{3}, \lambda_{4}} \frac{1}{b}q_{\text{succ}}\left\{ 1-\left(\tilde{\lambda}_{1}+\tilde{\lambda}_{2}\right) H\left(\frac{\tilde{\lambda}_{1}}{\tilde{\lambda}_{1}+\tilde{\lambda}_{2}}\right) \right .\\
		&\left .-\left(\tilde{\lambda}_{3}+\tilde{\lambda}_{4}\right) H\left(\frac{\tilde{\lambda}_{3}}{\tilde{\lambda}_{3}+\tilde{\lambda}_{4}}\right)-H\left(\tilde{\lambda}_{1}+\tilde{\lambda}_{2}\right)\right\}.
	\end{split}
\end{equation}  
\hfill 
When considering the practical model~\cite{2023Zhou}, Alice and Bob divide the raw keys $n_{z}$ into blocks of size $b$. After the AD step is successfully completed, 
the number of raw keys they retain are $n_{z}q_{\text{succ}}/b$, the number of single-photon bits are $\left (\underline{s}^{z}_{11}/n_{z}\right )^{b} n_{z}q_{\text{succ}}/b$, and the QBER in $Z$ basis can be changed from $E_{z}$ to $(E_{z})^{b}/q_{\text{succ}}$. Therefore, after executing the AD step, Eq.~(\ref{B12}) is amended as follows:   
   \hfill 
     \begin{equation}
   	\begin{aligned}
   		\tilde{R} \geq & \max _{b} \min _{\lambda_{1}, \lambda_{2}, \lambda_{3}, \lambda_{4}} \frac{1}{N}\frac{n_{z}}{b}q_{\text {succ}}\\ &\left\{{\left(\frac{\underline{s}^{z}_{0}}{n_{z}}\right)}^{b}+{\left(\frac{\underline{s}^{z}_{11}}{n_{z}}\right)}^{b}\left[1-\left(\tilde{\lambda}_{1}+
   		\tilde{\lambda}_{2}\right) 
   		H\left(\frac{\tilde{\lambda}_{1}}{\tilde{\lambda}_{1}+\tilde{\lambda}_{2}}\right)\right.\right . \\
   		&  \left. -\left(\tilde{\lambda}_{3}+\tilde{\lambda}_{4}\right) H\left(\frac{\tilde{\lambda}_{3}}{\tilde{\lambda}_{4}+\tilde{\lambda}_{4}}\right)\right]-fH\left(\tilde{E}_{z}\right)\\
   		&\left.-\frac{b}{n_{z}q_{\text{succ}}}\left (\text{log}_{2}\frac{2}{\varepsilon_{\text{cor}}}+2\text{log}_{2}\frac{2}{\varepsilon^{\prime}\hat{\varepsilon}}+2\text{log}_{2}\frac{1}{2\varepsilon_{\text{PA}}}\right )\right\},
   	\end{aligned}
   \end{equation}
subject to   	
	\begin{equation}	
	\begin{split}  		    	
		{\underline{\phi}^{z}_{11}}&{\le\lambda_{2}+\lambda_{4}\le 	\overline{\phi}^{z}_{11}},  \\  
	{\underline{e}^{z}_{11}}&{\le\lambda_{3}+\lambda_{4}\le\overline{e}^{z}_{11}},\\			 
		\tilde{E}_{z}&=\frac{\left(E_{z}\right)^{b}}{{\left(E_{z}\right)}^{b}+{\left(1-E_{z}\right)}^{b}}, 		
     	\end{split}
      \end{equation} 
where $\tilde{E}_{z}$ represents total error rate after AD postprocessing in $Z$ basis.

     %	\bibliography{AMDI-AD}

\begin{thebibliography}{76}%
	\makeatletter
	\providecommand \@ifxundefined [1]{%
		\@ifx{#1\undefined}
	}%
	\providecommand \@ifnum [1]{%
		\ifnum #1\expandafter \@firstoftwo
		\else \expandafter \@secondoftwo
		\fi
	}%
	\providecommand \@ifx [1]{%
		\ifx #1\expandafter \@firstoftwo
		\else \expandafter \@secondoftwo
		\fi
	}%
	\providecommand \natexlab [1]{#1}%
	\providecommand \enquote  [1]{``#1''}%
	\providecommand \bibnamefont  [1]{#1}%
	\providecommand \bibfnamefont [1]{#1}%
	\providecommand \citenamefont [1]{#1}%
	\providecommand \href@noop [0]{\@secondoftwo}%
	\providecommand \href [0]{\begingroup \@sanitize@url \@href}%
	\providecommand \@href[1]{\@@startlink{#1}\@@href}%
	\providecommand \@@href[1]{\endgroup#1\@@endlink}%
	\providecommand \@sanitize@url [0]{\catcode `\\12\catcode `\$12\catcode
		`\&12\catcode `\#12\catcode `\^12\catcode `\_12\catcode `\%12\relax}%
	\providecommand \@@startlink[1]{}%
	\providecommand \@@endlink[0]{}%
	\providecommand \url  [0]{\begingroup\@sanitize@url \@url }%
	\providecommand \@url [1]{\endgroup\@href {#1}{\urlprefix }}%
	\providecommand \urlprefix  [0]{URL }%
	\providecommand \Eprint [0]{\href }%
	\providecommand \doibase [0]{https://doi.org/}%
	\providecommand \selectlanguage [0]{\@gobble}%
	\providecommand \bibinfo  [0]{\@secondoftwo}%
	\providecommand \bibfield  [0]{\@secondoftwo}%
	\providecommand \translation [1]{[#1]}%
	\providecommand \BibitemOpen [0]{}%
	\providecommand \bibitemStop [0]{}%
	\providecommand \bibitemNoStop [0]{.\EOS\space}%
	\providecommand \EOS [0]{\spacefactor3000\relax}%
	\providecommand \BibitemShut  [1]{\csname bibitem#1\endcsname}%
	\let\auto@bib@innerbib\@empty
	%</preamble>
	\bibitem [{\citenamefont {Bennett}\ and\ \citenamefont
		{Brassard}(1984)}]{1984bennett}%
	\BibitemOpen
	\bibfield  {author} {\bibinfo {author} {\bibfnamefont {C.~H.}\ \bibnamefont
			{Bennett}}\ and\ \bibinfo {author} {\bibfnamefont {G.}~\bibnamefont
			{Brassard}},\ }\bibfield  {title} {\bibinfo {title} {Quantum cryptography:
			Public key distribution and coin tossing},\ }in\ \href@noop {} {\emph
		{\bibinfo {booktitle} {Proceedings of IEEE International Conference on
				Computers, Systems and Signal Processing, Bangalore, India}}}\ (\bibinfo
	{organization} {IEEE, New York},\ \bibinfo {year} {1984})\ pp.\ \bibinfo
	{pages} {175--179}\BibitemShut {NoStop}%
	\bibitem [{\citenamefont {Lo}\ \emph {et~al.}(2005)\citenamefont {Lo},
		\citenamefont {Ma},\ and\ \citenamefont {Chen}}]{2005lo_decoy}%
	\BibitemOpen
	\bibfield  {author} {\bibinfo {author} {\bibfnamefont {H.-K.}\ \bibnamefont
			{Lo}}, \bibinfo {author} {\bibfnamefont {X.}~\bibnamefont {Ma}},\ and\
		\bibinfo {author} {\bibfnamefont {K.}~\bibnamefont {Chen}},\ }\bibfield
	{title} {\bibinfo {title} {Decoy state quantum key distribution},\ }\href
	{https://doi.org/10.1103/PhysRevLett.94.230504} {\bibfield  {journal}
		{\bibinfo  {journal} {Phys. Rev. Lett.}\ }\textbf {\bibinfo {volume} {94}},\
		\bibinfo {pages} {230504} (\bibinfo {year} {2005})}\BibitemShut {NoStop}%
	\bibitem [{\citenamefont {Wang}(2005)}]{2005wang_beating}%
	\BibitemOpen
	\bibfield  {author} {\bibinfo {author} {\bibfnamefont {X.-B.}\ \bibnamefont
			{Wang}},\ }\bibfield  {title} {\bibinfo {title} {Beating the
			photon-number-splitting attack in practical quantum cryptography},\ }\href
	{https://doi.org/10.1103/PhysRevLett.94.230503} {\bibfield  {journal}
		{\bibinfo  {journal} {Phys. Rev. Lett.}\ }\textbf {\bibinfo {volume} {94}},\
		\bibinfo {pages} {230503} (\bibinfo {year} {2005})}\BibitemShut {NoStop}%
	\bibitem [{\citenamefont {Lim}\ \emph {et~al.}(2014)\citenamefont {Lim},
		\citenamefont {Curty}, \citenamefont {Walenta}, \citenamefont {Xu},\ and\
		\citenamefont {Zbinden}}]{2014Lim}%
	\BibitemOpen
	\bibfield  {author} {\bibinfo {author} {\bibfnamefont {C.~C.~W.}\
			\bibnamefont {Lim}}, \bibinfo {author} {\bibfnamefont {M.}~\bibnamefont
			{Curty}}, \bibinfo {author} {\bibfnamefont {N.}~\bibnamefont {Walenta}},
		\bibinfo {author} {\bibfnamefont {F.}~\bibnamefont {Xu}},\ and\ \bibinfo
		{author} {\bibfnamefont {H.}~\bibnamefont {Zbinden}},\ }\bibfield  {title}
	{\bibinfo {title} {Concise security bounds for practical decoy-state quantum
			key distribution},\ }\href {https://doi.org/10.1103/PhysRevA.89.022307}
	{\bibfield  {journal} {\bibinfo  {journal} {Phys. Rev. A}\ }\textbf {\bibinfo
			{volume} {89}},\ \bibinfo {pages} {022307} (\bibinfo {year}
		{2014})}\BibitemShut {NoStop}%
	\bibitem [{\citenamefont {Boaron}\ \emph {et~al.}(2018)\citenamefont {Boaron},
		\citenamefont {Boso}, \citenamefont {Rusca}, \citenamefont {Vulliez},
		\citenamefont {Autebert}, \citenamefont {Caloz}, \citenamefont {Perrenoud},
		\citenamefont {Gras}, \citenamefont {Bussi\`eres}, \citenamefont {Li},
		\citenamefont {Nolan}, \citenamefont {Martin},\ and\ \citenamefont
		{Zbinden}}]{2018Boaron}%
	\BibitemOpen
	\bibfield  {author} {\bibinfo {author} {\bibfnamefont {A.}~\bibnamefont
			{Boaron}}, \bibinfo {author} {\bibfnamefont {G.}~\bibnamefont {Boso}},
		\bibinfo {author} {\bibfnamefont {D.}~\bibnamefont {Rusca}}, \bibinfo
		{author} {\bibfnamefont {C.}~\bibnamefont {Vulliez}}, \bibinfo {author}
		{\bibfnamefont {C.}~\bibnamefont {Autebert}}, \bibinfo {author}
		{\bibfnamefont {M.}~\bibnamefont {Caloz}}, \bibinfo {author} {\bibfnamefont
			{M.}~\bibnamefont {Perrenoud}}, \bibinfo {author} {\bibfnamefont
			{G.}~\bibnamefont {Gras}}, \bibinfo {author} {\bibfnamefont {F.}~\bibnamefont
			{Bussi\`eres}}, \bibinfo {author} {\bibfnamefont {M.-J.}\ \bibnamefont {Li}},
		\bibinfo {author} {\bibfnamefont {D.}~\bibnamefont {Nolan}}, \bibinfo
		{author} {\bibfnamefont {A.}~\bibnamefont {Martin}},\ and\ \bibinfo {author}
		{\bibfnamefont {H.}~\bibnamefont {Zbinden}},\ }\bibfield  {title} {\bibinfo
		{title} {Secure quantum key distribution over 421 km of optical fiber},\
	}\href {https://doi.org/10.1103/PhysRevLett.121.190502} {\bibfield  {journal}
		{\bibinfo  {journal} {Phys. Rev. Lett.}\ }\textbf {\bibinfo {volume} {121}},\
		\bibinfo {pages} {190502} (\bibinfo {year} {2018})}\BibitemShut {NoStop}%
	\bibitem [{\citenamefont {Dynes}\ \emph {et~al.}(2019)\citenamefont {Dynes},
		\citenamefont {Wonfor}, \citenamefont {Tam}, \citenamefont {Sharpe},
		\citenamefont {Takahashi}, \citenamefont {Lucamarini}, \citenamefont {Plews},
		\citenamefont {Yuan}, \citenamefont {Dixon}, \citenamefont {Cho} \emph
		{et~al.}}]{2019Dynes}%
	\BibitemOpen
	\bibfield  {author} {\bibinfo {author} {\bibfnamefont {J.}~\bibnamefont
			{Dynes}}, \bibinfo {author} {\bibfnamefont {A.}~\bibnamefont {Wonfor}},
		\bibinfo {author} {\bibfnamefont {W.-S.}\ \bibnamefont {Tam}}, \bibinfo
		{author} {\bibfnamefont {A.}~\bibnamefont {Sharpe}}, \bibinfo {author}
		{\bibfnamefont {R.}~\bibnamefont {Takahashi}}, \bibinfo {author}
		{\bibfnamefont {M.}~\bibnamefont {Lucamarini}}, \bibinfo {author}
		{\bibfnamefont {A.}~\bibnamefont {Plews}}, \bibinfo {author} {\bibfnamefont
			{Z.}~\bibnamefont {Yuan}}, \bibinfo {author} {\bibfnamefont {A.}~\bibnamefont
			{Dixon}}, \bibinfo {author} {\bibfnamefont {J.}~\bibnamefont {Cho}}, \emph
		{et~al.},\ }\bibfield  {title} {\bibinfo {title} {Cambridge quantum
			network},\ }\href {https://doi.org/10.1038/s41534-019-0221-4} {\bibfield
		{journal} {\bibinfo  {journal} {npj Quantum Inf.}\ }\textbf {\bibinfo
			{volume} {5}},\ \bibinfo {pages} {101} (\bibinfo {year} {2019})}\BibitemShut
	{NoStop}%
	\bibitem [{\citenamefont {Gr{\"u}nenfelder}\ \emph {et~al.}(2020)\citenamefont
		{Gr{\"u}nenfelder}, \citenamefont {Boaron}, \citenamefont {Rusca},
		\citenamefont {Martin},\ and\ \citenamefont {Zbinden}}]{2020Gr}%
	\BibitemOpen
	\bibfield  {author} {\bibinfo {author} {\bibfnamefont {F.}~\bibnamefont
			{Gr{\"u}nenfelder}}, \bibinfo {author} {\bibfnamefont {A.}~\bibnamefont
			{Boaron}}, \bibinfo {author} {\bibfnamefont {D.}~\bibnamefont {Rusca}},
		\bibinfo {author} {\bibfnamefont {A.}~\bibnamefont {Martin}},\ and\ \bibinfo
		{author} {\bibfnamefont {H.}~\bibnamefont {Zbinden}},\ }\bibfield  {title}
	{\bibinfo {title} {Performance and security of 5 ghz repetition rate
			polarization-based quantum key distribution},\ }\href
	{https://doi.org/10.1063/5.0021468} {\bibfield  {journal} {\bibinfo
			{journal} {Appl. Phys. Lett}\ }\textbf {\bibinfo {volume} {117}},\ \bibinfo
		{pages} {144003} (\bibinfo {year} {2020})}\BibitemShut {NoStop}%
	\bibitem [{\citenamefont {Ma}\ \emph {et~al.}(2021)\citenamefont {Ma},
		\citenamefont {Liu}, \citenamefont {Huang}, \citenamefont {Chen},
		\citenamefont {Lin},\ and\ \citenamefont {Wei}}]{2021Madi-BB84}%
	\BibitemOpen
	\bibfield  {author} {\bibinfo {author} {\bibfnamefont {D.}~\bibnamefont
			{Ma}}, \bibinfo {author} {\bibfnamefont {X.}~\bibnamefont {Liu}}, \bibinfo
		{author} {\bibfnamefont {C.}~\bibnamefont {Huang}}, \bibinfo {author}
		{\bibfnamefont {H.}~\bibnamefont {Chen}}, \bibinfo {author} {\bibfnamefont
			{H.}~\bibnamefont {Lin}},\ and\ \bibinfo {author} {\bibfnamefont
			{K.}~\bibnamefont {Wei}},\ }\bibfield  {title} {\bibinfo {title} {Simple
			quantum key distribution using a stable transmitter-receiver scheme},\ }\href
	{https://doi.org/10.1364/OL.418851} {\bibfield  {journal} {\bibinfo
			{journal} {Opt. Lett.}\ }\textbf {\bibinfo {volume} {46}},\ \bibinfo {pages}
		{2152} (\bibinfo {year} {2021})}\BibitemShut {NoStop}%
	\bibitem [{\citenamefont {Scalcon}\ \emph {et~al.}(2022)\citenamefont
		{Scalcon}, \citenamefont {Agnesi}, \citenamefont {Avesani}, \citenamefont
		{Calderaro}, \citenamefont {Foletto}, \citenamefont {Stanco}, \citenamefont
		{Vallone},\ and\ \citenamefont {Villoresi}}]{2022Scalcon}%
	\BibitemOpen
	\bibfield  {author} {\bibinfo {author} {\bibfnamefont {D.}~\bibnamefont
			{Scalcon}}, \bibinfo {author} {\bibfnamefont {C.}~\bibnamefont {Agnesi}},
		\bibinfo {author} {\bibfnamefont {M.}~\bibnamefont {Avesani}}, \bibinfo
		{author} {\bibfnamefont {L.}~\bibnamefont {Calderaro}}, \bibinfo {author}
		{\bibfnamefont {G.}~\bibnamefont {Foletto}}, \bibinfo {author} {\bibfnamefont
			{A.}~\bibnamefont {Stanco}}, \bibinfo {author} {\bibfnamefont
			{G.}~\bibnamefont {Vallone}},\ and\ \bibinfo {author} {\bibfnamefont
			{P.}~\bibnamefont {Villoresi}},\ }\bibfield  {title} {\bibinfo {title}
		{Cross‐encoded quantum key distribution exploiting time‐bin and
			polarization states with qubit‐based synchronization},\ }\href
	{https://doi.org/10.1002/qute.202200051} {\bibfield  {journal} {\bibinfo
			{journal} {Adv. Quantum Technol.}\ }\textbf {\bibinfo {volume} {5}} (\bibinfo
		{year} {2022})}\BibitemShut {NoStop}%
	\bibitem [{\citenamefont {Li}\ \emph {et~al.}(2023{\natexlab{a}})\citenamefont
		{Li}, \citenamefont {Zhang}, \citenamefont {Tan}, \citenamefont {Lu},
		\citenamefont {Liao}, \citenamefont {Huang}, \citenamefont {Li},
		\citenamefont {Wang}, \citenamefont {Mao}, \citenamefont {Yan} \emph
		{et~al.}}]{2023LI_Wei}%
	\BibitemOpen
	\bibfield  {author} {\bibinfo {author} {\bibfnamefont {W.}~\bibnamefont
			{Li}}, \bibinfo {author} {\bibfnamefont {L.}~\bibnamefont {Zhang}}, \bibinfo
		{author} {\bibfnamefont {H.}~\bibnamefont {Tan}}, \bibinfo {author}
		{\bibfnamefont {Y.}~\bibnamefont {Lu}}, \bibinfo {author} {\bibfnamefont
			{S.-K.}\ \bibnamefont {Liao}}, \bibinfo {author} {\bibfnamefont
			{J.}~\bibnamefont {Huang}}, \bibinfo {author} {\bibfnamefont
			{H.}~\bibnamefont {Li}}, \bibinfo {author} {\bibfnamefont {Z.}~\bibnamefont
			{Wang}}, \bibinfo {author} {\bibfnamefont {H.-K.}\ \bibnamefont {Mao}},
		\bibinfo {author} {\bibfnamefont {B.}~\bibnamefont {Yan}}, \emph {et~al.},\
	}\bibfield  {title} {\bibinfo {title} {High-rate quantum key distribution
			exceeding 110 mb s--1},\ }\href {https://doi.org/10.1038/s41566-023-01166-4}
	{\bibfield  {journal} {\bibinfo  {journal} {Nat. Photon.}\ }\textbf {\bibinfo
			{volume} {17}},\ \bibinfo {pages} {416} (\bibinfo {year}
		{2023}{\natexlab{a}})}\BibitemShut {NoStop}%
	\bibitem [{\citenamefont {Wei}\ \emph {et~al.}(2023)\citenamefont {Wei},
		\citenamefont {Hu}, \citenamefont {Du}, \citenamefont {Hua}, \citenamefont
		{Zhao}, \citenamefont {Chen}, \citenamefont {Huang},\ and\ \citenamefont
		{Xiao}}]{2023wei}%
	\BibitemOpen
	\bibfield  {author} {\bibinfo {author} {\bibfnamefont {K.}~\bibnamefont
			{Wei}}, \bibinfo {author} {\bibfnamefont {X.}~\bibnamefont {Hu}}, \bibinfo
		{author} {\bibfnamefont {Y.}~\bibnamefont {Du}}, \bibinfo {author}
		{\bibfnamefont {X.}~\bibnamefont {Hua}}, \bibinfo {author} {\bibfnamefont
			{Z.}~\bibnamefont {Zhao}}, \bibinfo {author} {\bibfnamefont {Y.}~\bibnamefont
			{Chen}}, \bibinfo {author} {\bibfnamefont {C.}~\bibnamefont {Huang}},\ and\
		\bibinfo {author} {\bibfnamefont {X.}~\bibnamefont {Xiao}},\ }\bibfield
	{title} {\bibinfo {title} {Resource-efficient quantum key distribution with
			integrated silicon photonics},\ }\href {https://doi.org/10.1364/prj.482942}
	{\bibfield  {journal} {\bibinfo  {journal} {Photon. Res.}\ }\textbf {\bibinfo
			{volume} {11}},\ \bibinfo {pages} {1364} (\bibinfo {year}
		{2023})}\BibitemShut {NoStop}%
	\bibitem [{\citenamefont {Lo}\ \emph {et~al.}(2014)\citenamefont {Lo},
		\citenamefont {Curty},\ and\ \citenamefont {Tamaki}}]{2014Lo}%
	\BibitemOpen
	\bibfield  {author} {\bibinfo {author} {\bibfnamefont {H.-K.}\ \bibnamefont
			{Lo}}, \bibinfo {author} {\bibfnamefont {M.}~\bibnamefont {Curty}},\ and\
		\bibinfo {author} {\bibfnamefont {K.}~\bibnamefont {Tamaki}},\ }\bibfield
	{title} {\bibinfo {title} {Secure quantum key distribution},\ }\href
	{https://doi.org/10.1038/nphoton.2014.149} {\bibfield  {journal} {\bibinfo
			{journal} {Nat. Photon.}\ }\textbf {\bibinfo {volume} {8}},\ \bibinfo {pages}
		{595} (\bibinfo {year} {2014})}\BibitemShut {NoStop}%
	\bibitem [{\citenamefont {Xu}\ \emph {et~al.}(2020)\citenamefont {Xu},
		\citenamefont {Ma}, \citenamefont {Zhang}, \citenamefont {Lo},\ and\
		\citenamefont {Pan}}]{2020Xufeihu-Review}%
	\BibitemOpen
	\bibfield  {author} {\bibinfo {author} {\bibfnamefont {F.}~\bibnamefont
			{Xu}}, \bibinfo {author} {\bibfnamefont {X.}~\bibnamefont {Ma}}, \bibinfo
		{author} {\bibfnamefont {Q.}~\bibnamefont {Zhang}}, \bibinfo {author}
		{\bibfnamefont {H.-K.}\ \bibnamefont {Lo}},\ and\ \bibinfo {author}
		{\bibfnamefont {J.-W.}\ \bibnamefont {Pan}},\ }\bibfield  {title} {\bibinfo
		{title} {Secure quantum key distribution with realistic devices},\ }\href
	{https://doi.org/10.1103/RevModPhys.92.025002} {\bibfield  {journal}
		{\bibinfo  {journal} {Rev. Mod. Phys.}\ }\textbf {\bibinfo {volume} {92}},\
		\bibinfo {pages} {025002} (\bibinfo {year} {2020})}\BibitemShut {NoStop}%
	\bibitem [{\citenamefont {Lydersen}\ \emph {et~al.}(2010)\citenamefont
		{Lydersen}, \citenamefont {Wiechers}, \citenamefont {Wittmann}, \citenamefont
		{Elser}, \citenamefont {Skaar},\ and\ \citenamefont
		{Makarov}}]{2010Lydersen}%
	\BibitemOpen
	\bibfield  {author} {\bibinfo {author} {\bibfnamefont {L.}~\bibnamefont
			{Lydersen}}, \bibinfo {author} {\bibfnamefont {C.}~\bibnamefont {Wiechers}},
		\bibinfo {author} {\bibfnamefont {C.}~\bibnamefont {Wittmann}}, \bibinfo
		{author} {\bibfnamefont {D.}~\bibnamefont {Elser}}, \bibinfo {author}
		{\bibfnamefont {J.}~\bibnamefont {Skaar}},\ and\ \bibinfo {author}
		{\bibfnamefont {V.}~\bibnamefont {Makarov}},\ }\bibfield  {title} {\bibinfo
		{title} {Hacking commercial quantum cryptography systems by tailored bright
			illumination},\ }\href
	{https://doi.org/https://doi.org/10.1038/NPHOTON.2010.214} {\bibfield
		{journal} {\bibinfo  {journal} {Nat. Photon.}\ }\textbf {\bibinfo {volume}
			{4}},\ \bibinfo {pages} {686} (\bibinfo {year} {2010})}\BibitemShut {NoStop}%
	\bibitem [{\citenamefont {Wei}\ \emph {et~al.}(2019)\citenamefont {Wei},
		\citenamefont {Zhang}, \citenamefont {Tang}, \citenamefont {You},\ and\
		\citenamefont {Xu}}]{2019Wei-Attack}%
	\BibitemOpen
	\bibfield  {author} {\bibinfo {author} {\bibfnamefont {K.}~\bibnamefont
			{Wei}}, \bibinfo {author} {\bibfnamefont {W.}~\bibnamefont {Zhang}}, \bibinfo
		{author} {\bibfnamefont {Y.-L.}\ \bibnamefont {Tang}}, \bibinfo {author}
		{\bibfnamefont {L.}~\bibnamefont {You}},\ and\ \bibinfo {author}
		{\bibfnamefont {F.}~\bibnamefont {Xu}},\ }\bibfield  {title} {\bibinfo
		{title} {Implementation security of quantum key distribution due to
			polarization-dependent efficiency mismatch},\ }\href
	{https://doi.org/10.1103/PhysRevA.100.022325} {\bibfield  {journal} {\bibinfo
			{journal} {Phys. Rev. A}\ }\textbf {\bibinfo {volume} {100}},\ \bibinfo
		{pages} {022325} (\bibinfo {year} {2019})}\BibitemShut {NoStop}%
	\bibitem [{\citenamefont {Ye}\ \emph {et~al.}(2020)\citenamefont {Ye},
		\citenamefont {Li}, \citenamefont {Wang}, \citenamefont {Gao}, \citenamefont
		{Lu},\ and\ \citenamefont {Qian}}]{2020Ye}%
	\BibitemOpen
	\bibfield  {author} {\bibinfo {author} {\bibfnamefont {M.}~\bibnamefont
			{Ye}}, \bibinfo {author} {\bibfnamefont {J.-H.}\ \bibnamefont {Li}}, \bibinfo
		{author} {\bibfnamefont {Y.}~\bibnamefont {Wang}}, \bibinfo {author}
		{\bibfnamefont {P.}~\bibnamefont {Gao}}, \bibinfo {author} {\bibfnamefont
			{X.-X.}\ \bibnamefont {Lu}},\ and\ \bibinfo {author} {\bibfnamefont {Y.-J.}\
			\bibnamefont {Qian}},\ }\bibfield  {title} {\bibinfo {title} {Quantum key
			distribution system against the probabilistic faint after-gate attack},\
	}\href {https://doi.org/10.1088/1572-9494/abb7d8} {\bibfield  {journal}
		{\bibinfo  {journal} {Commun. Theor. Phys.}\ }\textbf {\bibinfo {volume}
			{72}},\ \bibinfo {pages} {115102} (\bibinfo {year} {2020})}\BibitemShut
	{NoStop}%
	\bibitem [{\citenamefont {Acheva}\ \emph {et~al.}(2023)\citenamefont {Acheva},
		\citenamefont {Zaitsev}, \citenamefont {Zavodilenko}, \citenamefont {Losev},
		\citenamefont {Huang},\ and\ \citenamefont {Makarov}}]{2023Acheva}%
	\BibitemOpen
	\bibfield  {author} {\bibinfo {author} {\bibfnamefont {P.}~\bibnamefont
			{Acheva}}, \bibinfo {author} {\bibfnamefont {K.}~\bibnamefont {Zaitsev}},
		\bibinfo {author} {\bibfnamefont {V.}~\bibnamefont {Zavodilenko}}, \bibinfo
		{author} {\bibfnamefont {A.}~\bibnamefont {Losev}}, \bibinfo {author}
		{\bibfnamefont {A.}~\bibnamefont {Huang}},\ and\ \bibinfo {author}
		{\bibfnamefont {V.}~\bibnamefont {Makarov}},\ }\bibfield  {title} {\bibinfo
		{title} {Automated verification of countermeasure against detector-control
			attack in quantum key distribution},\ }\href
	{https://doi.org/10.1140/epjqt/s40507-023-00178-x} {\bibfield  {journal}
		{\bibinfo  {journal} {EPJ Quantum Technol.}\ }\textbf {\bibinfo {volume}
			{10}},\ \bibinfo {pages} {1} (\bibinfo {year} {2023})}\BibitemShut {NoStop}%
	\bibitem [{\citenamefont {Tamaki}\ \emph {et~al.}(2014)\citenamefont {Tamaki},
		\citenamefont {Curty}, \citenamefont {Kato}, \citenamefont {Lo},\ and\
		\citenamefont {Azuma}}]{2014Tamaki}%
	\BibitemOpen
	\bibfield  {author} {\bibinfo {author} {\bibfnamefont {K.}~\bibnamefont
			{Tamaki}}, \bibinfo {author} {\bibfnamefont {M.}~\bibnamefont {Curty}},
		\bibinfo {author} {\bibfnamefont {G.}~\bibnamefont {Kato}}, \bibinfo {author}
		{\bibfnamefont {H.-K.}\ \bibnamefont {Lo}},\ and\ \bibinfo {author}
		{\bibfnamefont {K.}~\bibnamefont {Azuma}},\ }\bibfield  {title} {\bibinfo
		{title} {Loss-tolerant quantum cryptography with imperfect sources},\ }\href
	{https://link.aps.org/doi/10.1103/PhysRevA.90.052314} {\bibfield  {journal}
		{\bibinfo  {journal} {Phys. Rev. A}\ }\textbf {\bibinfo {volume} {90}},\
		\bibinfo {pages} {052314} (\bibinfo {year} {2014})}\BibitemShut {NoStop}%
	\bibitem [{\citenamefont {Ponosova}\ \emph {et~al.}(2022)\citenamefont
		{Ponosova}, \citenamefont {Ruzhitskaya}, \citenamefont {Chaiwongkhot},
		\citenamefont {Egorov}, \citenamefont {Makarov},\ and\ \citenamefont
		{Huang}}]{2022Anastasiya}%
	\BibitemOpen
	\bibfield  {author} {\bibinfo {author} {\bibfnamefont {A.}~\bibnamefont
			{Ponosova}}, \bibinfo {author} {\bibfnamefont {D.}~\bibnamefont
			{Ruzhitskaya}}, \bibinfo {author} {\bibfnamefont {P.}~\bibnamefont
			{Chaiwongkhot}}, \bibinfo {author} {\bibfnamefont {V.}~\bibnamefont
			{Egorov}}, \bibinfo {author} {\bibfnamefont {V.}~\bibnamefont {Makarov}},\
		and\ \bibinfo {author} {\bibfnamefont {A.}~\bibnamefont {Huang}},\ }\bibfield
	{title} {\bibinfo {title} {Protecting fiber-optic quantum key distribution
			sources against light-injection attacks},\ }\href
	{https://doi.org/10.1103/PRXQuantum.3.040307} {\bibfield  {journal} {\bibinfo
			{journal} {PRX Quantum}\ }\textbf {\bibinfo {volume} {3}},\ \bibinfo {pages}
		{040307} (\bibinfo {year} {2022})}\BibitemShut {NoStop}%
	\bibitem [{\citenamefont {Chen}\ \emph
		{et~al.}(2022{\natexlab{a}})\citenamefont {Chen}, \citenamefont {Huang},
		\citenamefont {Chen}, \citenamefont {He}, \citenamefont {Zhang},
		\citenamefont {Sun},\ and\ \citenamefont {Wei}}]{2022chen_ye}%
	\BibitemOpen
	\bibfield  {author} {\bibinfo {author} {\bibfnamefont {Y.}~\bibnamefont
			{Chen}}, \bibinfo {author} {\bibfnamefont {C.}~\bibnamefont {Huang}},
		\bibinfo {author} {\bibfnamefont {Z.}~\bibnamefont {Chen}}, \bibinfo {author}
		{\bibfnamefont {W.}~\bibnamefont {He}}, \bibinfo {author} {\bibfnamefont
			{C.}~\bibnamefont {Zhang}}, \bibinfo {author} {\bibfnamefont
			{S.}~\bibnamefont {Sun}},\ and\ \bibinfo {author} {\bibfnamefont
			{K.}~\bibnamefont {Wei}},\ }\bibfield  {title} {\bibinfo {title}
		{Experimental study of secure quantum key distribution with source and
			detection imperfections},\ }\href
	{https://doi.org/10.1103/PhysRevA.106.022614} {\bibfield  {journal} {\bibinfo
			{journal} {Phys. Rev. A}\ }\textbf {\bibinfo {volume} {106}},\ \bibinfo
		{pages} {022614} (\bibinfo {year} {2022}{\natexlab{a}})}\BibitemShut
	{NoStop}%
	\bibitem [{\citenamefont {Lo}\ \emph {et~al.}(2012)\citenamefont {Lo},
		\citenamefont {Curty},\ and\ \citenamefont {Qi}}]{2012Lo}%
	\BibitemOpen
	\bibfield  {author} {\bibinfo {author} {\bibfnamefont {H.-K.}\ \bibnamefont
			{Lo}}, \bibinfo {author} {\bibfnamefont {M.}~\bibnamefont {Curty}},\ and\
		\bibinfo {author} {\bibfnamefont {B.}~\bibnamefont {Qi}},\ }\bibfield
	{title} {\bibinfo {title} {Measurement-device-independent quantum key
			distribution},\ }\href {https://doi.org/10.1103/PhysRevLett.108.130503}
	{\bibfield  {journal} {\bibinfo  {journal} {Phys. Rev. Lett.}\ }\textbf
		{\bibinfo {volume} {108}},\ \bibinfo {pages} {130503} (\bibinfo {year}
		{2012})}\BibitemShut {NoStop}%
	\bibitem [{\citenamefont {Braunstein}\ and\ \citenamefont
		{Pirandola}(2012)}]{2012Braunstein}%
	\BibitemOpen
	\bibfield  {author} {\bibinfo {author} {\bibfnamefont {S.~L.}\ \bibnamefont
			{Braunstein}}\ and\ \bibinfo {author} {\bibfnamefont {S.}~\bibnamefont
			{Pirandola}},\ }\bibfield  {title} {\bibinfo {title} {Side-channel-free
			quantum key distribution},\ }\href
	{https://doi.org/10.1103/PhysRevLett.108.130502} {\bibfield  {journal}
		{\bibinfo  {journal} {Phys. Rev. Lett.}\ }\textbf {\bibinfo {volume} {108}},\
		\bibinfo {pages} {130502} (\bibinfo {year} {2012})}\BibitemShut {NoStop}%
	\bibitem [{\citenamefont {Yin}\ \emph {et~al.}(2016)\citenamefont {Yin},
		\citenamefont {Chen}, \citenamefont {Yu}, \citenamefont {Liu}, \citenamefont
		{You}, \citenamefont {Zhou}, \citenamefont {Chen}, \citenamefont {Mao},
		\citenamefont {Huang}, \citenamefont {Zhang}, \citenamefont {Chen},
		\citenamefont {Li}, \citenamefont {Nolan}, \citenamefont {Zhou},
		\citenamefont {Jiang}, \citenamefont {Wang}, \citenamefont {Zhang},
		\citenamefont {Wang},\ and\ \citenamefont {Pan}}]{2016Yin}%
	\BibitemOpen
	\bibfield  {author} {\bibinfo {author} {\bibfnamefont {H.-L.}\ \bibnamefont
			{Yin}}, \bibinfo {author} {\bibfnamefont {T.-Y.}\ \bibnamefont {Chen}},
		\bibinfo {author} {\bibfnamefont {Z.-W.}\ \bibnamefont {Yu}}, \bibinfo
		{author} {\bibfnamefont {H.}~\bibnamefont {Liu}}, \bibinfo {author}
		{\bibfnamefont {L.-X.}\ \bibnamefont {You}}, \bibinfo {author} {\bibfnamefont
			{Y.-H.}\ \bibnamefont {Zhou}}, \bibinfo {author} {\bibfnamefont {S.-J.}\
			\bibnamefont {Chen}}, \bibinfo {author} {\bibfnamefont {Y.}~\bibnamefont
			{Mao}}, \bibinfo {author} {\bibfnamefont {M.-Q.}\ \bibnamefont {Huang}},
		\bibinfo {author} {\bibfnamefont {W.-J.}\ \bibnamefont {Zhang}}, \bibinfo
		{author} {\bibfnamefont {H.}~\bibnamefont {Chen}}, \bibinfo {author}
		{\bibfnamefont {M.~J.}\ \bibnamefont {Li}}, \bibinfo {author} {\bibfnamefont
			{D.}~\bibnamefont {Nolan}}, \bibinfo {author} {\bibfnamefont
			{F.}~\bibnamefont {Zhou}}, \bibinfo {author} {\bibfnamefont {X.}~\bibnamefont
			{Jiang}}, \bibinfo {author} {\bibfnamefont {Z.}~\bibnamefont {Wang}},
		\bibinfo {author} {\bibfnamefont {Q.}~\bibnamefont {Zhang}}, \bibinfo
		{author} {\bibfnamefont {X.-B.}\ \bibnamefont {Wang}},\ and\ \bibinfo
		{author} {\bibfnamefont {J.-W.}\ \bibnamefont {Pan}},\ }\bibfield  {title}
	{\bibinfo {title} {Measurement-device-independent quantum key distribution
			over a 404 km optical fiber},\ }\href
	{https://doi.org/10.1103/PhysRevLett.117.190501} {\bibfield  {journal}
		{\bibinfo  {journal} {Phys. Rev. Lett.}\ }\textbf {\bibinfo {volume} {117}},\
		\bibinfo {pages} {190501} (\bibinfo {year} {2016})}\BibitemShut {NoStop}%
	\bibitem [{\citenamefont {Liu}\ \emph {et~al.}(2018)\citenamefont {Liu},
		\citenamefont {Wang}, \citenamefont {Ma},\ and\ \citenamefont
		{Sun}}]{2018Liu}%
	\BibitemOpen
	\bibfield  {author} {\bibinfo {author} {\bibfnamefont {H.}~\bibnamefont
			{Liu}}, \bibinfo {author} {\bibfnamefont {J.}~\bibnamefont {Wang}}, \bibinfo
		{author} {\bibfnamefont {H.}~\bibnamefont {Ma}},\ and\ \bibinfo {author}
		{\bibfnamefont {S.}~\bibnamefont {Sun}},\ }\bibfield  {title} {\bibinfo
		{title} {Polarization-multiplexing-based measurement-device-independent
			quantum key distribution without phase reference calibration},\ }\href
	{https://doi.org/10.1364/OPTICA.5.000902} {\bibfield  {journal} {\bibinfo
			{journal} {Optica}\ }\textbf {\bibinfo {volume} {5}},\ \bibinfo {pages} {902}
		(\bibinfo {year} {2018})}\BibitemShut {NoStop}%
	\bibitem [{\citenamefont {Wei}\ \emph {et~al.}(2020)\citenamefont {Wei},
		\citenamefont {Li}, \citenamefont {Tan}, \citenamefont {Li}, \citenamefont
		{Min}, \citenamefont {Zhang}, \citenamefont {Li}, \citenamefont {You},
		\citenamefont {Wang}, \citenamefont {Jiang}, \citenamefont {Chen},
		\citenamefont {Liao}, \citenamefont {Peng}, \citenamefont {Xu},\ and\
		\citenamefont {Pan}}]{2020Wei-MDI}%
	\BibitemOpen
	\bibfield  {author} {\bibinfo {author} {\bibfnamefont {K.}~\bibnamefont
			{Wei}}, \bibinfo {author} {\bibfnamefont {W.}~\bibnamefont {Li}}, \bibinfo
		{author} {\bibfnamefont {H.}~\bibnamefont {Tan}}, \bibinfo {author}
		{\bibfnamefont {Y.}~\bibnamefont {Li}}, \bibinfo {author} {\bibfnamefont
			{H.}~\bibnamefont {Min}}, \bibinfo {author} {\bibfnamefont {W.-J.}\
			\bibnamefont {Zhang}}, \bibinfo {author} {\bibfnamefont {H.}~\bibnamefont
			{Li}}, \bibinfo {author} {\bibfnamefont {L.}~\bibnamefont {You}}, \bibinfo
		{author} {\bibfnamefont {Z.}~\bibnamefont {Wang}}, \bibinfo {author}
		{\bibfnamefont {X.}~\bibnamefont {Jiang}}, \bibinfo {author} {\bibfnamefont
			{T.-Y.}\ \bibnamefont {Chen}}, \bibinfo {author} {\bibfnamefont {S.-K.}\
			\bibnamefont {Liao}}, \bibinfo {author} {\bibfnamefont {C.-Z.}\ \bibnamefont
			{Peng}}, \bibinfo {author} {\bibfnamefont {F.}~\bibnamefont {Xu}},\ and\
		\bibinfo {author} {\bibfnamefont {J.-W.}\ \bibnamefont {Pan}},\ }\bibfield
	{title} {\bibinfo {title} {High-speed measurement-device-independent quantum
			key distribution with integrated silicon photonics},\ }\href
	{https://doi.org/10.1103/PhysRevX.10.031030} {\bibfield  {journal} {\bibinfo
			{journal} {Phys. Rev. X}\ }\textbf {\bibinfo {volume} {10}},\ \bibinfo
		{pages} {031030} (\bibinfo {year} {2020})}\BibitemShut {NoStop}%
	\bibitem [{\citenamefont {Cao}\ \emph {et~al.}(2020)\citenamefont {Cao},
		\citenamefont {Luo}, \citenamefont {Wang}, \citenamefont {Zou}, \citenamefont
		{Yan}, \citenamefont {Cai}, \citenamefont {Zhang}, \citenamefont {Hu},
		\citenamefont {Jiang}, \citenamefont {Fan} \emph {et~al.}}]{2020Cao}%
	\BibitemOpen
	\bibfield  {author} {\bibinfo {author} {\bibfnamefont {L.}~\bibnamefont
			{Cao}}, \bibinfo {author} {\bibfnamefont {W.}~\bibnamefont {Luo}}, \bibinfo
		{author} {\bibfnamefont {Y.}~\bibnamefont {Wang}}, \bibinfo {author}
		{\bibfnamefont {J.}~\bibnamefont {Zou}}, \bibinfo {author} {\bibfnamefont
			{R.}~\bibnamefont {Yan}}, \bibinfo {author} {\bibfnamefont {H.}~\bibnamefont
			{Cai}}, \bibinfo {author} {\bibfnamefont {Y.}~\bibnamefont {Zhang}}, \bibinfo
		{author} {\bibfnamefont {X.}~\bibnamefont {Hu}}, \bibinfo {author}
		{\bibfnamefont {C.}~\bibnamefont {Jiang}}, \bibinfo {author} {\bibfnamefont
			{W.}~\bibnamefont {Fan}}, \emph {et~al.},\ }\bibfield  {title} {\bibinfo
		{title} {Chip-based measurement-device-independent quantum key distribution
			using integrated silicon photonic systems},\ }\href
	{https://doi.org/10.1103/PhysRevApplied.14.011001} {\bibfield  {journal}
		{\bibinfo  {journal} {Phys. Rev. Appl.}\ }\textbf {\bibinfo {volume} {14}},\
		\bibinfo {pages} {011001} (\bibinfo {year} {2020})}\BibitemShut {NoStop}%
	\bibitem [{\citenamefont {Fan-Yuan}\ \emph {et~al.}(2021)\citenamefont
		{Fan-Yuan}, \citenamefont {Lu}, \citenamefont {Wang}, \citenamefont {Yin},
		\citenamefont {He}, \citenamefont {Zhou}, \citenamefont {Teng}, \citenamefont
		{Chen}, \citenamefont {Guo},\ and\ \citenamefont {Han}}]{2021Fan}%
	\BibitemOpen
	\bibfield  {author} {\bibinfo {author} {\bibfnamefont {G.-J.}\ \bibnamefont
			{Fan-Yuan}}, \bibinfo {author} {\bibfnamefont {F.-Y.}\ \bibnamefont {Lu}},
		\bibinfo {author} {\bibfnamefont {S.}~\bibnamefont {Wang}}, \bibinfo {author}
		{\bibfnamefont {Z.-Q.}\ \bibnamefont {Yin}}, \bibinfo {author} {\bibfnamefont
			{D.-Y.}\ \bibnamefont {He}}, \bibinfo {author} {\bibfnamefont
			{Z.}~\bibnamefont {Zhou}}, \bibinfo {author} {\bibfnamefont {J.}~\bibnamefont
			{Teng}}, \bibinfo {author} {\bibfnamefont {W.}~\bibnamefont {Chen}}, \bibinfo
		{author} {\bibfnamefont {G.-C.}\ \bibnamefont {Guo}},\ and\ \bibinfo {author}
		{\bibfnamefont {Z.-F.}\ \bibnamefont {Han}},\ }\bibfield  {title} {\bibinfo
		{title} {Measurement-device-independent quantum key distribution for
			nonstandalone networks},\ }\href {https://doi.org/10.1364/PRJ.428309}
	{\bibfield  {journal} {\bibinfo  {journal} {Photon. Res.}\ }\textbf {\bibinfo
			{volume} {9}},\ \bibinfo {pages} {1881} (\bibinfo {year} {2021})}\BibitemShut
	{NoStop}%
	\bibitem [{\citenamefont {Woodward}\ \emph {et~al.}(2021)\citenamefont
		{Woodward}, \citenamefont {Lo}, \citenamefont {Pittaluga}, \citenamefont
		{Minder}, \citenamefont {Paraïso}, \citenamefont {Lucamarini}, \citenamefont
		{Yuan},\ and\ \citenamefont {Shields}}]{2021Woodward}%
	\BibitemOpen
	\bibfield  {author} {\bibinfo {author} {\bibfnamefont {R.~I.}\ \bibnamefont
			{Woodward}}, \bibinfo {author} {\bibfnamefont {Y.}~\bibnamefont {Lo}},
		\bibinfo {author} {\bibfnamefont {M.}~\bibnamefont {Pittaluga}}, \bibinfo
		{author} {\bibfnamefont {M.}~\bibnamefont {Minder}}, \bibinfo {author}
		{\bibfnamefont {T.}~\bibnamefont {Paraïso}}, \bibinfo {author}
		{\bibfnamefont {M.}~\bibnamefont {Lucamarini}}, \bibinfo {author}
		{\bibfnamefont {Z.}~\bibnamefont {Yuan}},\ and\ \bibinfo {author}
		{\bibfnamefont {A.}~\bibnamefont {Shields}},\ }\bibfield  {title} {\bibinfo
		{title} {Gigahertz measurement-device-independent quantum key distribution
			using directly modulated lasers},\ }\href
	{https://doi.org/https://doi.org/10.1038/s41534-021-00394-2} {\bibfield
		{journal} {\bibinfo  {journal} {npj Quantum Inf.}\ }\textbf {\bibinfo
			{volume} {7}},\ \bibinfo {pages} {58} (\bibinfo {year} {2021})}\BibitemShut
	{NoStop}%
	\bibitem [{\citenamefont {Lu}\ \emph {et~al.}(2022)\citenamefont {Lu},
		\citenamefont {Wang}, \citenamefont {Yin}, \citenamefont {Wang},
		\citenamefont {Wang}, \citenamefont {Fan-Yuan}, \citenamefont {Huang},
		\citenamefont {He}, \citenamefont {Chen}, \citenamefont {Zhou}, \citenamefont
		{Guo},\ and\ \citenamefont {Han}}]{2022Lu_MDI}%
	\BibitemOpen
	\bibfield  {author} {\bibinfo {author} {\bibfnamefont {F.-Y.}\ \bibnamefont
			{Lu}}, \bibinfo {author} {\bibfnamefont {Z.-H.}\ \bibnamefont {Wang}},
		\bibinfo {author} {\bibfnamefont {Z.-Q.}\ \bibnamefont {Yin}}, \bibinfo
		{author} {\bibfnamefont {S.}~\bibnamefont {Wang}}, \bibinfo {author}
		{\bibfnamefont {R.}~\bibnamefont {Wang}}, \bibinfo {author} {\bibfnamefont
			{G.-J.}\ \bibnamefont {Fan-Yuan}}, \bibinfo {author} {\bibfnamefont {X.-J.}\
			\bibnamefont {Huang}}, \bibinfo {author} {\bibfnamefont {D.-Y.}\ \bibnamefont
			{He}}, \bibinfo {author} {\bibfnamefont {W.}~\bibnamefont {Chen}}, \bibinfo
		{author} {\bibfnamefont {Z.}~\bibnamefont {Zhou}}, \bibinfo {author}
		{\bibfnamefont {G.-C.}\ \bibnamefont {Guo}},\ and\ \bibinfo {author}
		{\bibfnamefont {Z.-F.}\ \bibnamefont {Han}},\ }\bibfield  {title} {\bibinfo
		{title} {Unbalanced-basis-misalignment-tolerant
			measurement-device-independent quantum key distribution},\ }\href
	{https://doi.org/10.1364/OPTICA.454228} {\bibfield  {journal} {\bibinfo
			{journal} {Optica}\ }\textbf {\bibinfo {volume} {9}},\ \bibinfo {pages} {886}
		(\bibinfo {year} {2022})}\BibitemShut {NoStop}%
	\bibitem [{\citenamefont {Gu}\ \emph {et~al.}(2022)\citenamefont {Gu},
		\citenamefont {Cao}, \citenamefont {Fu}, \citenamefont {He}, \citenamefont
		{Yin}, \citenamefont {Yin},\ and\ \citenamefont {Chen}}]{2022Gu}%
	\BibitemOpen
	\bibfield  {author} {\bibinfo {author} {\bibfnamefont {J.}~\bibnamefont
			{Gu}}, \bibinfo {author} {\bibfnamefont {X.-Y.}\ \bibnamefont {Cao}},
		\bibinfo {author} {\bibfnamefont {Y.}~\bibnamefont {Fu}}, \bibinfo {author}
		{\bibfnamefont {Z.-W.}\ \bibnamefont {He}}, \bibinfo {author} {\bibfnamefont
			{Z.-J.}\ \bibnamefont {Yin}}, \bibinfo {author} {\bibfnamefont {H.-L.}\
			\bibnamefont {Yin}},\ and\ \bibinfo {author} {\bibfnamefont {Z.-B.}\
			\bibnamefont {Chen}},\ }\bibfield  {title} {\bibinfo {title} {Experimental
			measurement-device-independent type quantum key distribution with flawed and
			correlated sources},\ }\href {https://doi.org/10.1016/j.scib.2022.10.010}
	{\bibfield  {journal} {\bibinfo  {journal} {Sci. Bull.}\ }\textbf {\bibinfo
			{volume} {67}},\ \bibinfo {pages} {2167} (\bibinfo {year}
		{2022})}\BibitemShut {NoStop}%
	\bibitem [{\citenamefont {Liu}\ \emph {et~al.}(2023{\natexlab{a}})\citenamefont
		{Liu}, \citenamefont {Ma}, \citenamefont {Ding}, \citenamefont {Zhang},
		\citenamefont {Zhou},\ and\ \citenamefont {Wang}}]{2023Liu_MDI}%
	\BibitemOpen
	\bibfield  {author} {\bibinfo {author} {\bibfnamefont {J.-Y.}\ \bibnamefont
			{Liu}}, \bibinfo {author} {\bibfnamefont {X.}~\bibnamefont {Ma}}, \bibinfo
		{author} {\bibfnamefont {H.-J.}\ \bibnamefont {Ding}}, \bibinfo {author}
		{\bibfnamefont {C.-H.}\ \bibnamefont {Zhang}}, \bibinfo {author}
		{\bibfnamefont {X.-Y.}\ \bibnamefont {Zhou}},\ and\ \bibinfo {author}
		{\bibfnamefont {Q.}~\bibnamefont {Wang}},\ }\bibfield  {title} {\bibinfo
		{title} {Experimental demonstration of five-intensity
			measurement-device-independent quantum key distribution over 442 km},\ }\href
	{https://doi.org/10.1103/PhysRevA.108.022605} {\bibfield  {journal} {\bibinfo
			{journal} {Phys. Rev. A}\ }\textbf {\bibinfo {volume} {108}},\ \bibinfo
		{pages} {022605} (\bibinfo {year} {2023}{\natexlab{a}})}\BibitemShut
	{NoStop}%
	\bibitem [{\citenamefont {Takeoka}\ \emph {et~al.}(2014)\citenamefont
		{Takeoka}, \citenamefont {Guha},\ and\ \citenamefont {Wilde}}]{2014Takeoka}%
	\BibitemOpen
	\bibfield  {author} {\bibinfo {author} {\bibfnamefont {M.}~\bibnamefont
			{Takeoka}}, \bibinfo {author} {\bibfnamefont {S.}~\bibnamefont {Guha}},\ and\
		\bibinfo {author} {\bibfnamefont {M.~M.}\ \bibnamefont {Wilde}},\ }\bibfield
	{title} {\bibinfo {title} {Fundamental rate-loss tradeoff for optical quantum
			key distribution},\ }\href {https://doi.org/10.1038/ncomms6235} {\bibfield
		{journal} {\bibinfo  {journal} {Nat. Commun.}\ }\textbf {\bibinfo {volume}
			{5}},\ \bibinfo {pages} {5235} (\bibinfo {year} {2014})}\BibitemShut
	{NoStop}%
	\bibitem [{\citenamefont {Pirandola}\ \emph {et~al.}(2017)\citenamefont
		{Pirandola}, \citenamefont {Laurenza}, \citenamefont {Ottaviani},\ and\
		\citenamefont {Banchi}}]{2017Pirandola}%
	\BibitemOpen
	\bibfield  {author} {\bibinfo {author} {\bibfnamefont {S.}~\bibnamefont
			{Pirandola}}, \bibinfo {author} {\bibfnamefont {R.}~\bibnamefont {Laurenza}},
		\bibinfo {author} {\bibfnamefont {C.}~\bibnamefont {Ottaviani}},\ and\
		\bibinfo {author} {\bibfnamefont {L.}~\bibnamefont {Banchi}},\ }\bibfield
	{title} {\bibinfo {title} {Fundamental limits of repeaterless quantum
			communications},\ }\href {https://doi.org/10.1038/ncomms15043} {\bibfield
		{journal} {\bibinfo  {journal} {Nat. Commun.}\ }\textbf {\bibinfo {volume}
			{8}},\ \bibinfo {pages} {15043} (\bibinfo {year} {2017})}\BibitemShut
	{NoStop}%
	\bibitem [{\citenamefont {Lucamarini}\ \emph {et~al.}(2018)\citenamefont
		{Lucamarini}, \citenamefont {Yuan}, \citenamefont {Dynes},\ and\
		\citenamefont {Shields}}]{2018Lucamarini}%
	\BibitemOpen
	\bibfield  {author} {\bibinfo {author} {\bibfnamefont {M.}~\bibnamefont
			{Lucamarini}}, \bibinfo {author} {\bibfnamefont {Z.~L.}\ \bibnamefont
			{Yuan}}, \bibinfo {author} {\bibfnamefont {J.~F.}\ \bibnamefont {Dynes}},\
		and\ \bibinfo {author} {\bibfnamefont {A.~J.}\ \bibnamefont {Shields}},\
	}\bibfield  {title} {\bibinfo {title} {Overcoming the rate–distance limit
			of quantum key distribution without quantum repeaters},\ }\href
	{https://doi.org/10.1038/s41586-018-0066-6} {\bibfield  {journal} {\bibinfo
			{journal} {Nature}\ }\textbf {\bibinfo {volume} {557}},\ \bibinfo {pages}
		{400} (\bibinfo {year} {2018})}\BibitemShut {NoStop}%
	\bibitem [{\citenamefont {Ma}\ \emph {et~al.}(2018)\citenamefont {Ma},
		\citenamefont {Zeng},\ and\ \citenamefont {Zhou}}]{2018Ma}%
	\BibitemOpen
	\bibfield  {author} {\bibinfo {author} {\bibfnamefont {X.}~\bibnamefont
			{Ma}}, \bibinfo {author} {\bibfnamefont {P.}~\bibnamefont {Zeng}},\ and\
		\bibinfo {author} {\bibfnamefont {H.}~\bibnamefont {Zhou}},\ }\bibfield
	{title} {\bibinfo {title} {Phase-matching quantum key distribution},\ }\href
	{https://doi.org/10.1103/PhysRevX.8.031043} {\bibfield  {journal} {\bibinfo
			{journal} {Phys. Rev. X}\ }\textbf {\bibinfo {volume} {8}},\ \bibinfo {pages}
		{031043} (\bibinfo {year} {2018})}\BibitemShut {NoStop}%
	\bibitem [{\citenamefont {Wang}\ \emph {et~al.}(2018)\citenamefont {Wang},
		\citenamefont {Yu},\ and\ \citenamefont {Hu}}]{2018Wang}%
	\BibitemOpen
	\bibfield  {author} {\bibinfo {author} {\bibfnamefont {X.-B.}\ \bibnamefont
			{Wang}}, \bibinfo {author} {\bibfnamefont {Z.-W.}\ \bibnamefont {Yu}},\ and\
		\bibinfo {author} {\bibfnamefont {X.-L.}\ \bibnamefont {Hu}},\ }\bibfield
	{title} {\bibinfo {title} {Twin-field quantum key distribution with large
			misalignment error},\ }\href {https://doi.org/10.1103/PhysRevA.98.062323}
	{\bibfield  {journal} {\bibinfo  {journal} {Phys. Rev. A}\ }\textbf {\bibinfo
			{volume} {98}},\ \bibinfo {pages} {062323} (\bibinfo {year}
		{2018})}\BibitemShut {NoStop}%
	\bibitem [{\citenamefont {Curty}\ \emph {et~al.}(2019)\citenamefont {Curty},
		\citenamefont {Azuma},\ and\ \citenamefont {Lo}}]{2019Curty}%
	\BibitemOpen
	\bibfield  {author} {\bibinfo {author} {\bibfnamefont {M.}~\bibnamefont
			{Curty}}, \bibinfo {author} {\bibfnamefont {K.}~\bibnamefont {Azuma}},\ and\
		\bibinfo {author} {\bibfnamefont {H.-K.}\ \bibnamefont {Lo}},\ }\bibfield
	{title} {\bibinfo {title} {Simple security proof of twin-field type quantum
			key distribution protocol},\ }\href
	{https://doi.org/10.1038/s41534-019-0175-6} {\bibfield  {journal} {\bibinfo
			{journal} {npj Quantum Inf.}\ }\textbf {\bibinfo {volume} {5}},\ \bibinfo
		{pages} {64} (\bibinfo {year} {2019})}\BibitemShut {NoStop}%
	\bibitem [{\citenamefont {Wang}\ \emph {et~al.}(2019)\citenamefont {Wang},
		\citenamefont {He}, \citenamefont {Yin}, \citenamefont {Lu}, \citenamefont
		{Cui}, \citenamefont {Chen}, \citenamefont {Zhou}, \citenamefont {Guo},\ and\
		\citenamefont {Han}}]{2019Wangshuang-TF}%
	\BibitemOpen
	\bibfield  {author} {\bibinfo {author} {\bibfnamefont {S.}~\bibnamefont
			{Wang}}, \bibinfo {author} {\bibfnamefont {D.-Y.}\ \bibnamefont {He}},
		\bibinfo {author} {\bibfnamefont {Z.-Q.}\ \bibnamefont {Yin}}, \bibinfo
		{author} {\bibfnamefont {F.-Y.}\ \bibnamefont {Lu}}, \bibinfo {author}
		{\bibfnamefont {C.-H.}\ \bibnamefont {Cui}}, \bibinfo {author} {\bibfnamefont
			{W.}~\bibnamefont {Chen}}, \bibinfo {author} {\bibfnamefont {Z.}~\bibnamefont
			{Zhou}}, \bibinfo {author} {\bibfnamefont {G.-C.}\ \bibnamefont {Guo}},\ and\
		\bibinfo {author} {\bibfnamefont {Z.-F.}\ \bibnamefont {Han}},\ }\bibfield
	{title} {\bibinfo {title} {Beating the fundamental rate-distance limit in a
			proof-of-principle quantum key distribution system},\ }\href
	{https://doi.org/10.1103/PhysRevX.9.021046} {\bibfield  {journal} {\bibinfo
			{journal} {Phys. Rev. X}\ }\textbf {\bibinfo {volume} {9}},\ \bibinfo {pages}
		{021046} (\bibinfo {year} {2019})}\BibitemShut {NoStop}%
	\bibitem [{\citenamefont {Fang}\ \emph {et~al.}(2020)\citenamefont {Fang},
		\citenamefont {Zeng}, \citenamefont {Liu}, \citenamefont {Zou}, \citenamefont
		{Wu}, \citenamefont {Tang}, \citenamefont {Sheng}, \citenamefont {Xiang},
		\citenamefont {Zhang}, \citenamefont {Li} \emph {et~al.}}]{2020fang_PM}%
	\BibitemOpen
	\bibfield  {author} {\bibinfo {author} {\bibfnamefont {X.-T.}\ \bibnamefont
			{Fang}}, \bibinfo {author} {\bibfnamefont {P.}~\bibnamefont {Zeng}}, \bibinfo
		{author} {\bibfnamefont {H.}~\bibnamefont {Liu}}, \bibinfo {author}
		{\bibfnamefont {M.}~\bibnamefont {Zou}}, \bibinfo {author} {\bibfnamefont
			{W.}~\bibnamefont {Wu}}, \bibinfo {author} {\bibfnamefont {Y.-L.}\
			\bibnamefont {Tang}}, \bibinfo {author} {\bibfnamefont {Y.-J.}\ \bibnamefont
			{Sheng}}, \bibinfo {author} {\bibfnamefont {Y.}~\bibnamefont {Xiang}},
		\bibinfo {author} {\bibfnamefont {W.}~\bibnamefont {Zhang}}, \bibinfo
		{author} {\bibfnamefont {H.}~\bibnamefont {Li}}, \emph {et~al.},\ }\bibfield
	{title} {\bibinfo {title} {Implementation of quantum key distribution
			surpassing the linear rate-transmittance bound},\ }\href
	{https://doi.org/10.1038/s41566-020-0599-8} {\bibfield  {journal} {\bibinfo
			{journal} {Nat. Photon.}\ }\textbf {\bibinfo {volume} {14}},\ \bibinfo
		{pages} {422} (\bibinfo {year} {2020})}\BibitemShut {NoStop}%
	\bibitem [{\citenamefont {Pittaluga}\ \emph {et~al.}(2021)\citenamefont
		{Pittaluga}, \citenamefont {Minder}, \citenamefont {Lucamarini},
		\citenamefont {Sanzaro},\ and\ \citenamefont {Shields}}]{2021Pittaluga}%
	\BibitemOpen
	\bibfield  {author} {\bibinfo {author} {\bibfnamefont {M.}~\bibnamefont
			{Pittaluga}}, \bibinfo {author} {\bibfnamefont {M.}~\bibnamefont {Minder}},
		\bibinfo {author} {\bibfnamefont {M.}~\bibnamefont {Lucamarini}}, \bibinfo
		{author} {\bibfnamefont {M.}~\bibnamefont {Sanzaro}},\ and\ \bibinfo {author}
		{\bibfnamefont {A.~J.}\ \bibnamefont {Shields}},\ }\bibfield  {title}
	{\bibinfo {title} {600-km repeater-like quantum communications with dual-band
			stabilization},\ }\href {https://doi.org/10.1038/s41566-021-00811-0}
	{\bibfield  {journal} {\bibinfo  {journal} {Nat. Photon.}\ }\textbf {\bibinfo
			{volume} {15}},\ \bibinfo {pages} {530} (\bibinfo {year} {2021})}\BibitemShut
	{NoStop}%
	\bibitem [{\citenamefont {Wang}\ \emph
		{et~al.}(2022{\natexlab{a}})\citenamefont {Wang}, \citenamefont {Yin},
		\citenamefont {He}, \citenamefont {Chen}, \citenamefont {Wang}, \citenamefont
		{Ye}, \citenamefont {Zhou}, \citenamefont {Fan-Yuan}, \citenamefont {Wang},
		\citenamefont {Chen}, \citenamefont {Zhu}, \citenamefont {Morozov},
		\citenamefont {Divochiy}, \citenamefont {Zhou}, \citenamefont {Guo},\ and\
		\citenamefont {Han}}]{2022wang}%
	\BibitemOpen
	\bibfield  {author} {\bibinfo {author} {\bibfnamefont {S.}~\bibnamefont
			{Wang}}, \bibinfo {author} {\bibfnamefont {Z.-Q.}\ \bibnamefont {Yin}},
		\bibinfo {author} {\bibfnamefont {D.-Y.}\ \bibnamefont {He}}, \bibinfo
		{author} {\bibfnamefont {W.}~\bibnamefont {Chen}}, \bibinfo {author}
		{\bibfnamefont {R.-Q.}\ \bibnamefont {Wang}}, \bibinfo {author}
		{\bibfnamefont {P.}~\bibnamefont {Ye}}, \bibinfo {author} {\bibfnamefont
			{Y.}~\bibnamefont {Zhou}}, \bibinfo {author} {\bibfnamefont {G.-J.}\
			\bibnamefont {Fan-Yuan}}, \bibinfo {author} {\bibfnamefont {F.-X.}\
			\bibnamefont {Wang}}, \bibinfo {author} {\bibfnamefont {W.}~\bibnamefont
			{Chen}}, \bibinfo {author} {\bibfnamefont {Y.-G.}\ \bibnamefont {Zhu}},
		\bibinfo {author} {\bibfnamefont {P.~V.}\ \bibnamefont {Morozov}}, \bibinfo
		{author} {\bibfnamefont {A.~V.}\ \bibnamefont {Divochiy}}, \bibinfo {author}
		{\bibfnamefont {Z.}~\bibnamefont {Zhou}}, \bibinfo {author} {\bibfnamefont
			{G.-C.}\ \bibnamefont {Guo}},\ and\ \bibinfo {author} {\bibfnamefont {Z.-F.}\
			\bibnamefont {Han}},\ }\bibfield  {title} {\bibinfo {title} {Twin-field
			quantum key distribution over 830-km fibre},\ }\href
	{https://doi.org/10.1038/s41566-021-00928-2} {\bibfield  {journal} {\bibinfo
			{journal} {Nat. Photon.}\ }\textbf {\bibinfo {volume} {16}},\ \bibinfo
		{pages} {157} (\bibinfo {year} {2022}{\natexlab{a}})}\BibitemShut {NoStop}%
	\bibitem [{\citenamefont {Chen}\ \emph
		{et~al.}(2022{\natexlab{b}})\citenamefont {Chen}, \citenamefont {Zhang},
		\citenamefont {Liu}, \citenamefont {Jiang}, \citenamefont {Zhao},
		\citenamefont {Zhang}, \citenamefont {Chen}, \citenamefont {Li},
		\citenamefont {You}, \citenamefont {Wang}, \citenamefont {Chen},
		\citenamefont {Wang}, \citenamefont {Zhang},\ and\ \citenamefont
		{Pan}}]{2022Chen}%
	\BibitemOpen
	\bibfield  {author} {\bibinfo {author} {\bibfnamefont {J.-P.}\ \bibnamefont
			{Chen}}, \bibinfo {author} {\bibfnamefont {C.}~\bibnamefont {Zhang}},
		\bibinfo {author} {\bibfnamefont {Y.}~\bibnamefont {Liu}}, \bibinfo {author}
		{\bibfnamefont {C.}~\bibnamefont {Jiang}}, \bibinfo {author} {\bibfnamefont
			{D.-F.}\ \bibnamefont {Zhao}}, \bibinfo {author} {\bibfnamefont {W.-J.}\
			\bibnamefont {Zhang}}, \bibinfo {author} {\bibfnamefont {F.-X.}\ \bibnamefont
			{Chen}}, \bibinfo {author} {\bibfnamefont {H.}~\bibnamefont {Li}}, \bibinfo
		{author} {\bibfnamefont {L.-X.}\ \bibnamefont {You}}, \bibinfo {author}
		{\bibfnamefont {Z.}~\bibnamefont {Wang}}, \bibinfo {author} {\bibfnamefont
			{Y.}~\bibnamefont {Chen}}, \bibinfo {author} {\bibfnamefont {X.-B.}\
			\bibnamefont {Wang}}, \bibinfo {author} {\bibfnamefont {Q.}~\bibnamefont
			{Zhang}},\ and\ \bibinfo {author} {\bibfnamefont {J.-W.}\ \bibnamefont
			{Pan}},\ }\bibfield  {title} {\bibinfo {title} {Quantum key distribution over
			658 km fiber with distributed vibration sensing},\ }\href
	{https://doi.org/10.1103/PhysRevLett.128.180502} {\bibfield  {journal}
		{\bibinfo  {journal} {Phys. Rev. Lett.}\ }\textbf {\bibinfo {volume} {128}},\
		\bibinfo {pages} {180502} (\bibinfo {year} {2022}{\natexlab{b}})}\BibitemShut
	{NoStop}%
	\bibitem [{\citenamefont {Zhou}\ \emph
		{et~al.}(2023{\natexlab{a}})\citenamefont {Zhou}, \citenamefont {Lin},
		\citenamefont {Jing},\ and\ \citenamefont {Yuan}}]{2023Zhou_TF}%
	\BibitemOpen
	\bibfield  {author} {\bibinfo {author} {\bibfnamefont {L.}~\bibnamefont
			{Zhou}}, \bibinfo {author} {\bibfnamefont {J.}~\bibnamefont {Lin}}, \bibinfo
		{author} {\bibfnamefont {Y.}~\bibnamefont {Jing}},\ and\ \bibinfo {author}
		{\bibfnamefont {Z.}~\bibnamefont {Yuan}},\ }\bibfield  {title} {\bibinfo
		{title} {Twin-field quantum key distribution without optical frequency
			dissemination},\ }\href {https://doi.org/10.1038/s41467-023-36573-2}
	{\bibfield  {journal} {\bibinfo  {journal} {Nat. Commun.}\ }\textbf {\bibinfo
			{volume} {14}},\ \bibinfo {pages} {928} (\bibinfo {year}
		{2023}{\natexlab{a}})}\BibitemShut {NoStop}%
	\bibitem [{\citenamefont {Li}\ \emph {et~al.}(2023{\natexlab{b}})\citenamefont
		{Li}, \citenamefont {Zhang}, \citenamefont {Lu}, \citenamefont {Li},
		\citenamefont {Jiang}, \citenamefont {Liu}, \citenamefont {Huang},
		\citenamefont {Li}, \citenamefont {Wang}, \citenamefont {Wang}, \citenamefont
		{Zhang}, \citenamefont {You}, \citenamefont {Xu},\ and\ \citenamefont
		{Pan}}]{2023Li_TF}%
	\BibitemOpen
	\bibfield  {author} {\bibinfo {author} {\bibfnamefont {W.}~\bibnamefont
			{Li}}, \bibinfo {author} {\bibfnamefont {L.}~\bibnamefont {Zhang}}, \bibinfo
		{author} {\bibfnamefont {Y.}~\bibnamefont {Lu}}, \bibinfo {author}
		{\bibfnamefont {Z.-P.}\ \bibnamefont {Li}}, \bibinfo {author} {\bibfnamefont
			{C.}~\bibnamefont {Jiang}}, \bibinfo {author} {\bibfnamefont
			{Y.}~\bibnamefont {Liu}}, \bibinfo {author} {\bibfnamefont {J.}~\bibnamefont
			{Huang}}, \bibinfo {author} {\bibfnamefont {H.}~\bibnamefont {Li}}, \bibinfo
		{author} {\bibfnamefont {Z.}~\bibnamefont {Wang}}, \bibinfo {author}
		{\bibfnamefont {X.-B.}\ \bibnamefont {Wang}}, \bibinfo {author}
		{\bibfnamefont {Q.}~\bibnamefont {Zhang}}, \bibinfo {author} {\bibfnamefont
			{L.}~\bibnamefont {You}}, \bibinfo {author} {\bibfnamefont {F.}~\bibnamefont
			{Xu}},\ and\ \bibinfo {author} {\bibfnamefont {J.-W.}\ \bibnamefont {Pan}},\
	}\bibfield  {title} {\bibinfo {title} {Twin-field quantum key distribution
			without phase locking},\ }\href
	{https://doi.org/10.1103/physrevlett.130.250802} {\bibfield  {journal}
		{\bibinfo  {journal} {Phys. Rev. Lett.}\ }\textbf {\bibinfo {volume} {130}},\
		\bibinfo {pages} {250802} (\bibinfo {year} {2023}{\natexlab{b}})}\BibitemShut
	{NoStop}%
	\bibitem [{\citenamefont {Liu}\ \emph {et~al.}(2023{\natexlab{b}})\citenamefont
		{Liu}, \citenamefont {Zhang}, \citenamefont {Jiang}, \citenamefont {Chen},
		\citenamefont {Zhang}, \citenamefont {Pan}, \citenamefont {Ma}, \citenamefont
		{Dong}, \citenamefont {Xiong}, \citenamefont {Zhang}, \citenamefont {Li},
		\citenamefont {Wang}, \citenamefont {Wu}, \citenamefont {Chen}, \citenamefont
		{You}, \citenamefont {Wang}, \citenamefont {Zhang},\ and\ \citenamefont
		{Pan}}]{2023Liu}%
	\BibitemOpen
	\bibfield  {author} {\bibinfo {author} {\bibfnamefont {Y.}~\bibnamefont
			{Liu}}, \bibinfo {author} {\bibfnamefont {W.-J.}\ \bibnamefont {Zhang}},
		\bibinfo {author} {\bibfnamefont {C.}~\bibnamefont {Jiang}}, \bibinfo
		{author} {\bibfnamefont {J.-P.}\ \bibnamefont {Chen}}, \bibinfo {author}
		{\bibfnamefont {C.}~\bibnamefont {Zhang}}, \bibinfo {author} {\bibfnamefont
			{W.-X.}\ \bibnamefont {Pan}}, \bibinfo {author} {\bibfnamefont
			{D.}~\bibnamefont {Ma}}, \bibinfo {author} {\bibfnamefont {H.}~\bibnamefont
			{Dong}}, \bibinfo {author} {\bibfnamefont {J.-M.}\ \bibnamefont {Xiong}},
		\bibinfo {author} {\bibfnamefont {C.-J.}\ \bibnamefont {Zhang}}, \bibinfo
		{author} {\bibfnamefont {H.}~\bibnamefont {Li}}, \bibinfo {author}
		{\bibfnamefont {R.-C.}\ \bibnamefont {Wang}}, \bibinfo {author}
		{\bibfnamefont {J.}~\bibnamefont {Wu}}, \bibinfo {author} {\bibfnamefont
			{T.-Y.}\ \bibnamefont {Chen}}, \bibinfo {author} {\bibfnamefont
			{L.}~\bibnamefont {You}}, \bibinfo {author} {\bibfnamefont {X.-B.}\
			\bibnamefont {Wang}}, \bibinfo {author} {\bibfnamefont {Q.}~\bibnamefont
			{Zhang}},\ and\ \bibinfo {author} {\bibfnamefont {J.-W.}\ \bibnamefont
			{Pan}},\ }\bibfield  {title} {\bibinfo {title} {Experimental twin-field
			quantum key distribution over 1000 km fiber distance},\ }\href
	{https://doi.org/10.1103/PhysRevLett.130.210801} {\bibfield  {journal}
		{\bibinfo  {journal} {Phys. Rev. Lett.}\ }\textbf {\bibinfo {volume} {130}},\
		\bibinfo {pages} {210801} (\bibinfo {year} {2023}{\natexlab{b}})}\BibitemShut
	{NoStop}%
	\bibitem [{\citenamefont {Xie}\ \emph {et~al.}(2022)\citenamefont {Xie},
		\citenamefont {Lu}, \citenamefont {Weng}, \citenamefont {Cao}, \citenamefont
		{Jia}, \citenamefont {Bao}, \citenamefont {Wang}, \citenamefont {Fu},
		\citenamefont {Yin},\ and\ \citenamefont {Chen}}]{2022Xie}%
	\BibitemOpen
	\bibfield  {author} {\bibinfo {author} {\bibfnamefont {Y.-M.}\ \bibnamefont
			{Xie}}, \bibinfo {author} {\bibfnamefont {Y.-S.}\ \bibnamefont {Lu}},
		\bibinfo {author} {\bibfnamefont {C.-X.}\ \bibnamefont {Weng}}, \bibinfo
		{author} {\bibfnamefont {X.-Y.}\ \bibnamefont {Cao}}, \bibinfo {author}
		{\bibfnamefont {Z.-Y.}\ \bibnamefont {Jia}}, \bibinfo {author} {\bibfnamefont
			{Y.}~\bibnamefont {Bao}}, \bibinfo {author} {\bibfnamefont {Y.}~\bibnamefont
			{Wang}}, \bibinfo {author} {\bibfnamefont {Y.}~\bibnamefont {Fu}}, \bibinfo
		{author} {\bibfnamefont {H.-L.}\ \bibnamefont {Yin}},\ and\ \bibinfo {author}
		{\bibfnamefont {Z.-B.}\ \bibnamefont {Chen}},\ }\bibfield  {title} {\bibinfo
		{title} {Breaking the rate-loss bound of quantum key distribution with
			asynchronous two-photon interference},\ }\href
	{https://doi.org/10.1103/PRXQuantum.3.020315} {\bibfield  {journal} {\bibinfo
			{journal} {PRX Quantum}\ }\textbf {\bibinfo {volume} {3}},\ \bibinfo {pages}
		{020315} (\bibinfo {year} {2022})}\BibitemShut {NoStop}%
	\bibitem [{\citenamefont {Zeng}\ \emph {et~al.}(2022)\citenamefont {Zeng},
		\citenamefont {Zhou}, \citenamefont {Wu},\ and\ \citenamefont
		{Ma}}]{2022Zeng}%
	\BibitemOpen
	\bibfield  {author} {\bibinfo {author} {\bibfnamefont {P.}~\bibnamefont
			{Zeng}}, \bibinfo {author} {\bibfnamefont {H.}~\bibnamefont {Zhou}}, \bibinfo
		{author} {\bibfnamefont {W.}~\bibnamefont {Wu}},\ and\ \bibinfo {author}
		{\bibfnamefont {X.}~\bibnamefont {Ma}},\ }\bibfield  {title} {\bibinfo
		{title} {Mode-pairing quantum key distribution},\ }\href
	{https://doi.org/10.1103/10.1038/s41467-022-31534-7} {\bibfield  {journal}
		{\bibinfo  {journal} {Nat. Commun.}\ }\textbf {\bibinfo {volume} {13}},\
		\bibinfo {pages} {3903} (\bibinfo {year} {2022})}\BibitemShut {NoStop}%
	\bibitem [{\citenamefont {Zhu}\ \emph {et~al.}(2023{\natexlab{a}})\citenamefont
		{Zhu}, \citenamefont {Huang}, \citenamefont {Liu}, \citenamefont {Zeng},
		\citenamefont {Zou}, \citenamefont {Dai}, \citenamefont {Tang}, \citenamefont
		{Li}, \citenamefont {You}, \citenamefont {Wang}, \citenamefont {Chen},
		\citenamefont {Ma}, \citenamefont {Chen},\ and\ \citenamefont
		{Pan}}]{2023Zhu}%
	\BibitemOpen
	\bibfield  {author} {\bibinfo {author} {\bibfnamefont {H.-T.}\ \bibnamefont
			{Zhu}}, \bibinfo {author} {\bibfnamefont {Y.}~\bibnamefont {Huang}}, \bibinfo
		{author} {\bibfnamefont {H.}~\bibnamefont {Liu}}, \bibinfo {author}
		{\bibfnamefont {P.}~\bibnamefont {Zeng}}, \bibinfo {author} {\bibfnamefont
			{M.}~\bibnamefont {Zou}}, \bibinfo {author} {\bibfnamefont {Y.}~\bibnamefont
			{Dai}}, \bibinfo {author} {\bibfnamefont {S.}~\bibnamefont {Tang}}, \bibinfo
		{author} {\bibfnamefont {H.}~\bibnamefont {Li}}, \bibinfo {author}
		{\bibfnamefont {L.}~\bibnamefont {You}}, \bibinfo {author} {\bibfnamefont
			{Z.}~\bibnamefont {Wang}}, \bibinfo {author} {\bibfnamefont {Y.-A.}\
			\bibnamefont {Chen}}, \bibinfo {author} {\bibfnamefont {X.}~\bibnamefont
			{Ma}}, \bibinfo {author} {\bibfnamefont {T.-Y.}\ \bibnamefont {Chen}},\ and\
		\bibinfo {author} {\bibfnamefont {J.-W.}\ \bibnamefont {Pan}},\ }\bibfield
	{title} {\bibinfo {title} {Experimental mode-pairing
			measurement-device-independent quantum key distribution without global phase
			locking},\ }\href {https://doi.org/10.1103/PhysRevLett.130.030801} {\bibfield
		{journal} {\bibinfo  {journal} {Phys. Rev. Lett.}\ }\textbf {\bibinfo
			{volume} {130}},\ \bibinfo {pages} {030801} (\bibinfo {year}
		{2023}{\natexlab{a}})}\BibitemShut {NoStop}%
	\bibitem [{\citenamefont {Zhou}\ \emph
		{et~al.}(2023{\natexlab{b}})\citenamefont {Zhou}, \citenamefont {Lin},
		\citenamefont {Xie}, \citenamefont {Lu}, \citenamefont {Jing}, \citenamefont
		{Yin},\ and\ \citenamefont {Yuan}}]{2023Zhou}%
	\BibitemOpen
	\bibfield  {author} {\bibinfo {author} {\bibfnamefont {L.}~\bibnamefont
			{Zhou}}, \bibinfo {author} {\bibfnamefont {J.}~\bibnamefont {Lin}}, \bibinfo
		{author} {\bibfnamefont {Y.-M.}\ \bibnamefont {Xie}}, \bibinfo {author}
		{\bibfnamefont {Y.-S.}\ \bibnamefont {Lu}}, \bibinfo {author} {\bibfnamefont
			{Y.}~\bibnamefont {Jing}}, \bibinfo {author} {\bibfnamefont {H.-L.}\
			\bibnamefont {Yin}},\ and\ \bibinfo {author} {\bibfnamefont {Z.}~\bibnamefont
			{Yuan}},\ }\bibfield  {title} {\bibinfo {title} {Experimental quantum
			communication overcomes the rate-loss limit without global phase tracking},\
	}\href {https://doi.org/10.1103/PhysRevLett.130.250801} {\bibfield  {journal}
		{\bibinfo  {journal} {Phys. Rev. Lett.}\ }\textbf {\bibinfo {volume} {130}},\
		\bibinfo {pages} {250801} (\bibinfo {year} {2023}{\natexlab{b}})}\BibitemShut
	{NoStop}%
	\bibitem [{\citenamefont {Wang}\ \emph {et~al.}(2023)\citenamefont {Wang},
		\citenamefont {Wang}, \citenamefont {Yin}, \citenamefont {Wang},
		\citenamefont {Lu}, \citenamefont {Chen}, \citenamefont {He}, \citenamefont
		{Guo},\ and\ \citenamefont {Han}}]{2023wang}%
	\BibitemOpen
	\bibfield  {author} {\bibinfo {author} {\bibfnamefont {Z.-H.}\ \bibnamefont
			{Wang}}, \bibinfo {author} {\bibfnamefont {R.}~\bibnamefont {Wang}}, \bibinfo
		{author} {\bibfnamefont {Z.-Q.}\ \bibnamefont {Yin}}, \bibinfo {author}
		{\bibfnamefont {S.}~\bibnamefont {Wang}}, \bibinfo {author} {\bibfnamefont
			{F.-Y.}\ \bibnamefont {Lu}}, \bibinfo {author} {\bibfnamefont
			{W.}~\bibnamefont {Chen}}, \bibinfo {author} {\bibfnamefont {D.-Y.}\
			\bibnamefont {He}}, \bibinfo {author} {\bibfnamefont {G.-C.}\ \bibnamefont
			{Guo}},\ and\ \bibinfo {author} {\bibfnamefont {Z.-F.}\ \bibnamefont {Han}},\
	}\bibfield  {title} {\bibinfo {title} {Tight finite-key analysis for
			mode-pairing quantum key distribution},\ }\href
	{https://doi.org/10.1038/s42005-023-01382-y} {\bibfield  {journal} {\bibinfo
			{journal} {Commun. Phys.}\ }\textbf {\bibinfo {volume} {6}},\ \bibinfo
		{pages} {265} (\bibinfo {year} {2023})}\BibitemShut {NoStop}%
	\bibitem [{\citenamefont {Bai}\ \emph {et~al.}(2023)\citenamefont {Bai},
		\citenamefont {Xie}, \citenamefont {Yao}, \citenamefont {Yin},\ and\
		\citenamefont {Chen}}]{2023Bai}%
	\BibitemOpen
	\bibfield  {author} {\bibinfo {author} {\bibfnamefont {J.-L.}\ \bibnamefont
			{Bai}}, \bibinfo {author} {\bibfnamefont {Y.-M.}\ \bibnamefont {Xie}},
		\bibinfo {author} {\bibfnamefont {F.}~\bibnamefont {Yao}}, \bibinfo {author}
		{\bibfnamefont {H.-L.}\ \bibnamefont {Yin}},\ and\ \bibinfo {author}
		{\bibfnamefont {Z.-B.}\ \bibnamefont {Chen}},\ }\bibfield  {title} {\bibinfo
		{title} {Asynchronous measurement-device-independent quantum key distribution
			with hybrid source},\ }\href {https://doi.org/10.1364/OL.491511} {\bibfield
		{journal} {\bibinfo  {journal} {Opt. Lett.}\ }\textbf {\bibinfo {volume}
			{48}},\ \bibinfo {pages} {3551} (\bibinfo {year} {2023})}\BibitemShut
	{NoStop}%
	\bibitem [{\citenamefont {Xie}\ \emph {et~al.}(2023)\citenamefont {Xie},
		\citenamefont {Bai}, \citenamefont {Lu}, \citenamefont {Weng}, \citenamefont
		{Yin},\ and\ \citenamefont {Chen}}]{2023Xie}%
	\BibitemOpen
	\bibfield  {author} {\bibinfo {author} {\bibfnamefont {Y.-M.}\ \bibnamefont
			{Xie}}, \bibinfo {author} {\bibfnamefont {J.-L.}\ \bibnamefont {Bai}},
		\bibinfo {author} {\bibfnamefont {Y.-S.}\ \bibnamefont {Lu}}, \bibinfo
		{author} {\bibfnamefont {C.-X.}\ \bibnamefont {Weng}}, \bibinfo {author}
		{\bibfnamefont {H.-L.}\ \bibnamefont {Yin}},\ and\ \bibinfo {author}
		{\bibfnamefont {Z.-B.}\ \bibnamefont {Chen}},\ }\bibfield  {title} {\bibinfo
		{title} {Advantages of asynchronous measurement-device-independent quantum
			key distribution in intercity networks},\ }\href
	{https://doi.org/10.1103/PhysRevApplied.19.054070} {\bibfield  {journal}
		{\bibinfo  {journal} {Phys. Rev. Appl.}\ }\textbf {\bibinfo {volume} {19}},\
		\bibinfo {pages} {054070} (\bibinfo {year} {2023})}\BibitemShut {NoStop}%
	\bibitem [{\citenamefont {Liu}\ \emph {et~al.}(2023{\natexlab{c}})\citenamefont
		{Liu}, \citenamefont {Luo}, \citenamefont {Zhang},\ and\ \citenamefont
		{Wei}}]{2023Liu_x}%
	\BibitemOpen
	\bibfield  {author} {\bibinfo {author} {\bibfnamefont {X.}~\bibnamefont
			{Liu}}, \bibinfo {author} {\bibfnamefont {D.}~\bibnamefont {Luo}}, \bibinfo
		{author} {\bibfnamefont {Z.}~\bibnamefont {Zhang}},\ and\ \bibinfo {author}
		{\bibfnamefont {K.}~\bibnamefont {Wei}},\ }\bibfield  {title} {\bibinfo
		{title} {Mode-pairing quantum key distribution with advantage distillation},\
	}\href {https://doi.org/10.1103/PhysRevA.107.062613} {\bibfield  {journal}
		{\bibinfo  {journal} {Phys. Rev. A}\ }\textbf {\bibinfo {volume} {107}},\
		\bibinfo {pages} {062613} (\bibinfo {year} {2023}{\natexlab{c}})}\BibitemShut
	{NoStop}%
	\bibitem [{\citenamefont {Chen}\ \emph {et~al.}(2021)\citenamefont {Chen},
		\citenamefont {Zhang}, \citenamefont {Chen}, \citenamefont {Cai},
		\citenamefont {Liao}, \citenamefont {Zhang}, \citenamefont {Chen},
		\citenamefont {Yin}, \citenamefont {Ren}, \citenamefont {Chen} \emph
		{et~al.}}]{2021Chen_Yu}%
	\BibitemOpen
	\bibfield  {author} {\bibinfo {author} {\bibfnamefont {Y.-A.}\ \bibnamefont
			{Chen}}, \bibinfo {author} {\bibfnamefont {Q.}~\bibnamefont {Zhang}},
		\bibinfo {author} {\bibfnamefont {T.-Y.}\ \bibnamefont {Chen}}, \bibinfo
		{author} {\bibfnamefont {W.-Q.}\ \bibnamefont {Cai}}, \bibinfo {author}
		{\bibfnamefont {S.-K.}\ \bibnamefont {Liao}}, \bibinfo {author}
		{\bibfnamefont {J.}~\bibnamefont {Zhang}}, \bibinfo {author} {\bibfnamefont
			{K.}~\bibnamefont {Chen}}, \bibinfo {author} {\bibfnamefont {J.}~\bibnamefont
			{Yin}}, \bibinfo {author} {\bibfnamefont {J.-G.}\ \bibnamefont {Ren}},
		\bibinfo {author} {\bibfnamefont {Z.}~\bibnamefont {Chen}}, \emph {et~al.},\
	}\bibfield  {title} {\bibinfo {title} {An integrated space-to-ground quantum
			communication network over 4,600 kilometres},\ }\href
	{https://doi.org/10.1038/s41586-020-03093-8} {\bibfield  {journal} {\bibinfo
			{journal} {Nature}\ }\textbf {\bibinfo {volume} {589}},\ \bibinfo {pages}
		{214} (\bibinfo {year} {2021})}\BibitemShut {NoStop}%
	\bibitem [{\citenamefont {Maurer}(1993)}]{1993Maurer}%
	\BibitemOpen
	\bibfield  {author} {\bibinfo {author} {\bibfnamefont {U.}~\bibnamefont
			{Maurer}},\ }\bibfield  {title} {\bibinfo {title} {Secret key agreement by
			public discussion from common information},\ }\href
	{https://doi.org/10.1109/18.256484} {\bibfield  {journal} {\bibinfo
			{journal} {IEEE Trans. Inf. Theory}\ }\textbf {\bibinfo {volume} {39}},\
		\bibinfo {pages} {733} (\bibinfo {year} {1993})}\BibitemShut {NoStop}%
	\bibitem [{\citenamefont {Gottesman}\ and\ \citenamefont
		{Lo}(2003)}]{2003Gottesman}%
	\BibitemOpen
	\bibfield  {author} {\bibinfo {author} {\bibfnamefont {D.}~\bibnamefont
			{Gottesman}}\ and\ \bibinfo {author} {\bibfnamefont {H.-K.}\ \bibnamefont
			{Lo}},\ }\bibfield  {title} {\bibinfo {title} {Proof of security of quantum
			key distribution with two-way classical communications},\ }\href
	{https://doi.org/10.1109/TIT.2002.807289} {\bibfield  {journal} {\bibinfo
			{journal} {IEEE Trans. Inf. Theory}\ }\textbf {\bibinfo {volume} {49}},\
		\bibinfo {pages} {457} (\bibinfo {year} {2003})}\BibitemShut {NoStop}%
	\bibitem [{\citenamefont {Ma}\ \emph {et~al.}(2006)\citenamefont {Ma},
		\citenamefont {Fung}, \citenamefont {Dupuis}, \citenamefont {Chen},
		\citenamefont {Tamaki},\ and\ \citenamefont {Lo}}]{2006Ma}%
	\BibitemOpen
	\bibfield  {author} {\bibinfo {author} {\bibfnamefont {X.}~\bibnamefont
			{Ma}}, \bibinfo {author} {\bibfnamefont {C.-H.~F.}\ \bibnamefont {Fung}},
		\bibinfo {author} {\bibfnamefont {F.}~\bibnamefont {Dupuis}}, \bibinfo
		{author} {\bibfnamefont {K.}~\bibnamefont {Chen}}, \bibinfo {author}
		{\bibfnamefont {K.}~\bibnamefont {Tamaki}},\ and\ \bibinfo {author}
		{\bibfnamefont {H.-K.}\ \bibnamefont {Lo}},\ }\bibfield  {title} {\bibinfo
		{title} {Decoy-state quantum key distribution with two-way classical
			postprocessing},\ }\href {https://doi.org/10.1103/PhysRevA.74.032330}
	{\bibfield  {journal} {\bibinfo  {journal} {Phys. Rev. A}\ }\textbf {\bibinfo
			{volume} {74}},\ \bibinfo {pages} {032330} (\bibinfo {year}
		{2006})}\BibitemShut {NoStop}%
	\bibitem [{\citenamefont {Khatri}\ and\ \citenamefont
		{L\"utkenhaus}(2017)}]{2017Khatri}%
	\BibitemOpen
	\bibfield  {author} {\bibinfo {author} {\bibfnamefont {S.}~\bibnamefont
			{Khatri}}\ and\ \bibinfo {author} {\bibfnamefont {N.}~\bibnamefont
			{L\"utkenhaus}},\ }\bibfield  {title} {\bibinfo {title} {Numerical evidence
			for bound secrecy from two-way postprocessing in quantum key distribution},\
	}\href {https://doi.org/10.1103/PhysRevA.95.042320} {\bibfield  {journal}
		{\bibinfo  {journal} {Phys. Rev. A}\ }\textbf {\bibinfo {volume} {95}},\
		\bibinfo {pages} {042320} (\bibinfo {year} {2017})}\BibitemShut {NoStop}%
	\bibitem [{\citenamefont {Renner}(2008)}]{2008Renner}%
	\BibitemOpen
	\bibfield  {author} {\bibinfo {author} {\bibfnamefont {R.}~\bibnamefont
			{Renner}},\ }\bibfield  {title} {\bibinfo {title} {Security of quantum key
			distribution},\ }\href {https://doi.org/10.1103/10.1142/S0219749908003256}
	{\bibfield  {journal} {\bibinfo  {journal} {Int. J. Quantum Inf.}\ }\textbf
		{\bibinfo {volume} {06}},\ \bibinfo {pages} {1} (\bibinfo {year}
		{2008})}\BibitemShut {NoStop}%
	\bibitem [{\citenamefont {Tan}\ \emph {et~al.}(2020)\citenamefont {Tan},
		\citenamefont {Lim},\ and\ \citenamefont {Renner}}]{2020tan}%
	\BibitemOpen
	\bibfield  {author} {\bibinfo {author} {\bibfnamefont {E.~Y.-Z.}\
			\bibnamefont {Tan}}, \bibinfo {author} {\bibfnamefont {C.~C.-W.}\
			\bibnamefont {Lim}},\ and\ \bibinfo {author} {\bibfnamefont {R.}~\bibnamefont
			{Renner}},\ }\bibfield  {title} {\bibinfo {title} {Advantage distillation for
			device-independent quantum key distribution},\ }\href
	{https://doi.org/10.1103/PhysRevLett.124.020502} {\bibfield  {journal}
		{\bibinfo  {journal} {Phys. Rev. Lett.}\ }\textbf {\bibinfo {volume} {124}},\
		\bibinfo {pages} {020502} (\bibinfo {year} {2020})}\BibitemShut {NoStop}%
	\bibitem [{\citenamefont {Hu}\ \emph {et~al.}(2023)\citenamefont {Hu},
		\citenamefont {Zhang},\ and\ \citenamefont {Li}}]{2023Hu}%
	\BibitemOpen
	\bibfield  {author} {\bibinfo {author} {\bibfnamefont {L.-W.}\ \bibnamefont
			{Hu}}, \bibinfo {author} {\bibfnamefont {C.-M.}\ \bibnamefont {Zhang}},\ and\
		\bibinfo {author} {\bibfnamefont {H.-W.}\ \bibnamefont {Li}},\ }\bibfield
	{title} {\bibinfo {title} {Practical measurement-device-independent quantum
			key distribution with advantage distillation},\ }\href
	{https://doi.org/https://doi.org/10.1007/s11128-022-03810-4} {\bibfield
		{journal} {\bibinfo  {journal} {Quantum Inf. Process.}\ }\textbf {\bibinfo
			{volume} {22}},\ \bibinfo {pages} {77} (\bibinfo {year} {2023})}\BibitemShut
	{NoStop}%
	\bibitem [{\citenamefont {Zhu}\ \emph {et~al.}(2023{\natexlab{b}})\citenamefont
		{Zhu}, \citenamefont {Zhang}, \citenamefont {Wang},\ and\ \citenamefont
		{Li}}]{2023ZhuRFI}%
	\BibitemOpen
	\bibfield  {author} {\bibinfo {author} {\bibfnamefont {J.-R.}\ \bibnamefont
			{Zhu}}, \bibinfo {author} {\bibfnamefont {C.-M.}\ \bibnamefont {Zhang}},
		\bibinfo {author} {\bibfnamefont {R.}~\bibnamefont {Wang}},\ and\ \bibinfo
		{author} {\bibfnamefont {H.-W.}\ \bibnamefont {Li}},\ }\bibfield  {title}
	{\bibinfo {title} {Reference-frame-independent quantum key distribution with
			advantage distillation},\ }\href {https://doi.org/10.1364/OL.480427}
	{\bibfield  {journal} {\bibinfo  {journal} {Opt. Lett.}\ }\textbf {\bibinfo
			{volume} {48}},\ \bibinfo {pages} {542} (\bibinfo {year}
		{2023}{\natexlab{b}})}\BibitemShut {NoStop}%
	\bibitem [{\citenamefont {Jiang}\ \emph {et~al.}(2023)\citenamefont {Jiang},
		\citenamefont {Wang}, \citenamefont {Li}, \citenamefont {Lu}, \citenamefont
		{Hao}, \citenamefont {Zhou},\ and\ \citenamefont {Bao}}]{2023Jiang}%
	\BibitemOpen
	\bibfield  {author} {\bibinfo {author} {\bibfnamefont {X.-L.}\ \bibnamefont
			{Jiang}}, \bibinfo {author} {\bibfnamefont {Y.}~\bibnamefont {Wang}},
		\bibinfo {author} {\bibfnamefont {J.-J.}\ \bibnamefont {Li}}, \bibinfo
		{author} {\bibfnamefont {Y.-F.}\ \bibnamefont {Lu}}, \bibinfo {author}
		{\bibfnamefont {C.-P.}\ \bibnamefont {Hao}}, \bibinfo {author} {\bibfnamefont
			{C.}~\bibnamefont {Zhou}},\ and\ \bibinfo {author} {\bibfnamefont {W.-S.}\
			\bibnamefont {Bao}},\ }\bibfield  {title} {\bibinfo {title} {Improving the
			performance of reference-frame-independent quantum key distribution with
			advantage distillation technology},\ }\href
	{https://doi.org/10.1364/OE.480570} {\bibfield  {journal} {\bibinfo
			{journal} {Opt. Express}\ }\textbf {\bibinfo {volume} {31}},\ \bibinfo
		{pages} {9196} (\bibinfo {year} {2023})}\BibitemShut {NoStop}%
	\bibitem [{\citenamefont {Li}\ \emph {et~al.}(2022)\citenamefont {Li},
		\citenamefont {Zhang}, \citenamefont {Jiang},\ and\ \citenamefont
		{Cai}}]{2022Li}%
	\BibitemOpen
	\bibfield  {author} {\bibinfo {author} {\bibfnamefont {H.-W.}\ \bibnamefont
			{Li}}, \bibinfo {author} {\bibfnamefont {C.-M.}\ \bibnamefont {Zhang}},
		\bibinfo {author} {\bibfnamefont {M.-S.}\ \bibnamefont {Jiang}},\ and\
		\bibinfo {author} {\bibfnamefont {Q.-Y.}\ \bibnamefont {Cai}},\ }\bibfield
	{title} {\bibinfo {title} {Improving the performance of practical decoy-state
			quantum key distribution with advantage distillation technology},\ }\href
	{https://doi.org/10.1103/10.1038/s42005-022-00831-4} {\bibfield  {journal}
		{\bibinfo  {journal} {Commun. Phys.}\ }\textbf {\bibinfo {volume} {5}},\
		\bibinfo {pages} {53} (\bibinfo {year} {2022})}\BibitemShut {NoStop}%
	\bibitem [{\citenamefont {Li}\ \emph {et~al.}(2023{\natexlab{c}})\citenamefont
		{Li}, \citenamefont {Wang}, \citenamefont {Zhang},\ and\ \citenamefont
		{Cai}}]{2022LITF}%
	\BibitemOpen
	\bibfield  {author} {\bibinfo {author} {\bibfnamefont {H.-W.}\ \bibnamefont
			{Li}}, \bibinfo {author} {\bibfnamefont {R.-Q.}\ \bibnamefont {Wang}},
		\bibinfo {author} {\bibfnamefont {C.-M.}\ \bibnamefont {Zhang}},\ and\
		\bibinfo {author} {\bibfnamefont {Q.-Y.}\ \bibnamefont {Cai}},\ }\bibfield
	{title} {\bibinfo {title} {Improving the performance of twin-field quantum
			key distribution with advantage distillation technology},\ }\href
	{https://doi.org/10.22331/q-2023-12-06-1201} {\bibfield  {journal} {\bibinfo
			{journal} {Quantum}\ }\textbf {\bibinfo {volume} {7}},\ \bibinfo {pages}
		{1201} (\bibinfo {year} {2023}{\natexlab{c}})}\BibitemShut {NoStop}%
	\bibitem [{\citenamefont {Wang}\ \emph
		{et~al.}(2022{\natexlab{b}})\citenamefont {Wang}, \citenamefont {Zhang},
		\citenamefont {Yin}, \citenamefont {Li}, \citenamefont {Wang}, \citenamefont
		{Chen}, \citenamefont {Guo},\ and\ \citenamefont {Han}}]{2022Wang_R}%
	\BibitemOpen
	\bibfield  {author} {\bibinfo {author} {\bibfnamefont {R.-Q.}\ \bibnamefont
			{Wang}}, \bibinfo {author} {\bibfnamefont {C.-M.}\ \bibnamefont {Zhang}},
		\bibinfo {author} {\bibfnamefont {Z.-Q.}\ \bibnamefont {Yin}}, \bibinfo
		{author} {\bibfnamefont {H.-W.}\ \bibnamefont {Li}}, \bibinfo {author}
		{\bibfnamefont {S.}~\bibnamefont {Wang}}, \bibinfo {author} {\bibfnamefont
			{W.}~\bibnamefont {Chen}}, \bibinfo {author} {\bibfnamefont {G.-C.}\
			\bibnamefont {Guo}},\ and\ \bibinfo {author} {\bibfnamefont {Z.-F.}\
			\bibnamefont {Han}},\ }\bibfield  {title} {\bibinfo {title} {Phase-matching
			quantum key distribution with advantage distillation},\ }\href
	{https://doi.org/https://doi.org/10.1088/1367-2630/ac8115} {\bibfield
		{journal} {\bibinfo  {journal} {New J. Phys.}\ }\textbf {\bibinfo {volume}
			{24}},\ \bibinfo {pages} {073049} (\bibinfo {year}
		{2022}{\natexlab{b}})}\BibitemShut {NoStop}%
	\bibitem [{\citenamefont {Zhang}\ \emph {et~al.}(2023)\citenamefont {Zhang},
		\citenamefont {Liu}, \citenamefont {Ding}, \citenamefont {Zhou},
		\citenamefont {Zhang},\ and\ \citenamefont {Wang}}]{2023zhang}%
	\BibitemOpen
	\bibfield  {author} {\bibinfo {author} {\bibfnamefont {K.}~\bibnamefont
			{Zhang}}, \bibinfo {author} {\bibfnamefont {J.}~\bibnamefont {Liu}}, \bibinfo
		{author} {\bibfnamefont {H.}~\bibnamefont {Ding}}, \bibinfo {author}
		{\bibfnamefont {X.}~\bibnamefont {Zhou}}, \bibinfo {author} {\bibfnamefont
			{C.}~\bibnamefont {Zhang}},\ and\ \bibinfo {author} {\bibfnamefont
			{Q.}~\bibnamefont {Wang}},\ }\bibfield  {title} {\bibinfo {title} {Asymmetric
			measurement-device-independent quantum key distribution through advantage
			distillation},\ }\href {https://doi.org/https://doi.org/10.3390/e25081174}
	{\bibfield  {journal} {\bibinfo  {journal} {Entropy}\ }\textbf {\bibinfo
			{volume} {25}},\ \bibinfo {pages} {1174} (\bibinfo {year}
		{2023})}\BibitemShut {NoStop}%
	\bibitem [{\citenamefont {Zhou}\ \emph {et~al.}(2024)\citenamefont {Zhou},
		\citenamefont {Wang}, \citenamefont {Zhang}, \citenamefont {Yin},
		\citenamefont {Wang}, \citenamefont {Wang}, \citenamefont {Chen},
		\citenamefont {Guo},\ and\ \citenamefont {Han}}]{2024ZhouYao}%
	\BibitemOpen
	\bibfield  {author} {\bibinfo {author} {\bibfnamefont {Y.}~\bibnamefont
			{Zhou}}, \bibinfo {author} {\bibfnamefont {R.-Q.}\ \bibnamefont {Wang}},
		\bibinfo {author} {\bibfnamefont {C.-M.}\ \bibnamefont {Zhang}}, \bibinfo
		{author} {\bibfnamefont {Z.-Q.}\ \bibnamefont {Yin}}, \bibinfo {author}
		{\bibfnamefont {Z.-H.}\ \bibnamefont {Wang}}, \bibinfo {author}
		{\bibfnamefont {S.}~\bibnamefont {Wang}}, \bibinfo {author} {\bibfnamefont
			{W.}~\bibnamefont {Chen}}, \bibinfo {author} {\bibfnamefont {G.-C.}\
			\bibnamefont {Guo}},\ and\ \bibinfo {author} {\bibfnamefont {Z.-F.}\
			\bibnamefont {Han}},\ }\bibfield  {title} {\bibinfo {title}
		{Sending-or-not-sending twin-field quantum key distribution with advantage
			distillation},\ }\href {https://doi.org/10.1103/PhysRevApplied.21.014036}
	{\bibfield  {journal} {\bibinfo  {journal} {Phys. Rev. Appl.}\ }\textbf
		{\bibinfo {volume} {21}},\ \bibinfo {pages} {014036} (\bibinfo {year}
		{2024})}\BibitemShut {NoStop}%
	\bibitem [{\citenamefont {Li}\ and\ \citenamefont {Wei}(2022)}]{2022Lizijian}%
	\BibitemOpen
	\bibfield  {author} {\bibinfo {author} {\bibfnamefont {Z.}~\bibnamefont
			{Li}}\ and\ \bibinfo {author} {\bibfnamefont {K.}~\bibnamefont {Wei}},\
	}\bibfield  {title} {\bibinfo {title} {Improving parameter optimization in
			decoy-state quantum key distribution},\ }\href
	{https://doi.org/10.1155/2022/9717591} {\bibfield  {journal} {\bibinfo
			{journal} {Quantum Eng.}\ }\textbf {\bibinfo {volume} {2022}},\ \bibinfo
		{pages} {9717591} (\bibinfo {year} {2022})}\BibitemShut {NoStop}%
	\bibitem [{\citenamefont {Zhang}\ \emph {et~al.}(2024)\citenamefont {Zhang},
		\citenamefont {Wang}, \citenamefont {Wu}, \citenamefont {Zhu}, \citenamefont
		{Wang},\ and\ \citenamefont {Li}}]{2024zhang}%
	\BibitemOpen
	\bibfield  {author} {\bibinfo {author} {\bibfnamefont {C.-M.}\ \bibnamefont
			{Zhang}}, \bibinfo {author} {\bibfnamefont {Z.}~\bibnamefont {Wang}},
		\bibinfo {author} {\bibfnamefont {Y.-D.}\ \bibnamefont {Wu}}, \bibinfo
		{author} {\bibfnamefont {J.-R.}\ \bibnamefont {Zhu}}, \bibinfo {author}
		{\bibfnamefont {R.}~\bibnamefont {Wang}},\ and\ \bibinfo {author}
		{\bibfnamefont {H.-W.}\ \bibnamefont {Li}},\ }\bibfield  {title} {\bibinfo
		{title} {Discrete-phase-randomized twin-field quantum key distribution with
			advantage distillation},\ }\href
	{https://doi.org/10.1103/PhysRevA.109.052432} {\bibfield  {journal} {\bibinfo
			{journal} {Phys. Rev. A}\ }\textbf {\bibinfo {volume} {109}},\ \bibinfo
		{pages} {052432} (\bibinfo {year} {2024})}\BibitemShut {NoStop}%
	\bibitem [{\citenamefont {Curty}\ \emph {et~al.}(2014)\citenamefont {Curty},
		\citenamefont {Xu}, \citenamefont {Cui}, \citenamefont {Lim}, \citenamefont
		{Tamaki},\ and\ \citenamefont {Lo}}]{2014curty}%
	\BibitemOpen
	\bibfield  {author} {\bibinfo {author} {\bibfnamefont {M.}~\bibnamefont
			{Curty}}, \bibinfo {author} {\bibfnamefont {F.}~\bibnamefont {Xu}}, \bibinfo
		{author} {\bibfnamefont {W.}~\bibnamefont {Cui}}, \bibinfo {author}
		{\bibfnamefont {C.~C.~W.}\ \bibnamefont {Lim}}, \bibinfo {author}
		{\bibfnamefont {K.}~\bibnamefont {Tamaki}},\ and\ \bibinfo {author}
		{\bibfnamefont {H.-K.}\ \bibnamefont {Lo}},\ }\bibfield  {title} {\bibinfo
		{title} {Finite-key analysis for measurement-device-independent quantum key
			distribution},\ }\href {https://doi.org/10.1038/ncomms4732} {\bibfield
		{journal} {\bibinfo  {journal} {Nat. Commun.}\ }\textbf {\bibinfo {volume}
			{5}},\ \bibinfo {pages} {3732} (\bibinfo {year} {2014})}\BibitemShut
	{NoStop}%
	\bibitem [{\citenamefont {Tomamichel}\ \emph {et~al.}(2011)\citenamefont
		{Tomamichel}, \citenamefont {Schaffner}, \citenamefont {Smith},\ and\
		\citenamefont {Renner}}]{2011_tomamichel}%
	\BibitemOpen
	\bibfield  {author} {\bibinfo {author} {\bibfnamefont {M.}~\bibnamefont
			{Tomamichel}}, \bibinfo {author} {\bibfnamefont {C.}~\bibnamefont
			{Schaffner}}, \bibinfo {author} {\bibfnamefont {A.}~\bibnamefont {Smith}},\
		and\ \bibinfo {author} {\bibfnamefont {R.}~\bibnamefont {Renner}},\
	}\bibfield  {title} {\bibinfo {title} {Leftover hashing against quantum side
			information},\ }\href {https://doi.org/10.1109/tit.2011.2158473} {\bibfield
		{journal} {\bibinfo  {journal} {IEEE Trans. Inf. Theory}\ }\textbf {\bibinfo
			{volume} {57}},\ \bibinfo {pages} {5524} (\bibinfo {year}
		{2011})}\BibitemShut {NoStop}%
	\bibitem [{\citenamefont {Tomamichel}\ \emph {et~al.}(2012)\citenamefont
		{Tomamichel}, \citenamefont {Lim}, \citenamefont {Gisin},\ and\ \citenamefont
		{Renner}}]{2012tomamichel}%
	\BibitemOpen
	\bibfield  {author} {\bibinfo {author} {\bibfnamefont {M.}~\bibnamefont
			{Tomamichel}}, \bibinfo {author} {\bibfnamefont {C.~C.~W.}\ \bibnamefont
			{Lim}}, \bibinfo {author} {\bibfnamefont {N.}~\bibnamefont {Gisin}},\ and\
		\bibinfo {author} {\bibfnamefont {R.}~\bibnamefont {Renner}},\ }\bibfield
	{title} {\bibinfo {title} {Tight finite-key analysis for quantum
			cryptography},\ }\href {https://doi.org/10.1109/tit.2013.2238656} {\bibfield
		{journal} {\bibinfo  {journal} {Nat. Commun.}\ }\textbf {\bibinfo {volume}
			{3}},\ \bibinfo {pages} {634} (\bibinfo {year} {2012})}\BibitemShut {NoStop}%
	\bibitem [{\citenamefont {Vitanov}\ \emph {et~al.}(2013)\citenamefont
		{Vitanov}, \citenamefont {Dupuis}, \citenamefont {Tomamichel},\ and\
		\citenamefont {Renner}}]{2013Vitanov}%
	\BibitemOpen
	\bibfield  {author} {\bibinfo {author} {\bibfnamefont {A.}~\bibnamefont
			{Vitanov}}, \bibinfo {author} {\bibfnamefont {F.}~\bibnamefont {Dupuis}},
		\bibinfo {author} {\bibfnamefont {M.}~\bibnamefont {Tomamichel}},\ and\
		\bibinfo {author} {\bibfnamefont {R.}~\bibnamefont {Renner}},\ }\bibfield
	{title} {\bibinfo {title} {Chain rules for smooth min- and max-entropies},\
	}\href {https://doi.org/10.1109/tit.2013.2238656} {\bibfield  {journal}
		{\bibinfo  {journal} {IEEE Trans. Inf. Theory}\ }\textbf {\bibinfo {volume}
			{59}},\ \bibinfo {pages} {2603} (\bibinfo {year} {2013})}\BibitemShut
	{NoStop}%
	\bibitem [{\citenamefont {Tomamichel}\ and\ \citenamefont
		{Renner}(2011)}]{2011Tomamichel}%
	\BibitemOpen
	\bibfield  {author} {\bibinfo {author} {\bibfnamefont {M.}~\bibnamefont
			{Tomamichel}}\ and\ \bibinfo {author} {\bibfnamefont {R.}~\bibnamefont
			{Renner}},\ }\bibfield  {title} {\bibinfo {title} {Uncertainty relation for
			smooth entropies},\ }\href {https://doi.org/10.1103/PhysRevLett.106.110506}
	{\bibfield  {journal} {\bibinfo  {journal} {Phys. Rev. Lett.}\ }\textbf
		{\bibinfo {volume} {106}},\ \bibinfo {pages} {110506} (\bibinfo {year}
		{2011})}\BibitemShut {NoStop}%
	\bibitem [{\citenamefont {Yin}\ \emph {et~al.}(2020)\citenamefont {Yin},
		\citenamefont {Zhou}, \citenamefont {Gu}, \citenamefont {Xie}, \citenamefont
		{Lu},\ and\ \citenamefont {Chen}}]{2020Yin}%
	\BibitemOpen
	\bibfield  {author} {\bibinfo {author} {\bibfnamefont {H.-L.}\ \bibnamefont
			{Yin}}, \bibinfo {author} {\bibfnamefont {M.-G.}\ \bibnamefont {Zhou}},
		\bibinfo {author} {\bibfnamefont {J.}~\bibnamefont {Gu}}, \bibinfo {author}
		{\bibfnamefont {Y.-M.}\ \bibnamefont {Xie}}, \bibinfo {author} {\bibfnamefont
			{Y.-S.}\ \bibnamefont {Lu}},\ and\ \bibinfo {author} {\bibfnamefont {Z.-B.}\
			\bibnamefont {Chen}},\ }\bibfield  {title} {\bibinfo {title} {Tight security
			bounds for decoy-state quantum key distribution},\ }\href
	{https://doi.org/https://doi.org/10.1038/s41598-020-71107-6} {\bibfield
		{journal} {\bibinfo  {journal} {Sci. Rep.}\ }\textbf {\bibinfo {volume}
			{10}},\ \bibinfo {pages} {14312} (\bibinfo {year} {2020})}\BibitemShut
	{NoStop}%
\end{thebibliography}

%apsrev4-2.bst 2019-01-14 (MD) hand-edited version of apsrev4-1.bst
%Control: key (0)
%Control: author (8) initials jnrlst
%Control: editor formatted (1) identically to author
%Control: production of article title (0) allowed
%Control: page (0) single
%Control: year (1) truncated
%Control: production of eprint (0) enabled
%

        \end{document}